\documentclass[preprintnumbers,aps,superscriptaddress,nofootinbib,tightenlines,floatfix,11pt]{revtex4}

\usepackage{amsmath,amssymb}
\usepackage{graphicx}
\usepackage{bm}
\usepackage{comment}
\usepackage{color}
\usepackage{hyperref}
\usepackage{subfigure}
\usepackage{units}
\usepackage{lipsum}
\usepackage{natbib}

\newcommand{\gsim}{\raisebox{-0.7ex}{$\stackrel{\textstyle >}{\sim}$ }}
\newcommand{\lsim}{\raisebox{-0.7ex}{$\stackrel{\textstyle <}{\sim}$ }}

\def\pislash{ {\pi\hskip-0.6em /} }
\def\nopi{ {\rm EFT}(\pislash) }

\newcount\hour \newcount\hourminute \newcount\minute 
\hour=\time \divide \hour by 60
\hourminute=\hour \multiply \hourminute by 60
\minute=\time \advance \minute by -\hourminute
\newcommand{\mydate}{\ \today \ - \number\hour :\number\minute}

\begin{document}
\dimen\footins=5\baselineskip\relax

\preprint{\vbox{
\hbox{INT-PUB-14-006}
}}

\title{
Finite-Volume Electromagnetic Corrections to the Masses of
\\
Mesons, Baryons and Nuclei}

\author{Zohreh Davoudi{\footnote{\tt davoudi@uw.edu}}
}
\affiliation{Department of Physics, University of Washington,
 Box 351560, Seattle, WA 98195, USA}
\affiliation{Institute for Nuclear Theory, Box 351550, Seattle, WA 98195-1550, USA}

\author{Martin J. Savage{\footnote{\tt mjs5@uw.edu}}}
\affiliation{Department of Physics, University of Washington,
 Box 351560, Seattle, WA 98195, USA}
\affiliation{Institute for Nuclear Theory, Box 351550, Seattle, WA 98195-1550, USA}

\date{\mydate}

\begin{abstract}
\noindent
Now that Lattice QCD calculations are beginning to include QED, it is important to 
better understand how hadronic properties are modified by  finite-volume QED effects. 
They are known to exhibit power-law scaling with volume, in contrast to the exponential 
behavior of finite-volume strong interaction effects.
We use non-relativistic effective field theories describing the low-momentum behavior of 
hadrons
to determine the finite-volume QED corrections to the masses of  mesons, baryons and nuclei
out to ${\cal O}\left(1/{\rm L}^4\right)$ in a volume expansion, where ${\rm L}$ is the spatial extent of the cubic volume.
This generalizes the previously determined  expansion for mesons, and 
extends it  by two orders in $1/{\rm L}$ to
include contributions from the 
charge radius, magnetic moment and polarizabilities of the hadron.
We make an observation about direct calculations of the muon $g-2$ in a finite volume.
\end{abstract}
\maketitle

\section{Introduction}
\noindent
Lattice Quantum Chromodynamics (LQCD) 
has matured to the point where basic properties of the light hadrons are being calculated at the physical 
pion mass~ \cite{Aoki:2009ix, Durr:2010vn, Arthur:2012opa, Aoki:2012st, Durr:2013goa}.
In some instances, the up- and down-quark masses and quenched Quantum Electrodynamics (QED) have been included in an effort to 
precisely postdict the observed isospin splittings in the 
spectrum of hadrons~\cite{Blum:2007cy,Basak:2008na,Blum:2010ym,Portelli:2010yn,Portelli:2012pn,Aoki:2012st,deDivitiis:2013xla,Borsanyi:2013lga,Drury:2013sfa, Borsanyi:2014jba}.
While naively appearing to be a simple extension of pure LQCD calculations,
there are subtleties associated with including
QED. 
In particular, Gauss's law and Ampere's law cannot be satisfied 
when the electromagnetic gauge field is subject to periodic boundary conditions (PBCs)~\cite{Hilf1983412,Duncan:1996xy,Hayakawa:2008an}. 
However, a uniform background charge density can be introduced to circumvent this problem and  restore these laws. 
This is equivalent to removing the zero modes of the photon in a finite-volume (FV) calculation,
which does not change the infinite-volume value of calculated quantities.
One-loop level calculations in chiral perturbation theory ($\chi$PT)
and partially-quenched $\chi$PT ($PQ\chi$PT) have been performed~\cite{Hayakawa:2008an}
to determine the leading FV modifications to the mass of  mesons induced by constraining QED to a cubic volume subject to 
PBCs.~\footnote{
Vector dominance~\cite{PhysRevLett.62.1343}
has been previously used to model the low-momentum contributions to the
FV electromagnetic mass splittings of the pseudo-scalar 
mesons, see Refs. \cite{Duncan:1996xy, Blum:2007cy}.
}
Due to the photon being massless, the FV QED 
corrections to the mass of the $\pi^+$ are predicted to be an expansion in powers of the volume, 
and have been determined to be of the form 
$\delta m_{\pi^+}\sim 1/{\rm L} + 2/(m_{\pi^+} {\rm L}^2) + \cdots $,
where ${\rm L}$ is the spatial extent of the cubic volume.
As the spatial extents of present-day gauge-field configurations at the physical pion mass are not large, with $m_\pi {\rm L}\lsim 4$,
the exponentially suppressed strong interaction FV effects,  ${\cal O}\left( e^{- m_\pi {\rm L}}\right)$,  
are not negligible for precision studies of hadrons, and 
when QED is included, the power-law corrections, although suppressed by $\alpha_e$, are expected to be important, 
particularly in mass splittings.

In this work, we return to the issue of calculating  FV QED effects, and show that non-relativistic effective field theories (NREFTs)
provide a straightforward way to calculate such corrections to the properties of hadrons.  
With these EFTs, the FV mass shift of  
mesons, baryons and nuclei are calculated 
out to ${\cal O}\left(1/{\rm L}^4\right)$ in the 
$1/{\rm L}$ expansion, 
including contributions from their charge radii, magnetic moments and polarizabilities.
The NREFTs have the advantage that the coefficients of  operators coupling to the electromagnetic field 
are directly related,
order by order in the $\alpha_e$,
to the electromagnetic moments of the hadrons (in the continuum limit), 
as opposed to a 
perturbative estimate thereof (as is the case in $\chi$PT).
For protons and neutrons, the NREFT is the well-established 
non-relativistic QED (NRQED)~\cite{Isgur:1989vq,Isgur:1989ed,Jenkins:1990jv,Jenkins:1991ne,Thacker:1990bm, Labelle:1992hd,Manohar:1997qy,Luke:1997ys, Hill:2011wy},  
modified to include the finite extent of the charge and current densities~\cite{Chen:1999tn}.
Including multi-nucleon interactions, this framework has been 
used extensively to describe the low-energy behavior of nucleons and nuclear interactions, $\nopi$,
along with their interactions with electromagnetic fields~\cite{Kaplan:1998tg, Kaplan:1998we, Chen:1999tn, Butler:1999sv, Butler:2000zp, Butler:2001jj},
and is straightforwardly generalized to hadrons and nuclei with arbitrary angular momentum.
LQCD calculations performed with background electromagnetic fields  are currently making use of these NREFTs 
to extract  the properties of hadrons, including magnetic moments and polarizabilities \cite{Martinelli:1982cb, Fiebig:1988en, Bernard:1982yu, Lee:2005ds, Christensen:2004ca, Lee:2005dq, Engelhardt:2007ub, Detmold:2009dx, Alexandru:2009id, Detmold:2010ts, Primer:2013pva, Lee:2013lxa}.

\section{Finite-Volume QED}
\noindent
The issues complicating the inclusion of QED in FV calculations 
with PBCs are well 
known,
the most glaring of which is the inability to preserve Gauss's law~ \cite{Duncan:1996xy,Blum:2007cy,Hayakawa:2008an}, 
which relates the electric flux penetrating 
any closed surface to the charge enclosed by the surface,  and Ampere's Law, which relates the integral of the 
magnetic field around a closed loop to the current penetrating the loop.
An obvious way to see the problem is to consider the electric field along the axes of the cubic volume 
(particularly at the surface) associated with a point charge at the center.
Restating the discussions of Ref.~\cite{Hayakawa:2008an}, the variation of the QED action is,
for a fermion of charge $e Q$,
\begin{eqnarray}
\delta S & = & 
\int\ d^4x\ 
\left[ \ 
\partial_\mu F^{\mu\nu} (x)
\ -\ 
e\ Q\ \overline{\psi} (x)\gamma^\nu \psi  (x)
\ \right]
\ \delta \left(A_\nu (x)\right)
\nonumber\\
& = & 
\int  dt\ 
{1\over {\rm L}^3}\ \sum_{\bf q}\ 
\delta\left(\tilde A_\nu (t,{\bf q})\right)
\int_{{\rm L}^3}\ d^3 {\bf x}\ 
e^{i{\bf q}\cdot  {\bf x}}\ 
\left[\ 
\partial_\mu F^{\mu\nu} (t,{\bf x})
\ -\ e\ Q\ \overline{\psi}  (t,{\bf x})\gamma^\nu \psi  (t,{\bf x})
\ \right]
\ ,
\label{eq:gauss}
\end{eqnarray}
where 
$\tilde A_\nu (t,{\bf q})$ is the spatial Fourier transform of $A_\nu (t,{\bf x})$, and
$e=|e|$ is the magnitude of the electronic charge.
For simplicity, here and in what follows, 
we assume the time direction of the FV to be infinite~\footnote{
In practice, there are thermal effects in LQCD calculations due to the finite extent of the time direction.}
 while the spatial directions are of length ${\rm L}$.
Eq. (\ref{eq:gauss}) leads to 
$\partial_\mu F^{\mu\nu} = e Q \overline{\psi} \gamma^\nu \psi $ 
for 
$\delta S=0$ and hence Gauss's Law and Ampere's Law. 
This can be modified to $\partial_\mu F^{\mu\nu} = e Q \overline{\psi} \gamma^\nu \psi  + b^\nu$ simply by omitting the 
spatial zero modes of 
$A_\mu$, i.e.  $\tilde A_\nu (t,{\bf 0}) = 0$, 
or more generally by setting $\delta \tilde A_\nu (t,{\bf 0}) = 0$,
where $b^\nu$ is some uniform background charge 
distribution~\cite{Portelli:2010yn}.~\footnote{
The introduction of a uniformly charged background is a technique that has 
been used extensively to include electromagnetic interactions into calculations of many-body systems, 
such as nuclear matter and condensed matter, see for example Ref.~\cite{2000physics..12024C}.
} 
This readily eliminates the relation between the electric flux penetrating a closed surface and the inserted charge,
and the analogous relation between the magnetic field and 
current.~\footnote{For a discussion about including QED with C-PBCs (anti-PBCs), see Ref.~\cite{Kronfeld:1992ae}.}
Ensuring this constraint is preserved under gauge transformations, 
$A_\mu (t,{\bf x})\rightarrow A^\prime_\mu (t,{\bf x})  = A_\mu (t,{\bf x}) + \partial_\mu \Lambda(t,{\bf x})$, where 
$\Lambda$ is a periodic function in the spatial volume,
requires
$ \partial_0 \tilde\Lambda(t,{\bf 0})=0$, where $\tilde\Lambda (t,{\bf q})$ is the Fourier transform of $\Lambda (t,{\bf x})$.
Modes with ${\bf q}\ne {\bf 0}$ are subject to the standard gauge-fixing conditions, and in LQCD calculations it is
sometimes convenient to work in Coulomb gauge,
${\bm\nabla}\cdot {\bf A}=0$.
This is because of the asymmetry between the spatial and temporal directions that is present in most ensembles of gauge field configurations,
along with the fact that the photon fields are generated in momentum space as opposed to position space \cite{Blum:2007cy}.

In infinite volume, the Coulomb potential energy between charges $eQ$ is well known to be 
$U(r) = \frac{\alpha_e Q^2}{r}$, where $\alpha_e=e^2/4\pi$ is the QED fine-structure constant,
while in a cubic spatial volume 
with the zero modes removed, it is
\begin{eqnarray} 
U( {\bf r} ,{\rm L}) 
& = &  
{\alpha_e Q^2 \over \pi {\rm L}}
\sum_{{\bf n}\ne {\bf 0}}
{1\over |{\bf n}|^2}
e^{i 2\pi {\bf n}\cdot {\bf r}\over {\rm L}}
\nonumber\\
& = & 
{\alpha_e Q^2\over \pi {\rm L}}
\left[-1 + 
\sum_{{\bf n}\ne {\bf 0}}
{e^{-|{\bf n}|^2}\over |{\bf n}|^2}
e^{i 2\pi {\bf n}\cdot {\bf r}\over {\rm L}}
+
\sum_{\bf p}\ \int_0^1\ dt\ \left({\pi\over t}\right)^{3/2}
e^{ -  {\pi^2 |{\bf p}-{\bf r}/{\rm L} |^2\over t}  }
\ \right]
\ \ \ ,
\label{eq:Vgreen}
\end{eqnarray}
where 
${\bf n}$ and $\mathbf{p}$ are triplets of integers.
The latter,
exponentially accelerated,  expression in Eq.~(\ref{eq:Vgreen}) is obtained from the former using the Poisson summation formula.
\begin{figure}[t]
  \centering
     \includegraphics[scale=0.55]{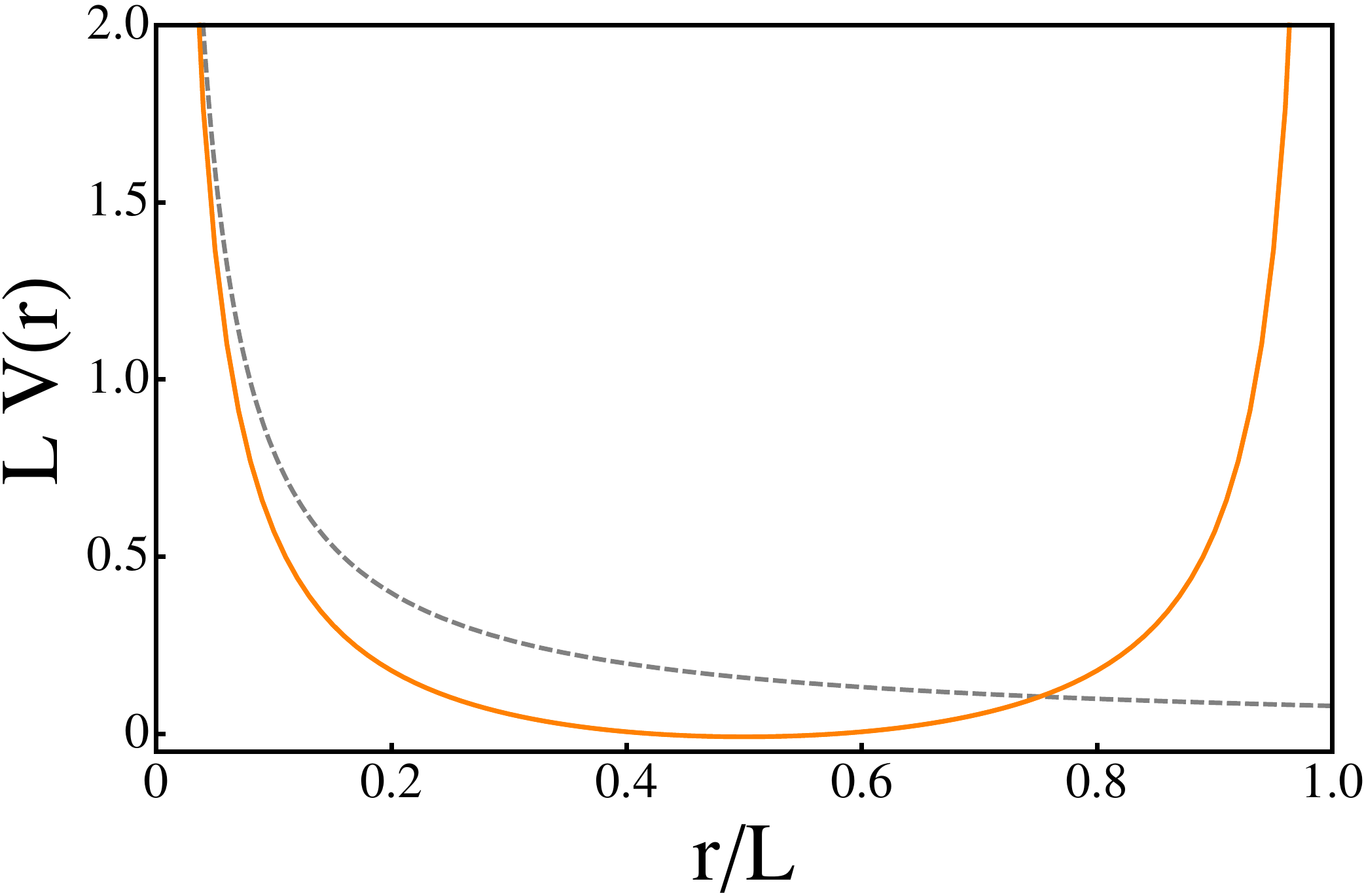}     
     \caption{The FV potential energy between two charges with $Qe=1$ along one of the axes of a cubic volume of spatial extent ${\rm L}$ (solid orange curve), 
     obtained from Eq.~(\protect\ref{eq:Vgreen}), and the corresponding infinite-volume Coulomb potential energy (dashed gray curve).
       }
  \label{fig:pot}
\end{figure}
The  FV potential energy between two charges with $Qe  = 1$, and the
corresponding infinite-volume Coulomb potential energy are shown in Fig.~\ref{fig:pot}.

In the next sections, we construct non-relativistic EFTs to allow for order-by-order calculations of the FV QED modifications
to the energy of hadrons in 
the continuum limit of
LQCD calculations, going beyond the first two orders in the $1/{\rm L}$ expansion that have 
been determined previously.
While these EFTs permit calculations to any given precision, including quantum fluctuations, some of 
the results that will be presented can be determined simply without the EFTs; 
a demonstration of which is the self-energy of a uniformly charged, rigid and fixed,  sphere in a FV. 
In this textbook case, the self-energy can be determined directly by integrating the interaction 
between infinitesimal volumes of the charge density, as governed by the modified Coulomb potential, Eq.~(\ref{eq:Vgreen}), 
over the sphere of radius $R$.  It is straightforward to show that the self-energy can be written in an expansion of $R/{\rm L}$,
\begin{eqnarray}
\label{ChargedSphere}
U^{\rm sphere} (R,{\rm L}) & = & 
{3\over 5} {(Qe)^2\over 4 \pi R}\ 
\ +\ 
{(Qe)^2\over 8\pi {\rm L}}\ c_1
\ +\ 
{(Qe)^2\over 10 {\rm L}} \left({R\over {\rm L}}\right)^2\ 
 +\ \cdots
\ \ \ ,
\end{eqnarray}
where $c_1 =-2.83729$~\cite{Luscher:1986pf, Hasenfratz:1989pk, Luscher:1990ux}. 
The leading contribution is the well-known result for a uniformly charged sphere,
while the second term, the 
leading order (LO) FV correction, is independent of the structure of the charge distribution. 
This suggests that it is also valid for a point particle; a result that proves to be valid 
for the corrections to the masses of single particles calculated with $\chi$PT and with the 
NREFTs presented in this work. 
It is simply the modification to the Coulomb self-energy of a point charge. 
The third term can be written as ${(Qe)^2} \langle r^2\rangle /  6 {\rm L}^3$, 
where $ \langle r^2\rangle=\frac{3}{5}R^2$ is the mean-squared radius of the sphere, and 
reproduces the charge-radius contributions determined with the NREFTs, 
as will be shown in the next section.

\section{ Scalar NRQED for Mesons and $J=0$ Nuclei}
\noindent
LQCD calculations including QED have been largely 
focused on the masses of the pions and kaons in an effort to extract the values of electromagnetic counterterms of $\chi$PT, thus
we begin by considering the FV corrections to the masses of scalar hadrons.
In the limit where the volume of space is much larger than that of the hadron, keeping in mind that only the zero modes are being excluded from the photon fields, 
the FV corrections to the mass of the hadron will have a power-law dependence upon ${\rm L}$, and  vanish as ${\rm L}\rightarrow\infty$.
As the modifications to the self-energy arise from the infrared behavior of the theory, 
low-energy EFT provides a tool 
with which to systematically determine the FV effects in an expansion in one or more small parameters.

Using the methods developed to describe heavy-quark and heavy-hadron 
systems~\cite{Isgur:1989vq,Isgur:1989ed,Jenkins:1990jv,Jenkins:1991ne,Thacker:1990bm,Labelle:1992hd,Manohar:1997qy,Luke:1997ys,Chen:1999tn,Beane:2007es,Lee:2013lxa},
the Lagrange density describing the low-energy dynamics of a charged composite scalar particle, $\phi$, with charge $eQ$ 
can be written as an expansion in $1/m_{\phi}$ and in the  scale of compositeness,
\begin{eqnarray}
{\cal L}_\phi
& = & 
\phi^\dagger \left[\ 
iD_0
\ +\ {|{\bf D}|^2\over 2 m_\phi}
\ + \ { |{\bf D}|^4 \over 8 m_\phi^3}\ 
\ +\   {e \langle r^2\rangle_\phi\over 6}\ {\bm\nabla}\cdot {\bf E}
\ +\ 2 \pi \tilde\alpha_E^{(\phi)}  |{\bf E}|^2
\ +\ 2\pi \tilde\beta_M^{(\phi)}  |{\bf B}|^2
\right.\nonumber\\
&&\left.
\qquad \qquad 
\ +\ i  e  c_M\ { \{ D^i , ({\bm\nabla}\times {\bf B})^i \} \over  8 m_\phi^3}
\ +\  \cdots
\ \right] \phi
\ ,
\label{eq:scalarLag}
\end{eqnarray}
where $m_{\phi}$ is the mass of the particle, the covariant derivative is $D_\mu = \partial_\mu + i e \hat Q A_\mu$ with $\hat Q$ the charge operator.
$\langle r^2\rangle_\phi$ is the mean-squared charge radius of the $\phi$,
and we have performed the standard field redefinition to the NR normalization of states,
$\phi\rightarrow \phi/\sqrt{2 m_\phi}$.
The remaining coefficients of operators involving the 
electric, ${\bf E}$, and magnetic, ${\bf B}$, fields,
have been determined by matching this EFT to  scalar QED, to yield
\begin{eqnarray}
\tilde\alpha_E^{(\phi)}  & = & 
\alpha_E^{(\phi)}  -  {\alpha_e Q\over 3 m_\phi} \langle r^2\rangle_\phi 
\ \ ,\ \ 
\tilde\beta_M^{(\phi)}  \ = \
\beta_M^{(\phi)}  
\ \ ,\ \ 
c_M\ =\  {2\over 3} m_\phi^2 \langle r^2\rangle_\phi 
\ ,
\label{eq:scalarpols}
\end{eqnarray}
where $\alpha_E^{(\phi)} , \beta_M^{(\phi)}$ are the electric and magnetic polarizabilities of the 
$\phi$.~\footnote{
The presence of a charge-radius dependent term in the coefficient of the 
electric polarizability  indicates a subtlety in using this EFT  to describe hadrons in a 
background electric field~\cite{Lee:2013lxa}. 
Such contributions can be canceled by including redundant operators in the EFT Lagrange density 
when matching to S-matrix elements. 
Since a classical uniform electric field modifies the equations of motion, such operators must  be retained in the 
Lagrange density and their coefficients matched directly to Green functions. 
}
These coefficients will be modified at higher orders in perturbation theory, 
starting at $\mathcal{O}(\alpha_e)$.
They will also be modified by terms that are exponentially suppressed by  compositeness length scales, e.g. 
$\sim e^{-m_\pi {\rm L}}$ for QCD.
The ellipses denote terms that are higher order in 
derivatives acting on the  fields, with  coefficients dictated by the mass and compositeness scale 
-- the chiral symmetry breaking scale, $\Lambda_\chi$, for mesons and baryons.
For  one-body observables,  terms beyond 
$\phi^\dagger i \partial_0 \phi$ are treated in perturbation theory, providing  a systematic expansion in $1/{\rm L}$.

\begin{figure}[!ht]
  \centering
     \includegraphics[scale=0.230]{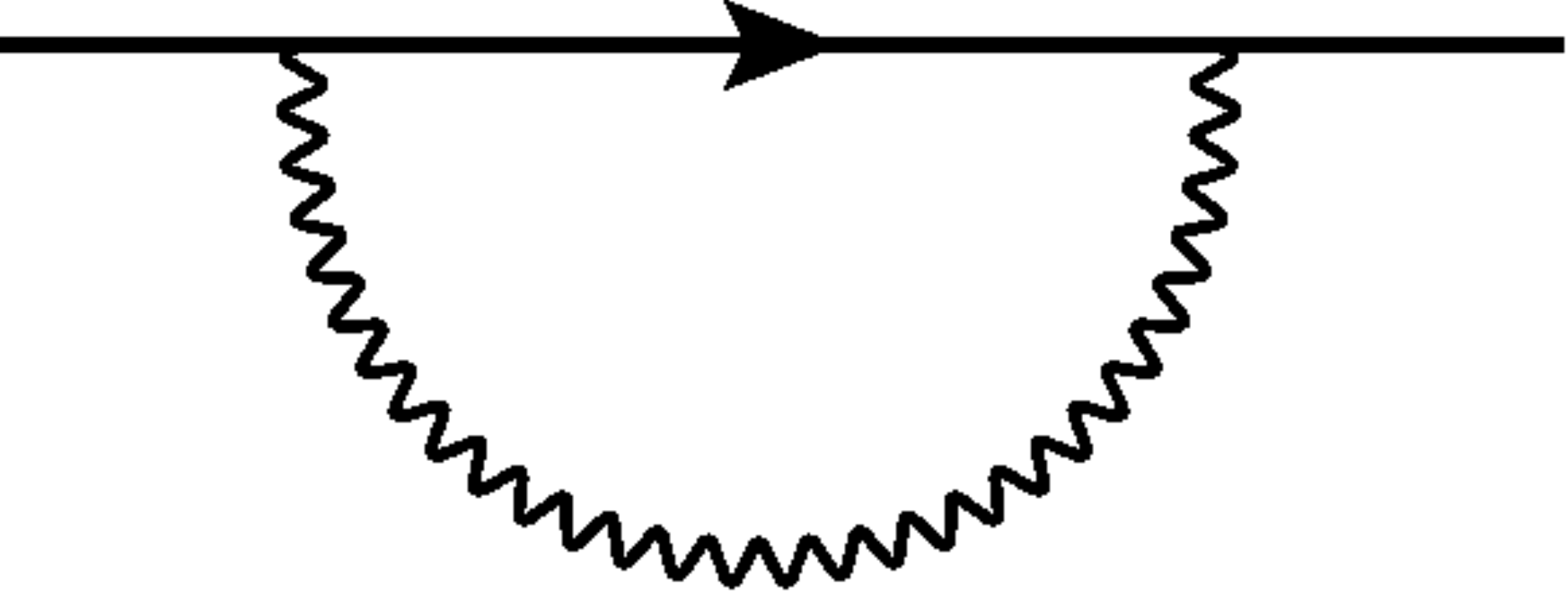}     
     \caption{The one-loop diagram providing the LO, $\mathcal{O}(\alpha_e/{\rm L})$, 
     FV correction to the mass of a charged scalar particle. 
     The solid straight line denotes a scalar particle, while the wavy line denotes a photon.
     }
  \label{fig:LO}
\end{figure}
The LO, $\mathcal{O}(\alpha_e/{\rm L})$, 
correction to the 
mass of a charged scalar particle in FV, $\delta m_\phi$,
is from  the one-loop diagram shown in 
Fig.~\ref{fig:LO}.
While most simply calculated in Coulomb gauge, the diagram can be calculated in any gauge and,
in agreement with previous determinations~\cite{Hayakawa:2008an},   is
\begin{eqnarray}
\delta m_\phi^{({\rm LO})}
& = & 
 {\alpha_e Q^2\over 2\pi {\rm L}}\ 
 \hat{\sum_{ {\bf n}\ne {\bf 0}}}\ 
 {1\over |{\bf n}|^2}
\ =\ 
{\alpha_e Q^2\over 2 {\rm L}}\ c_1
  \ \ \ ,
\label{eq:scalarLO}
\end{eqnarray}
with $c_1 = -2.83729$. 
The sum, $\hat{\sum}$, represents the difference between the sum over the FV modes and the 
infinite-volume integral, e.g.,
\begin{eqnarray}
{1\over {\rm L}^3}\hat{\sum_{{\bf k}\ne {\bf 0}}}\ 
f({\bf k})
&\equiv &
{1\over {\rm L}^3}\sum_{{\bf k}\ne {\bf 0}}\ f({\bf k})
\ -\ 
\int {d^3{\bf k}\over (2\pi)^3}\ f({\bf k})
  \ \ \ ,
\label{eq:FVsumint}
\end{eqnarray}
for an arbitrary function $f({\bf k})$,
and is therefore finite.
This shift is a power law in $1/{\rm L}$ as expected, and provides a reduction in the mass of the hadron.
As the infinite-volume Coulomb interaction increases the mass, and the FV result is obtained 
from the modes that satisfy the PBCs (minus the zero modes), the sign of the correction is also expected.
The result in Eq.~(\ref{eq:scalarLO}) is nothing more than the difference between the FV and infinite-volume 
contribution to the Coulomb self-energy of a charged point particle, as seen from 
Eq.~(\ref{eq:Vgreen}), $U({\bf 0},{\rm L})/2$.

\begin{figure}[!ht]
\begin{center}
\subfigure[]{
\includegraphics[scale=0.25]{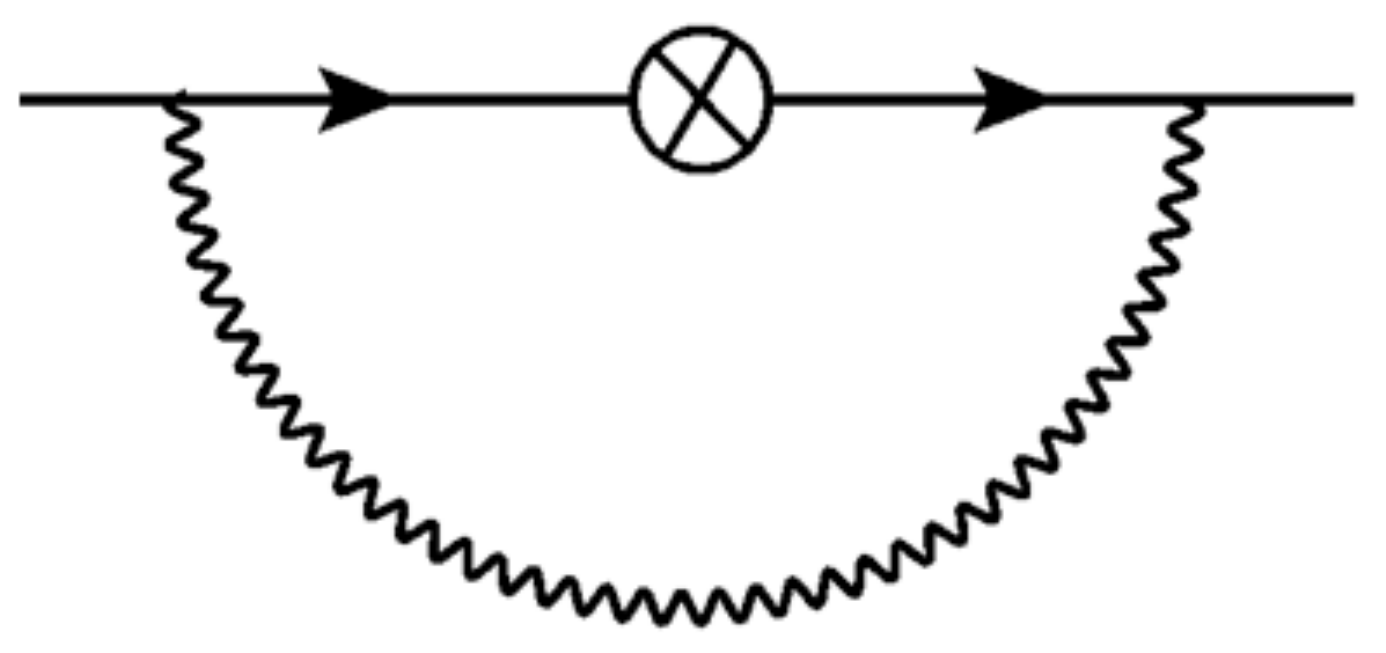}}
\subfigure[]{
\includegraphics[scale=0.23]{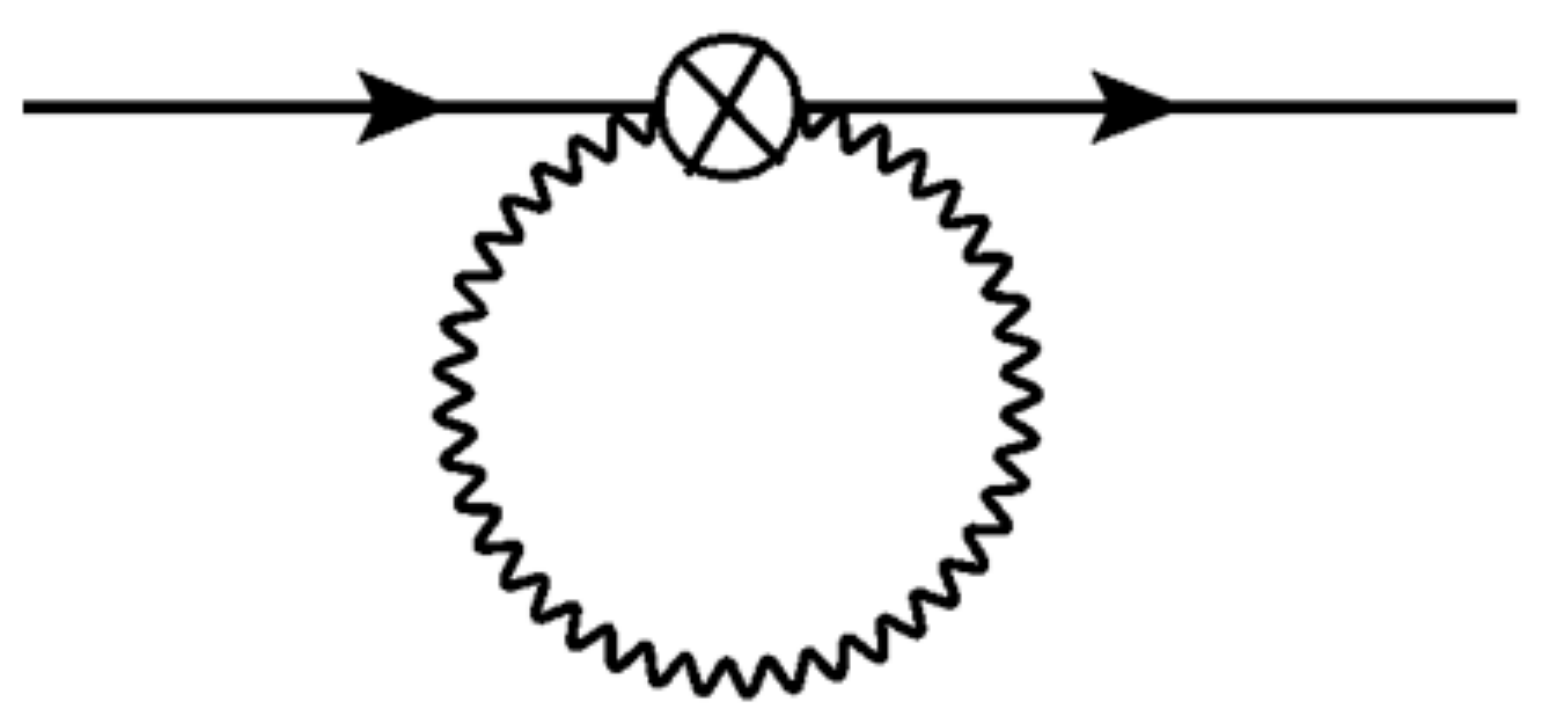}}
\subfigure[]{
\includegraphics[scale=0.25]{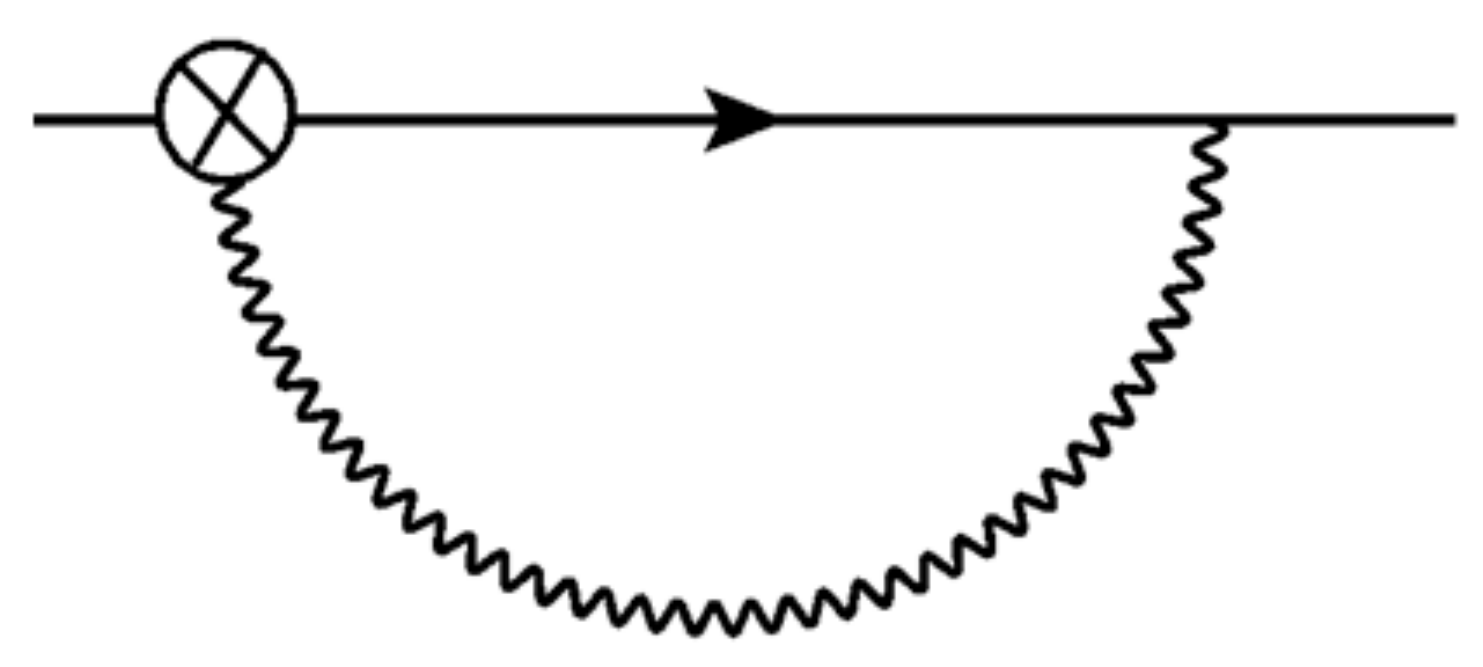}}
\subfigure[]{
\includegraphics[scale=0.25]{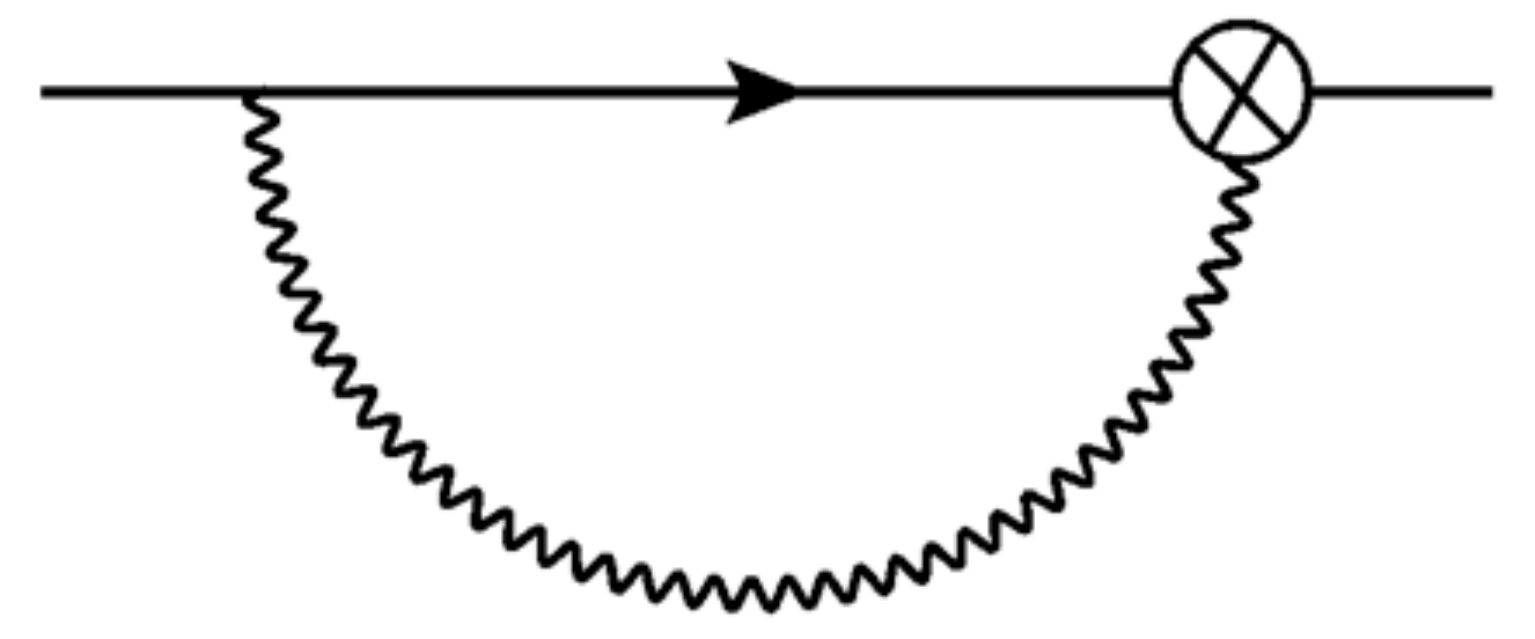}}
\caption{Diagrams contributing at NLO,
$\mathcal{O}(\alpha_e/m_{\phi}{\rm L}^2)$, in the ${1/ {\rm L}}$ expansion.
The crossed circle denotes an insertion of  the $|{\bf D}|^2/2 m_\phi$ 
      operator in the scalar QED Lagrange density, Eq.~(\protect\ref{eq:scalarLag}).
}
\label{fig:mall}
\end{center}
\end{figure}
The next-to-LO (NLO) contribution,  ${\cal O}\left(\alpha_e/{\rm L}^2\right)$, 
arises from a single insertion of the 
$|{\bf D}|^2/2 m_\phi$ operator in Eq.~(\protect\ref{eq:scalarLag}) into the one-loop diagrams shown in Fig.~\ref{fig:mall}.
The contribution from each of these diagrams depends upon the choice of gauge, however the sum is gauge independent,~\footnote{
The sums appearing at LO and NLO are 
\begin{eqnarray}
 \hat{\sum_{ {\bf n}\ne {\bf 0}}}\  {1\over |{\bf n}|} & = & c_1
 \ \ ,\ \ 
  \hat{\sum_{ {\bf n}\ne {\bf 0}}}\  {1\over |{\bf n}|^2} \ = \ \pi\  c_1
  \ \ \ .
  \nonumber
\end{eqnarray}
} 
\begin{eqnarray}
\delta m_\phi^{({\rm NLO})}
& = & 
 {\alpha_e Q^2\over m_\phi {\rm L}^2 }\ 
 \hat{\sum_{{\bf n}\ne {\bf 0}}}\ 
 {1\over |{\bf n}|}
\ =\ 
{\alpha_e Q^2\over m_\phi {\rm L}^2 }\ c_1
  \ \ \ .
\label{eq:scalarNLO}
\end{eqnarray}
This NLO recoil correction agrees with previous calculations~\cite{Hayakawa:2008an,deDivitiis:2013xla},
and is the highest order in the $1/{\rm L}$ expansion 
to which these FV effects have been previously 
determined.~\footnote{
The $\mathcal{O}(\alpha_e)$ calculations of Ref.~\cite{Hayakawa:2008an} at NLO in $\chi$PT and PQ$\chi$PT  do not 
include the full contributions from the meson charge radius and polarizabilities, but are perturbatively close. 
This is in contrast to the NREFT calculations presented in this work
where the low-energy coefficients are matched to these quantities 
order by order in  $ \alpha_e$,
and provide the  result at any given order in  $1/{\rm L}$  as an expansion in $ \alpha_e$.
}

\begin{figure}[!ht]
\begin{center}
\subfigure[]{
\includegraphics[scale=0.22]{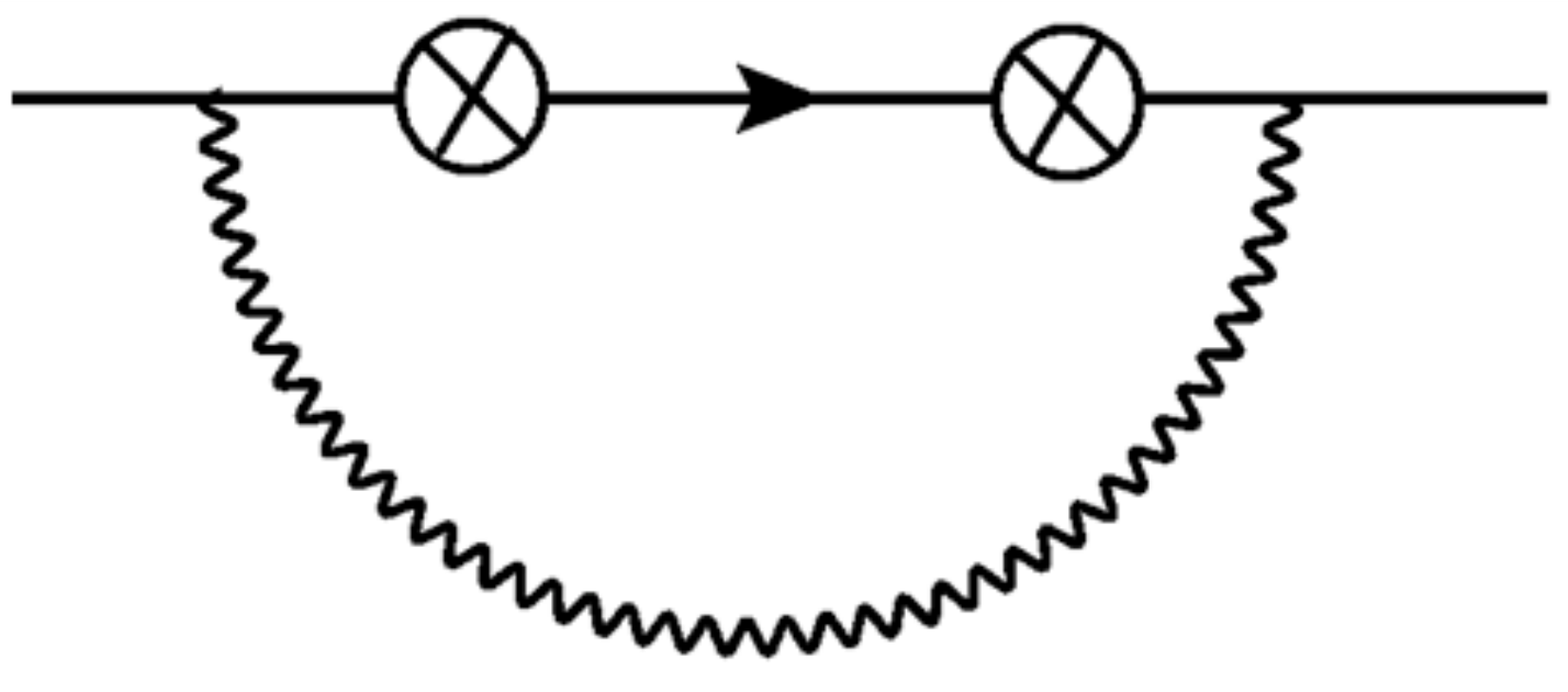}}
\subfigure[]{
\includegraphics[scale=0.22]{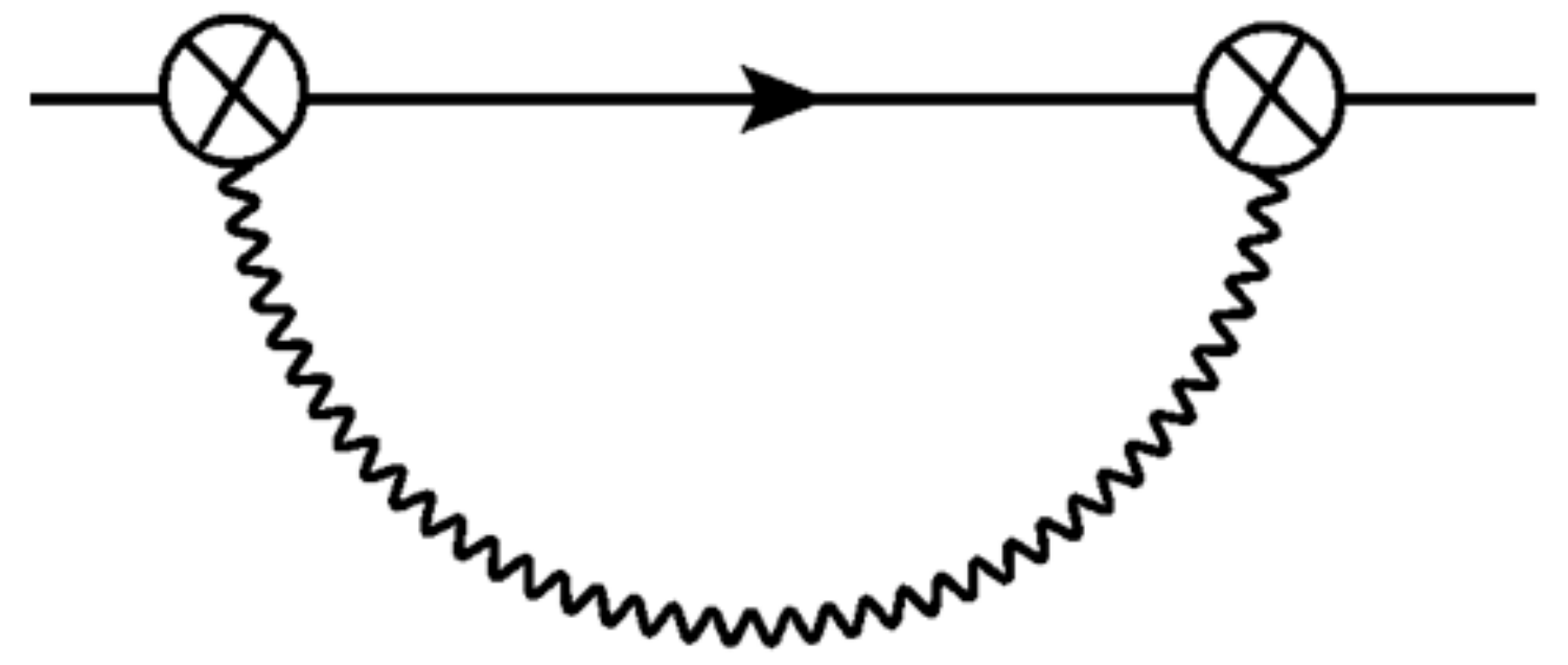}}
\subfigure[]{
\includegraphics[scale=0.22]{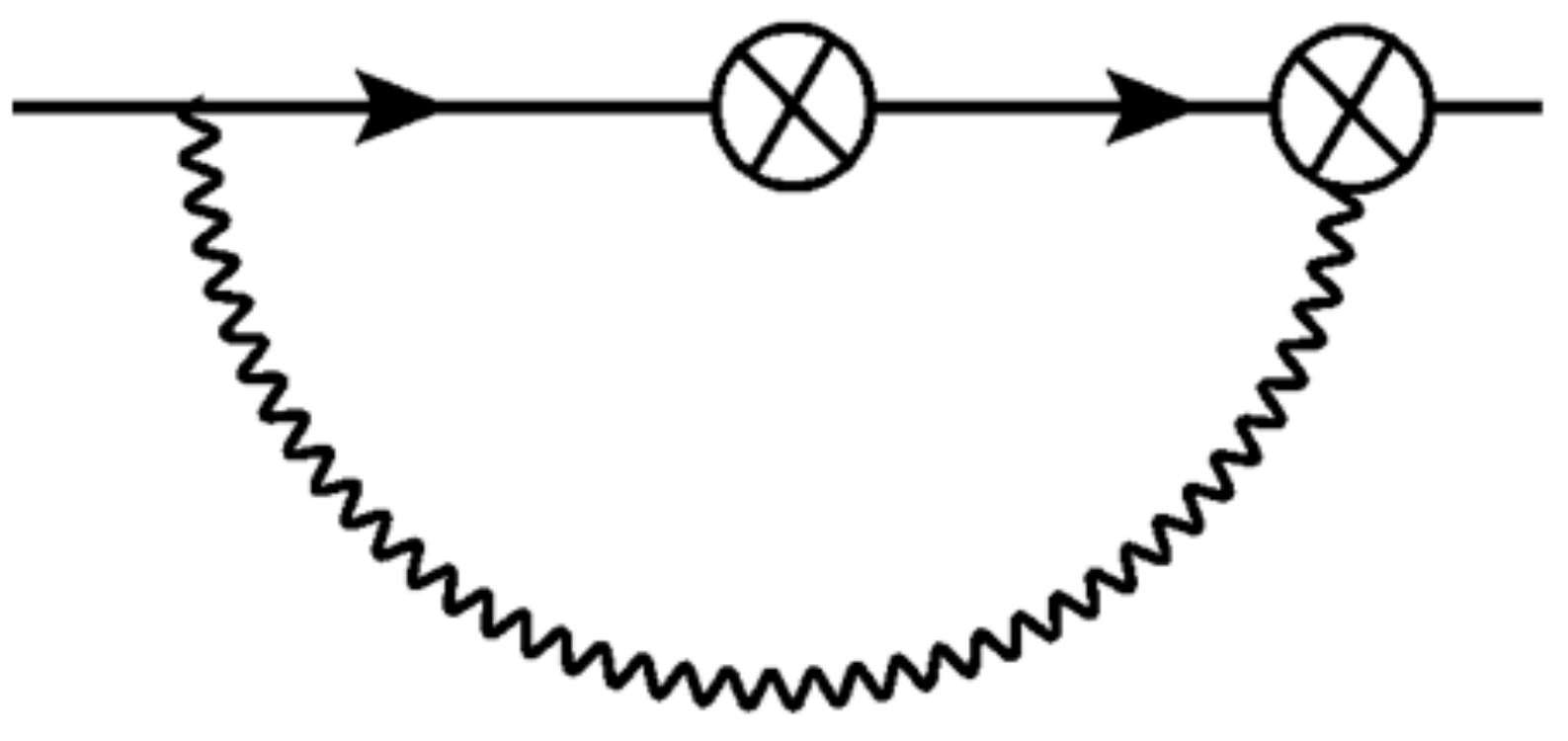}}
\subfigure[]{
\includegraphics[scale=0.22]{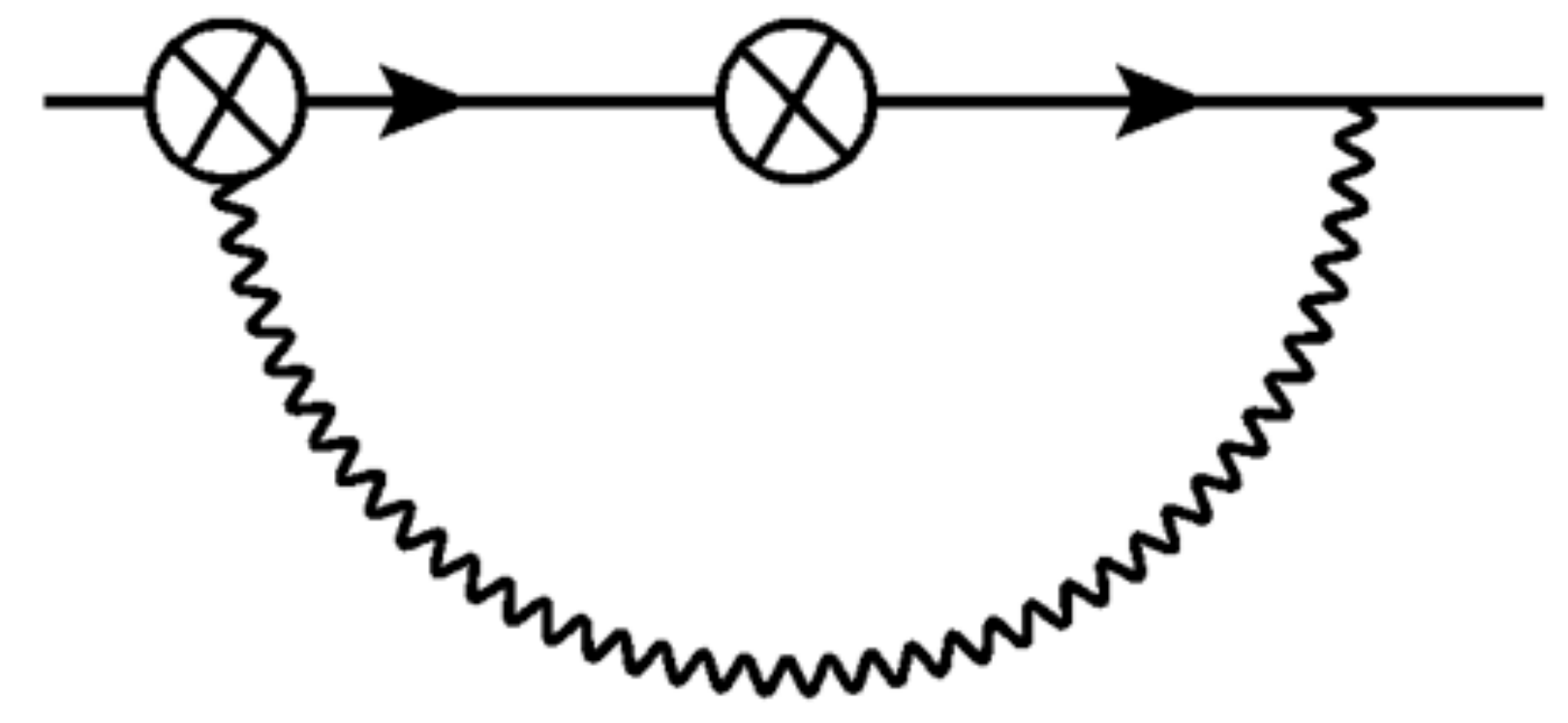}}
\subfigure[]{
\includegraphics[scale=0.195]{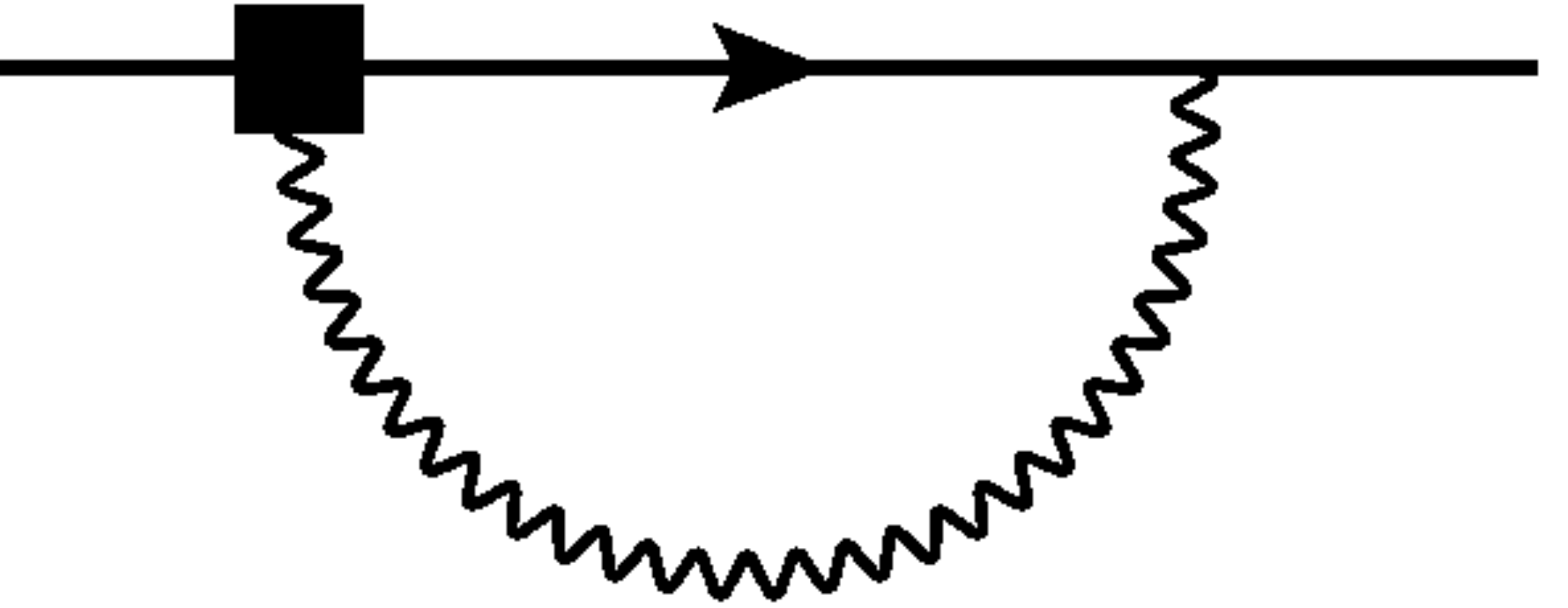}}
\subfigure[]{
\includegraphics[scale=0.195]{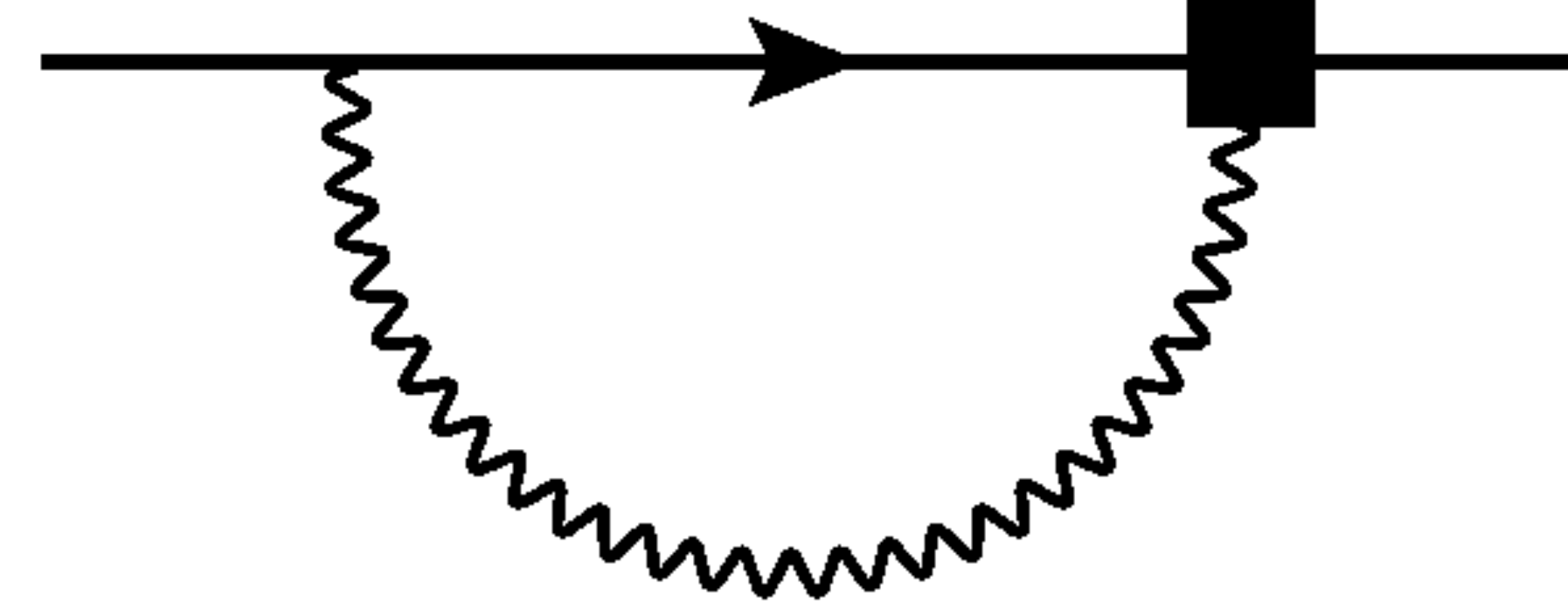}}
\caption{
(a-d) One-loop diagrams giving rise to the recoil corrections of $\mathcal{O}(\alpha_e/m_\phi^2 {\rm L}^3)$. 
The crossed circle denotes an insertion of the $|{\bf D}|^2/2 m_\phi$ operator. 
(e,f) One-loop diagrams providing the leading contribution from the charge radius of the scalar hadron, 
$\sim \alpha_e \langle r^2\rangle_\phi/{\rm L}^3$. 
The solid square denotes an insertion of the charge-radius 
operator in the scalar Lagrange density, Eq.~(\protect\ref{eq:scalarLag}).
}
\label{fig:m2}
\end{center}
\end{figure}
At next-to-next-to-LO
(N$^2$LO), ${\cal O}\left(\alpha_e/{\rm L}^3\right)$, 
there are potentially two contributions -- one is a recoil correction of the form $\sim  \alpha_e /m_\phi^2 {\rm L}^3$ 
and one is from the charge radius, $\sim \alpha_e  \langle r^2\rangle_\phi/{\rm L}^3$.
An evaluation of the one-loop diagrams giving rise to the recoil contributions, Fig.~\ref{fig:m2}(a-d), shows that
while individual diagrams are generally non-zero for a given gauge, their sum vanishes in 
any gauge.  
Therefore, there are no contributions of the form $\alpha_e/m_\phi^2 {\rm L}^3$ to the mass of $\phi$.
In contrast, the leading contribution from the charge radius of the scalar particle, resulting from the one-loop diagrams shown in Fig.~\ref{fig:m2}(e,f)
gives a contribution of the form
\begin{eqnarray}
\delta m^{({\rm N^2LO})}_\phi & = & 
-{2\pi\alpha_e  Q\over 3 {\rm L}^3}\  \langle r^2 \rangle_\phi \ 
\hat{\sum_{{\bf n}\ne {\bf 0}}} ~ 1
\ =\ 
+{2\pi\alpha_e Q\over 3 {\rm L}^3}\  \langle r^2 \rangle_\phi 
\ \ \ ,
\label{eq:CRscalar}
\end{eqnarray}
where  $\hat{\sum\limits_{\bf n} }~1 = 0$.

\begin{figure}[!ht]
\begin{center}
\subfigure[]{
\includegraphics[scale=0.2375]{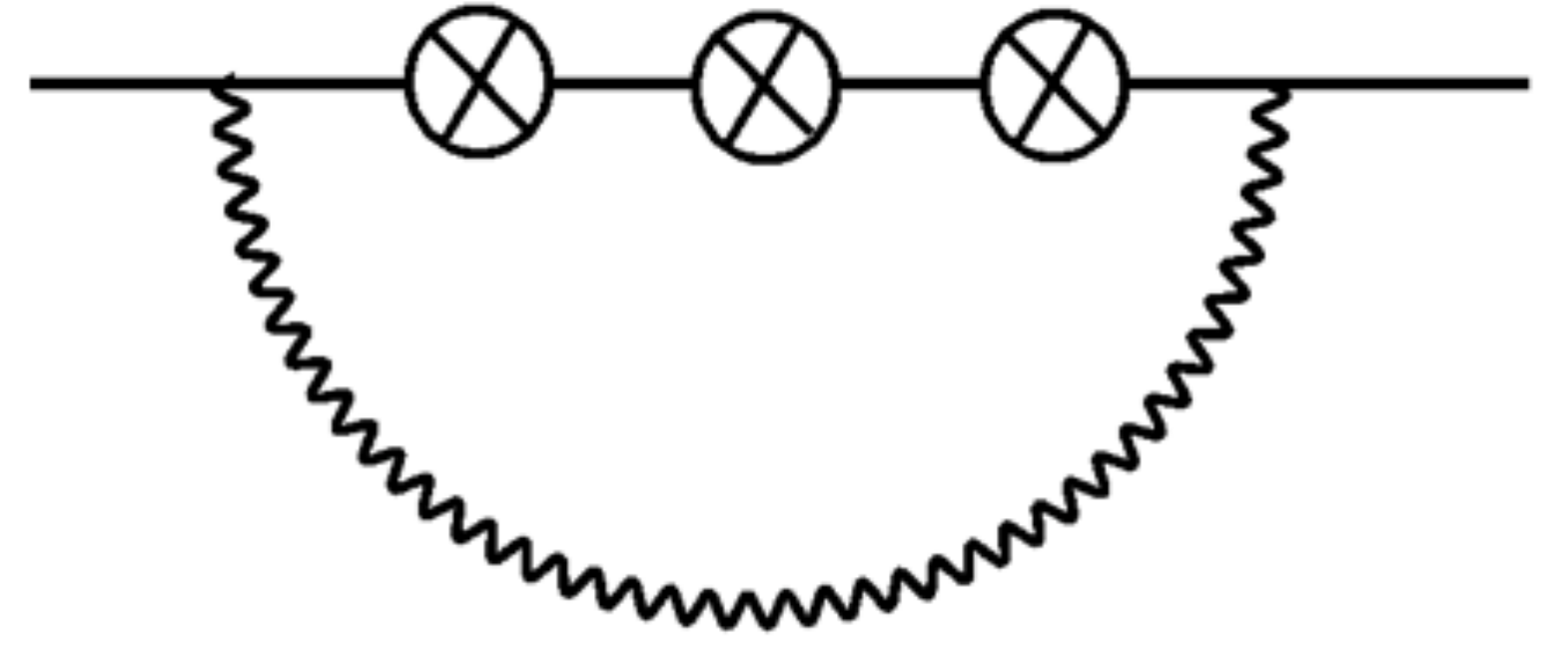}}
\subfigure[]{
\includegraphics[scale=0.2375]{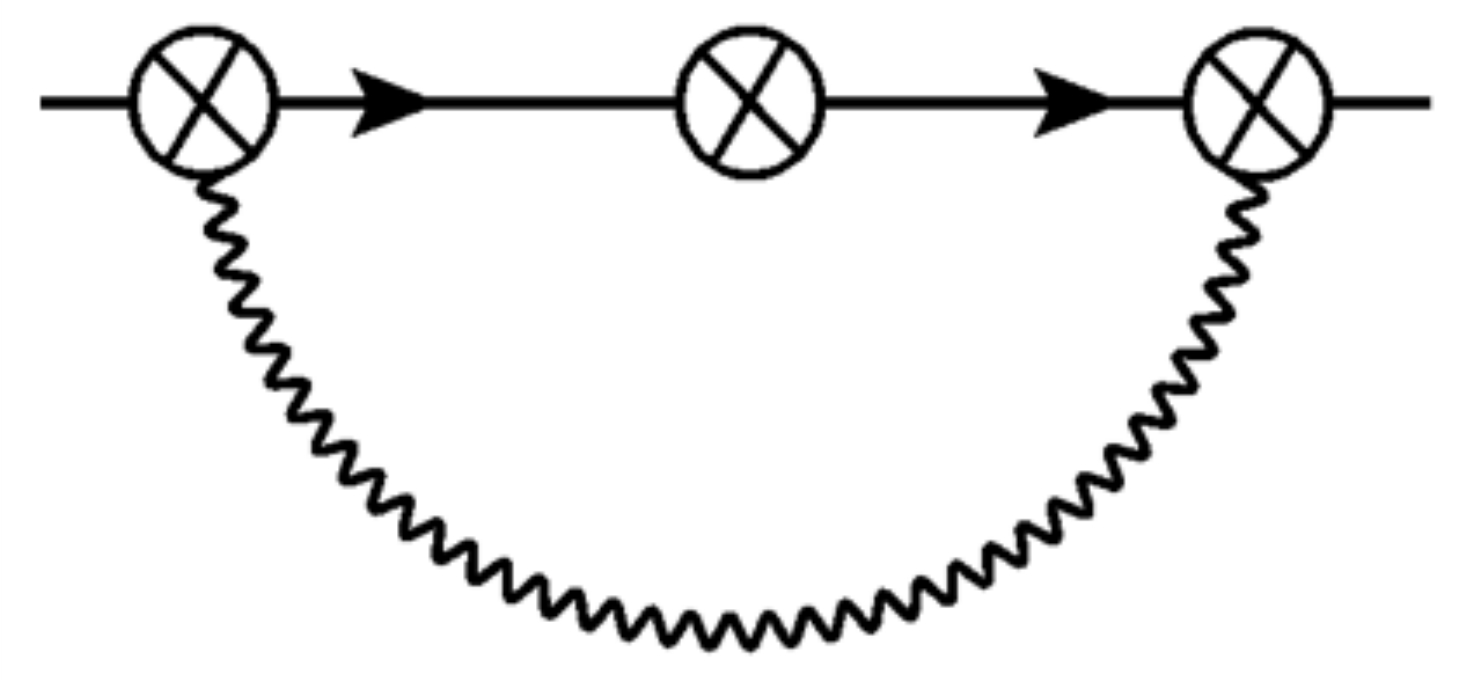}}
\subfigure[]{
\includegraphics[scale=0.2375]{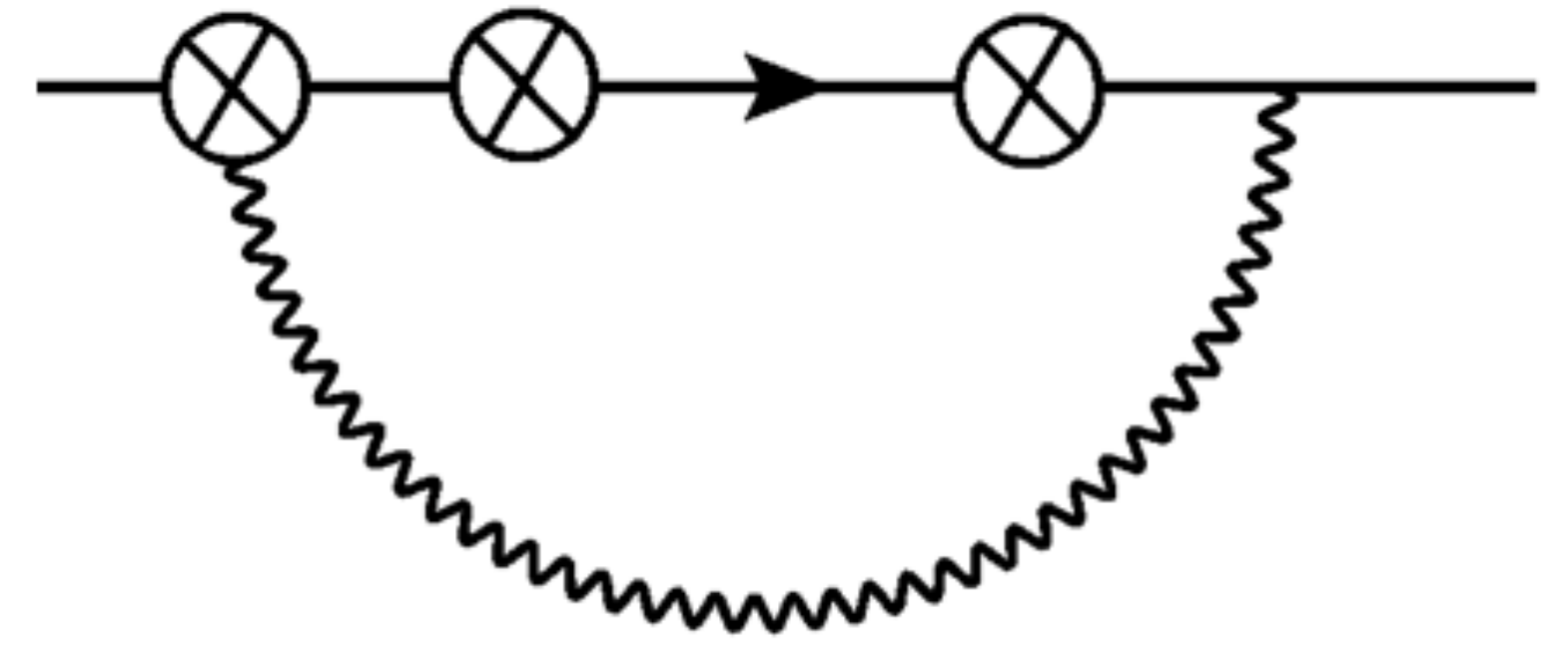}}
\subfigure[]{
\includegraphics[scale=0.2375]{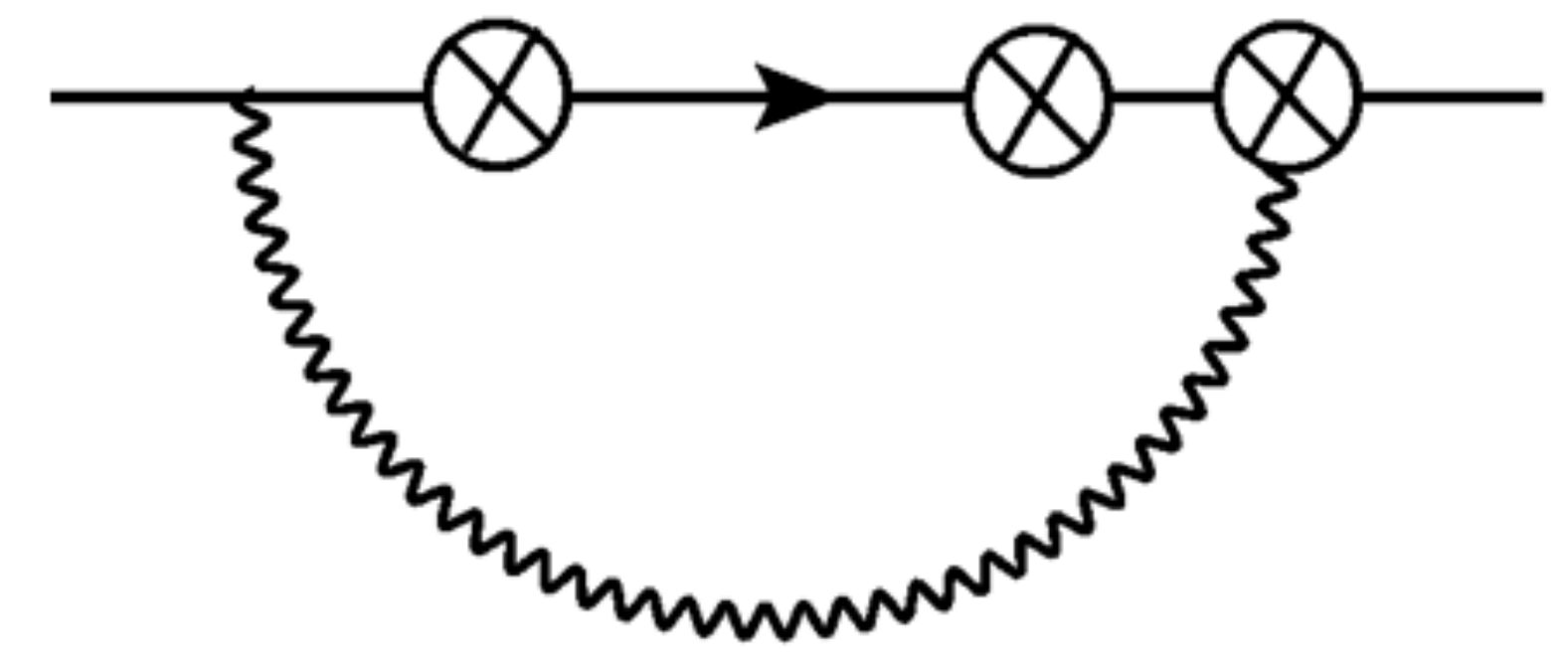}}
\subfigure[]{
\includegraphics[scale=0.230]{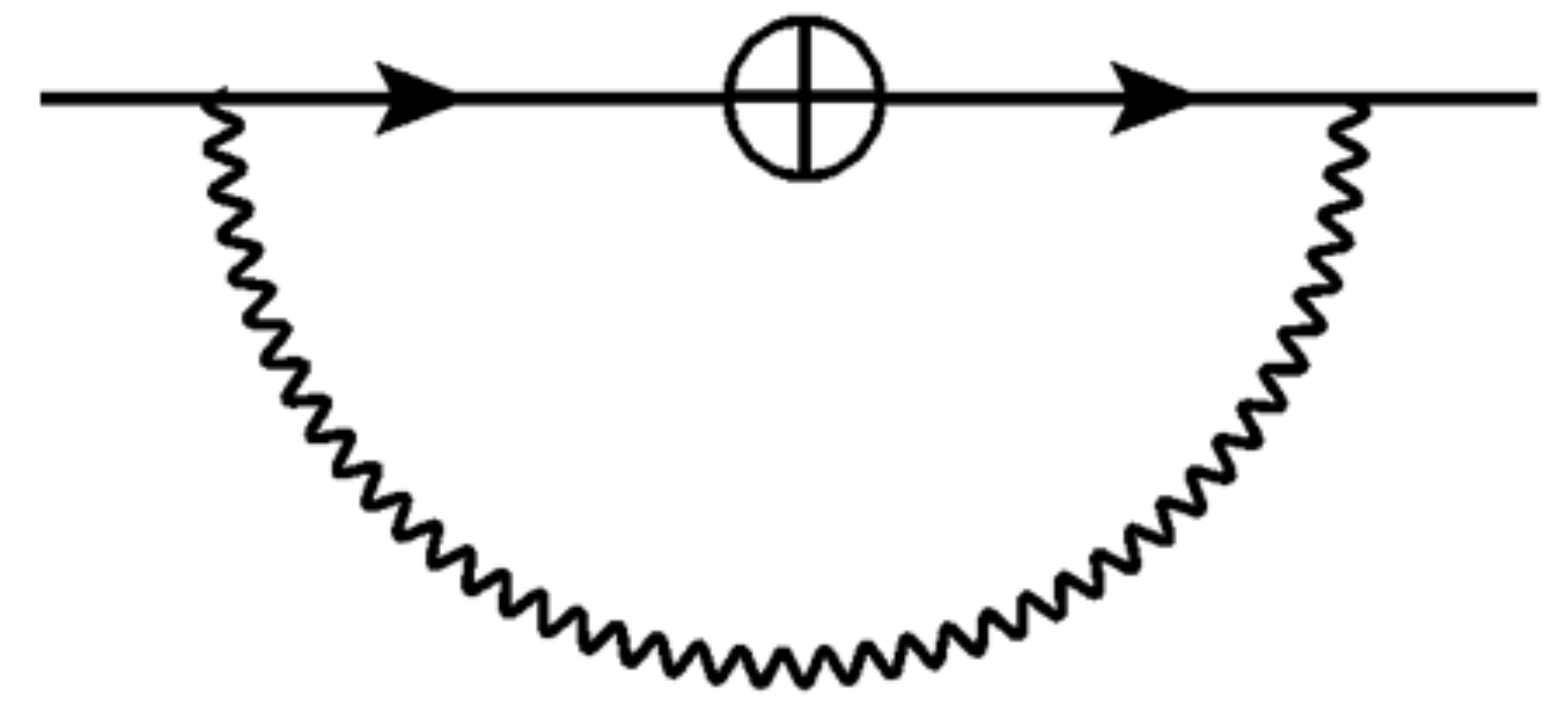}}
\subfigure[]{
\includegraphics[scale=0.230]{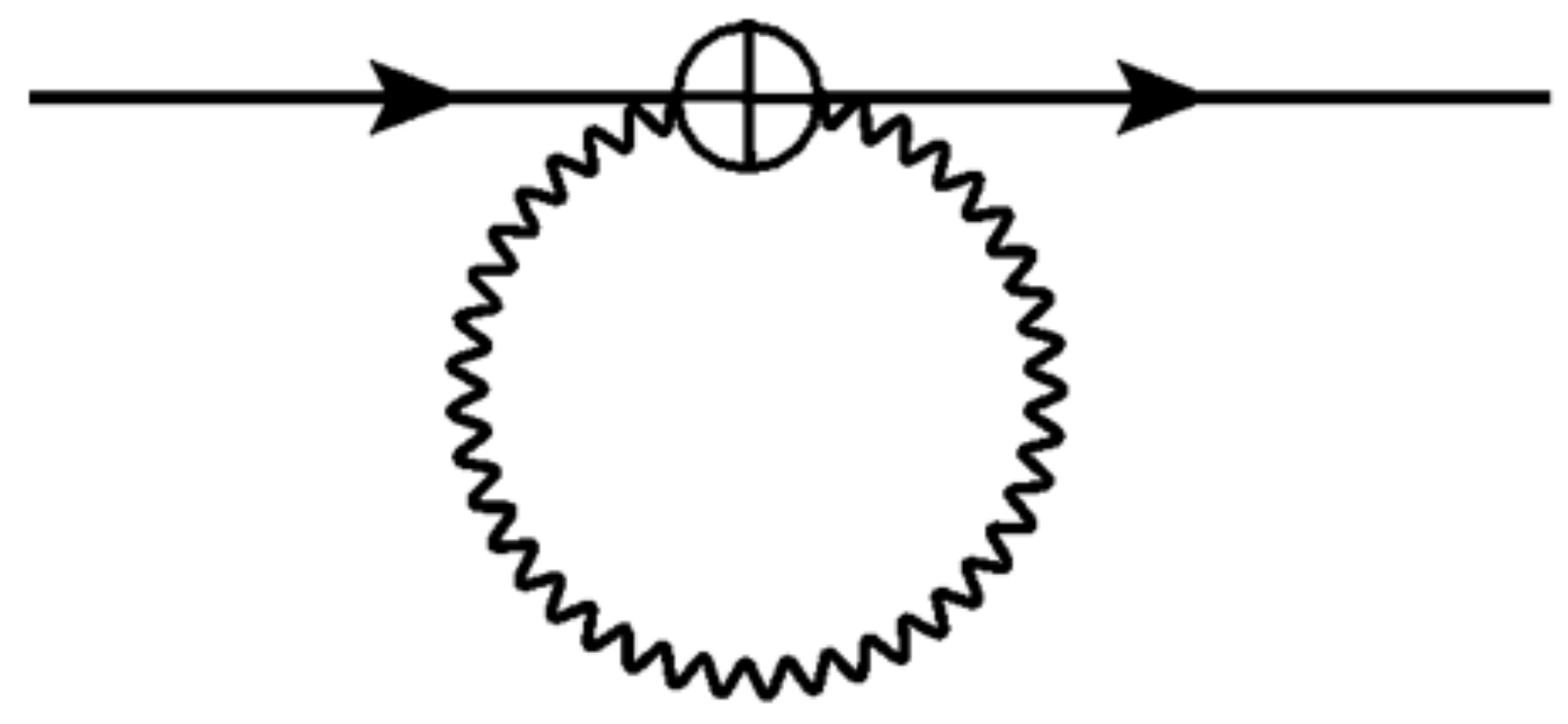}}
\subfigure[]{
\includegraphics[scale=0.155]{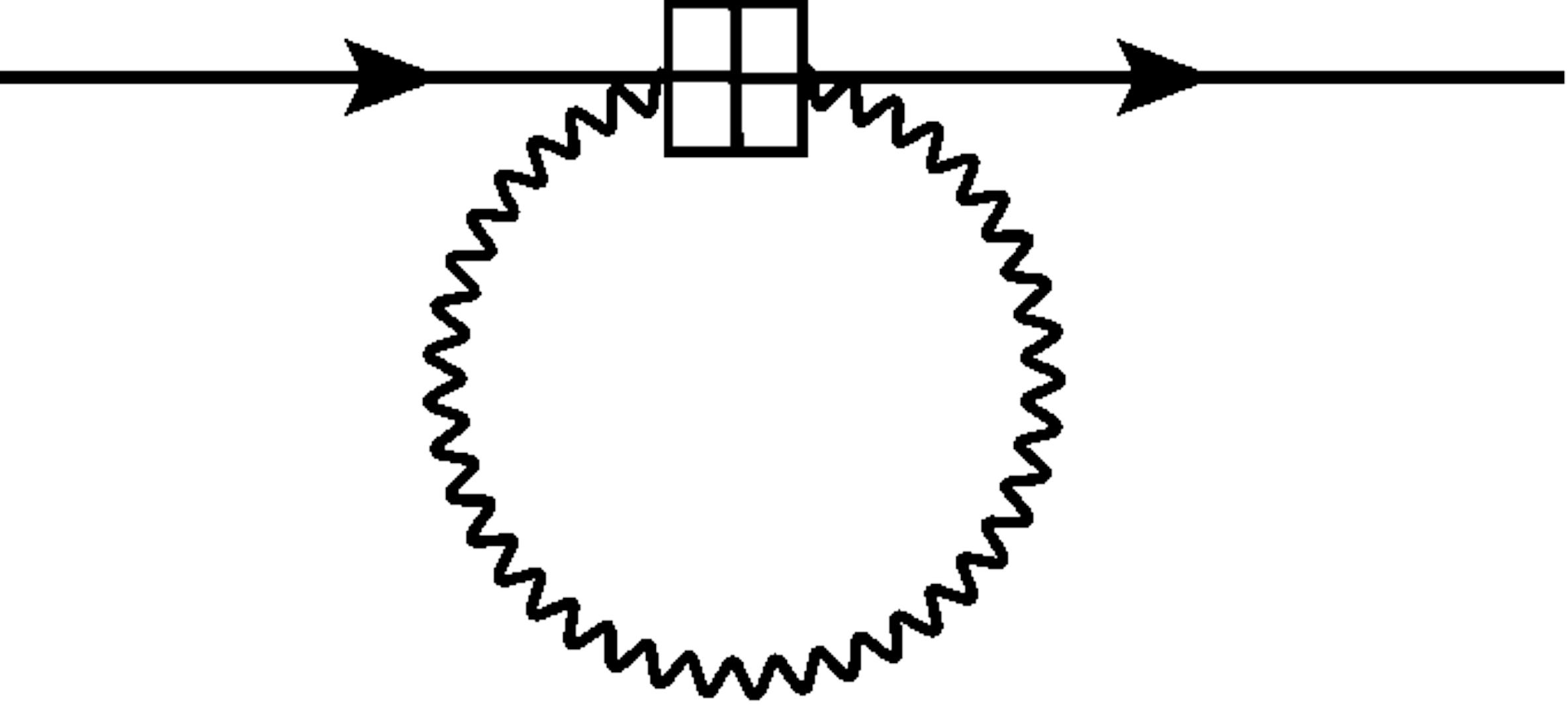}}
\caption{
One-loop diagrams contributing to the FV corrections to the mass of a scalar hadron at 
N$^3$LO, ${\cal O}\left(1/{\rm L}^4\right)$.
Diagrams (a-d) involve  three insertions of the $|{\bf D}|^2/2 m_\phi$ operator (crossed circles) 
in the scalar QED Lagrange density in Eq.~(\protect\ref{eq:scalarLag}), 
while (e,f) involve one insertion of the $|{\bf D}|^4/8 m_\phi^3$ operator (the sun cross), 
giving a $\mathcal{O}(\alpha_e/m_\phi^3 {\rm L}^4)$ correction. 
Diagram (g) involves an  insertion of $\tilde{\alpha}_E^{(\phi)} \ |{\bf E}|^2$ and $\tilde{\beta}_M^{(\phi)}\  |{\bf B}|^2$, 
operators (crossed square), 
contributing terms of the form 
$\sim (\alpha_E+\beta_M)/{\rm L}^4$ and $\sim \alpha_e\langle r^2 \rangle_\phi/m_{\phi}{\rm L}^4)$.
A diagram analogous to (g) provides the leading contribution from the $c_M$ operator at   $\mathcal{O}(\alpha_e/m_\phi {\rm L}^4)$.
}
\label{fig:m3A}
\end{center}
\end{figure}

At N$^3$LO, ${\cal O}\left(\alpha_e/{\rm L}^4\right)$, 
there are potentially three contributions: 
recoil corrections, $\sim \alpha_e/m_\phi^3 {\rm L}^4$, 
contributions from the electric and magnetic polarizability operators, 
$\sim \tilde\alpha_E^{(\phi)}/{\rm L}^4$ , $\tilde\beta_M^{(\phi)} /{\rm L}^4$, 
and contributions from the $c_M$ operator, Eq. (\ref{eq:scalarLag}). 
There are two distinct sets of  recoil corrections at this order. 
One set is from diagrams involving three insertions of the $|{\bf D}|^2/2 m_\phi$ operator, 
as shown in Fig.~\ref{fig:m3A}(a-d), 
and the other is from a single insertion of the $|{\bf D}|^4/8 m_\phi^3$ operator, shown in Fig.~\ref{fig:m3A}(e,f). 
The sum of diagrams contributing to each set vanishes, and so there are no contributions of the form $\alpha_e/m_\phi^3 {\rm L}^4$.
The other contributions, which include the electric and magnetic polarizabilities, arise from the one-loop diagrams shown in Fig.~\ref{fig:m3A}(g). A straightforward evaluation yields  a mass shift of 
\begin{eqnarray}
\delta m_\phi^{({\rm N^3LO}; \tilde{\alpha},\tilde{\beta})} 
&& = 
- {4\pi^2\over {\rm L}^4}\ \left( \tilde{\alpha}_E^{(\phi)} + \tilde{\beta}_M^{(\phi)} \right)\ 
\hat{\sum_{{\bf n}\ne {\bf 0}}}\ |{\bf n}|\nonumber\\
&& = 
- {4\pi^2\over {\rm L}^4}\ \left( \alpha_E^{(\phi)} + \beta_M^{(\phi)} \right)\ c_{-1}
+{4\pi^2 \alpha_e Q \over 3 m_\phi {\rm L}^4} \ 
\langle r^2 \rangle_\phi \ c_{-1}
\ \ \ ,
\end{eqnarray}
where the regularized sum is the same as that contributing to the energy density  associated with the Casimir effect,
and is
$c_{-1} = -0.266596$ \cite{Hasenfratz:1989pk}.
A similar calculation yields the contribution from the $c_M$ operator, 
\begin{eqnarray}
\delta m_\phi ^{({\rm N^3LO}; c_M)} 
& = & 
+{4\pi^2 \alpha_e Q \over 3 m_\phi {\rm L}^4} \ 
\langle r^2 \rangle_\phi \ c_{-1}
\ \ \ .
\end{eqnarray}

\begin{figure}[!ht]
  \centering
     \includegraphics[scale=0.365]{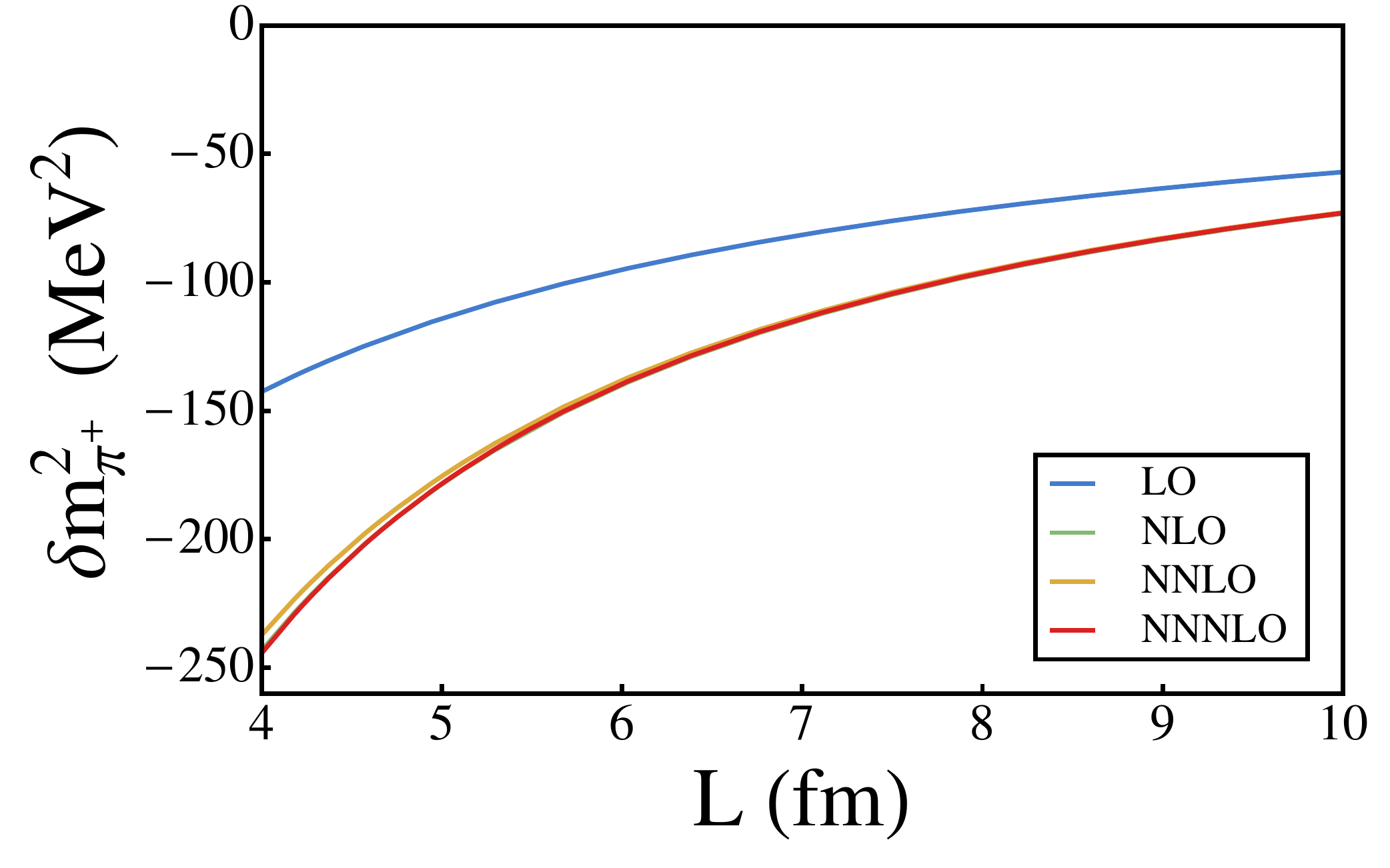}     \qquad
     \includegraphics[scale=0.365]{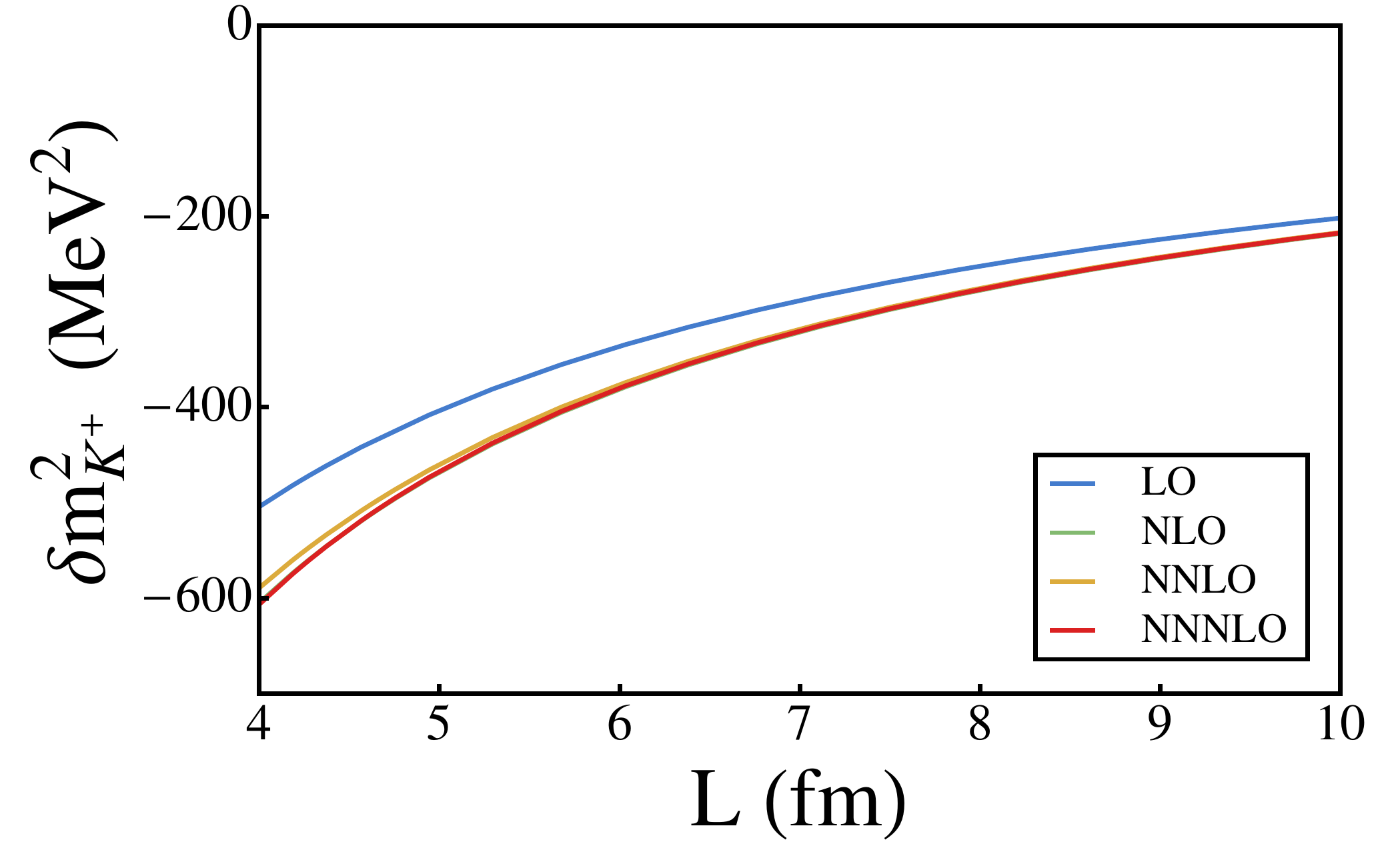}     
     \caption{The FV QED correction to  the mass squared of a charged pion (left panel) and kaon (right panel) 
     at rest in a FV at the physical pion mass.
     The leading contribution is due to  their electric charge, and scales as $1/{\rm L}$.
     The $1\sigma$-uncertainty bands associated with each order in the expansion are determined from the uncertainties in the experimental and theoretical inputs.
    }
  \label{fig:chargedkaonpionmsquare}
\end{figure}

\begin{figure}[!ht]
  \centering
     \includegraphics[scale=0.365]{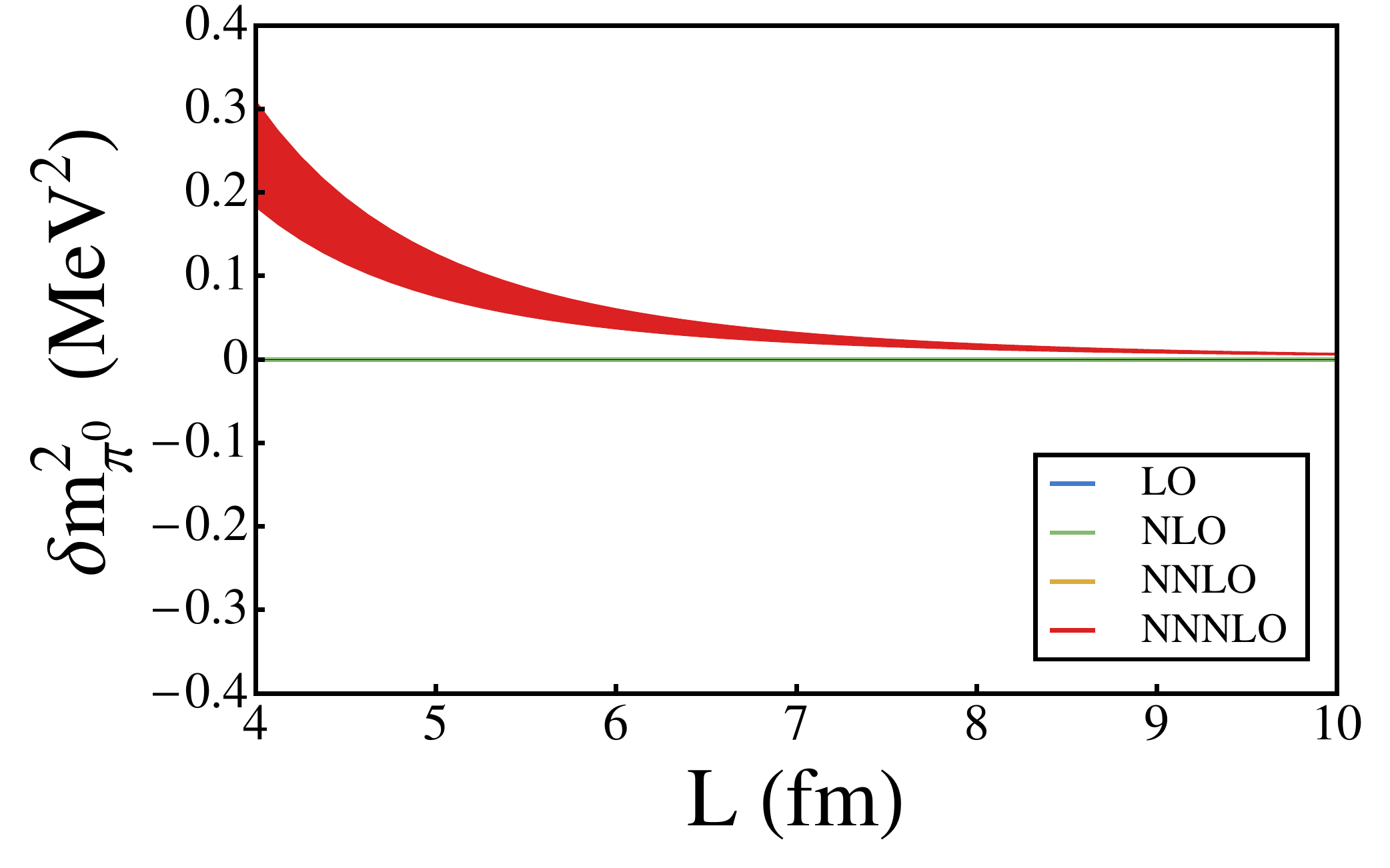}     \qquad
     \includegraphics[scale=0.365]{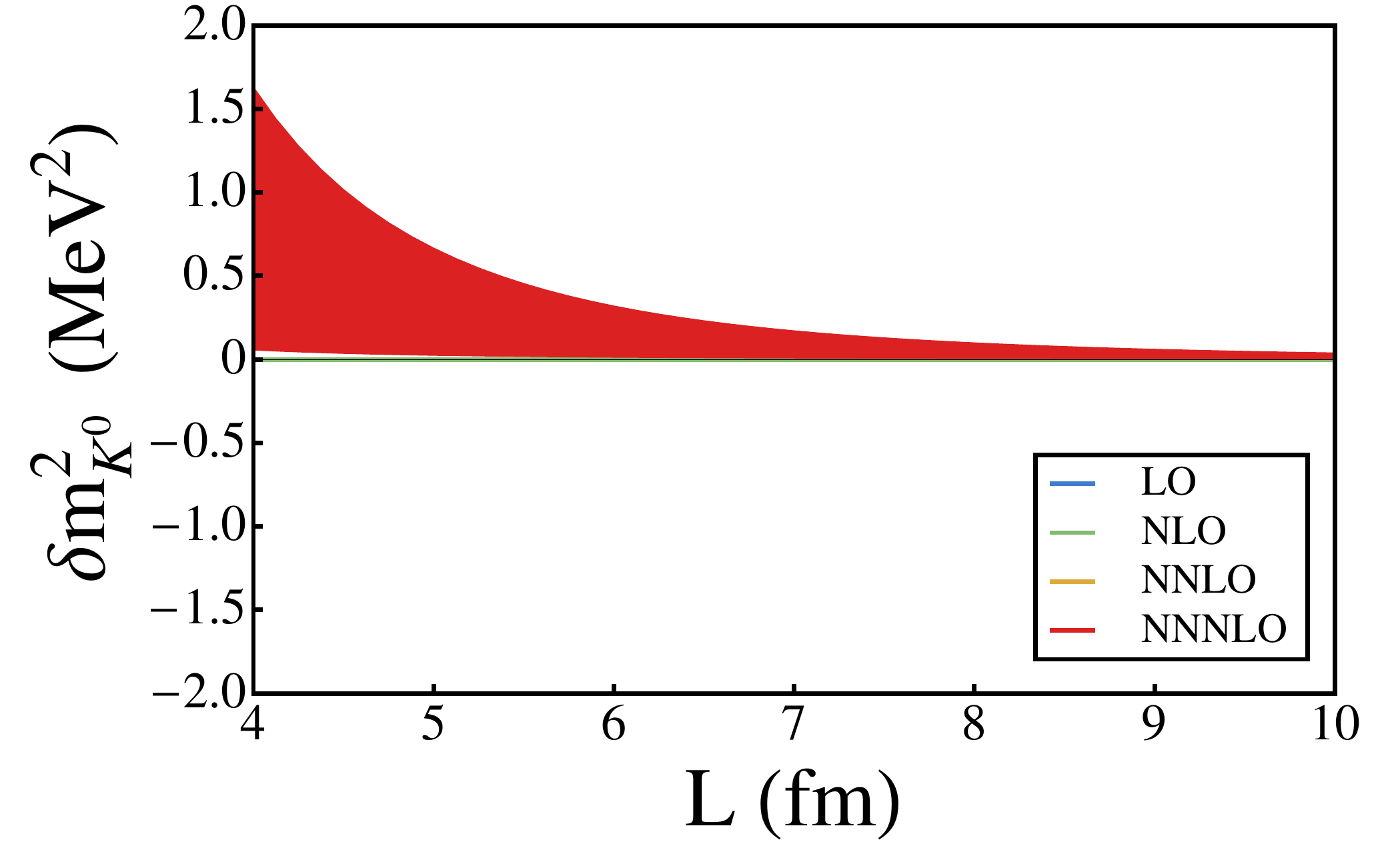}     
     \caption{The FV QED correction to  the mass squared of a neutral pion (left panel) and kaon (right panel) at rest in a FV
     at the physical pion mass.
     The leading contributions are from their polarizabilities, and scale as $1/{\rm L}^4$.
     The $1\sigma$-uncertainty bands associated with each order in the expansion are determined from the uncertainties in the experimental and theoretical inputs.
    }
  \label{fig:neutralkaonpionmsquare}
\end{figure}

Collecting  the contributions up to N$^3$LO,
the mass shift of a composite scalar particle in the $1/{\rm L}$ expansion is
\begin{eqnarray}
\delta m_\phi & = & 
{\alpha_e Q^2\over 2 {\rm L}} c_1 
\left( 1 + {2\over m_\phi {\rm L}} \right)
+ {2\pi \alpha_e  Q\over 3 {\rm L}^3}  \left(1+ {4\pi\over m_\phi {\rm L}} c_{-1} \right) \langle r^2 \rangle_\phi 
-  {4\pi^2\over {\rm L}^4}\ \left( \alpha_E^{(\phi)} + \beta_M^{(\phi)} \right) c_{-1}
\ .
\end{eqnarray}
Therefore, for the charged and neutral pions, the mass shifts are 
\begin{eqnarray}
\delta m_{\pi^+}& = & 
{\alpha_e \over 2 {\rm L}} c_1 
\left( 1 + {2\over m_{\pi^+} {\rm L}} \right)
+ {2\pi \alpha_e \over 3 {\rm L}^3} \left(1+ {4\pi\over m_{\pi^+} {\rm L}} c_{-1} \right) \langle r^2 \rangle_{\pi^+}
-  {4\pi^2\over {\rm L}^4} \left( \alpha_E^{(\pi^+)} + \beta_M^{(\pi^+)} \right) c_{-1},
\nonumber\\
\delta m_{\pi^0}& = & 
\ -\  {4\pi^2\over {\rm L}^4}\ \left( \alpha_E^{(\pi^0)} + \beta_M^{(\pi^0)} \right)\ c_{-1}
\ ,
\end{eqnarray}
where  potential complications due to the electromagnetic decay of the $\pi^0$ via the anomaly  have been neglected .
The shifts of the charged and neutral kaons have the same form, with $m_{\pi^{\pm ,0}} \rightarrow m_{K^{\pm ,0}}$,
$ \langle r^2 \rangle_{\pi^+} \rightarrow \langle r^2 \rangle_{K^+}$, 
$\alpha_E^{(\pi^{\pm , 0})}\rightarrow \alpha_E^{(K^{\pm , 0})}$
and 
$\beta_E^{(\pi^{\pm , 0})}\rightarrow \beta_E^{(K^{\pm , 0})}$. With the experimental constraints on the charge radii 
and polarizabilities of the pions and kaons, 
numerical estimates of the FV corrections can be performed at N$^3$LO.
The LO and NLO contributions are dictated by  only  the charge and mass of the meson.
The N$^2$LO contribution depends upon the charge and charge radius, which, for the charged mesons, are known experimentally to 
be~\cite{Beringer:1900zz},
\begin{eqnarray}
\sqrt{ \langle r^2 \rangle}_{\pi^+} 
& = & 
0.672\pm 0.008~{\rm fm}
\ \ ,\ \ 
\sqrt{ \langle r^2 \rangle}_{K^+} 
\ = \ 
0.560\pm 0.031~{\rm fm}
\ .
\end{eqnarray}
The N$^3$LO contribution from the electric and magnetic polarizabilities of the mesons depends upon their sum.
The Baldin sum rule determines the charged pion combination, while the result of a two-loop $\chi$PT calculation is used for the neutral pion 
combination~\cite{Holstein:2013kia},
\begin{eqnarray}
 \alpha_E^{(\pi^+)} + \beta_M^{(\pi^+)} &  = & 
 \left(0.39\pm 0.04\right)\times 10^{-4}~{\rm fm}^3
 \ ,\ 
 \alpha_E^{(\pi^0)} + \beta_M^{(\pi^0)}  =  
 \left(1.1\pm 0.3\right)\times 10^{-4}~{\rm fm}^3
 \ .
\end{eqnarray}
Unfortunately, little is known about the polarizabilities of the kaons, and so 
naive dimensional analysis is used to provide an estimate 
of their contribution~\cite{Holstein:2013kia}, 
$ \alpha_E^{(K^+)} + \beta_M^{(K^+)}$, $ \alpha_E^{(K^0)} + \beta_M^{(K^0)} = \left(1\pm 1\right)\times 10^{-4}~{\rm fm}^3$.
With these values, along with their experimentally measured masses, 
the expected FV corrections to the charged meson masses are shown in 
Fig.~\ref{fig:chargedkaonpionmsquare} and to the neutral meson masses in 
Fig.~\ref{fig:neutralkaonpionmsquare}.~\footnote{
When comparing with  previous results one should note that the squared mass shift of the $\pi^+$, as an example, due to  
FV QED is 
\begin{eqnarray}
\delta m_{\pi^+}^2 
& = & 
\left(
m_{\pi^+} + \delta m_{\pi^+} 
\right)^2 - m_{\pi^+}^2 
\ =\ 2 m_{\pi^+}  \ \delta m_{\pi^+} \ +\ {\cal O}(\alpha_e^2)
\ ,
\nonumber
\end{eqnarray}
As is evident, the leading contribution to the mass squared scales as $1/{\rm L}$, 
contrary to a recent suggestion in the literature~\cite{Portelli:2012pn} of $1/{\rm L}^2$. 
Note that the quantity shown in 
Fig.~\ref{fig:chargedkaonpionmsquare} and Fig.~\ref{fig:neutralkaonpionmsquare} 
is $\delta m_\phi^2$ as opposed to $\delta m_\phi$,
as it is this that enters into the determination of the light-quark masses from LQCD calculations.
}

In a volume with ${\rm L}=4~{\rm fm}$, the FV QED mass shift of a charged meson  is approximately $0.5~{\rm MeV}$.
Figure~\ref{fig:chargedkaonpionmsquare} shows that for volumes with ${\rm L}\gsim 4~{\rm fm}$,
the meson charge is responsible for essentially all of the FV modifications, with their compositeness 
making only a  small contribution, i.e. the differences between the NLO and N$^2$LO mass shifts are small.
For the neutral mesons, the  contribution from the polarizabilities is very small, but with substantial uncertainty.
It is worth re-emphasizing that in forming these estimates of the QED power-law corrections, 
 exponential corrections  of the form $e^{-m_\pi {\rm L}}$ have been neglected.

\section{NRQED for the Baryons and $J={1\over 2}$ Nuclei}
\begin{figure}[!ht]
\begin{center}
\includegraphics[scale=0.175]{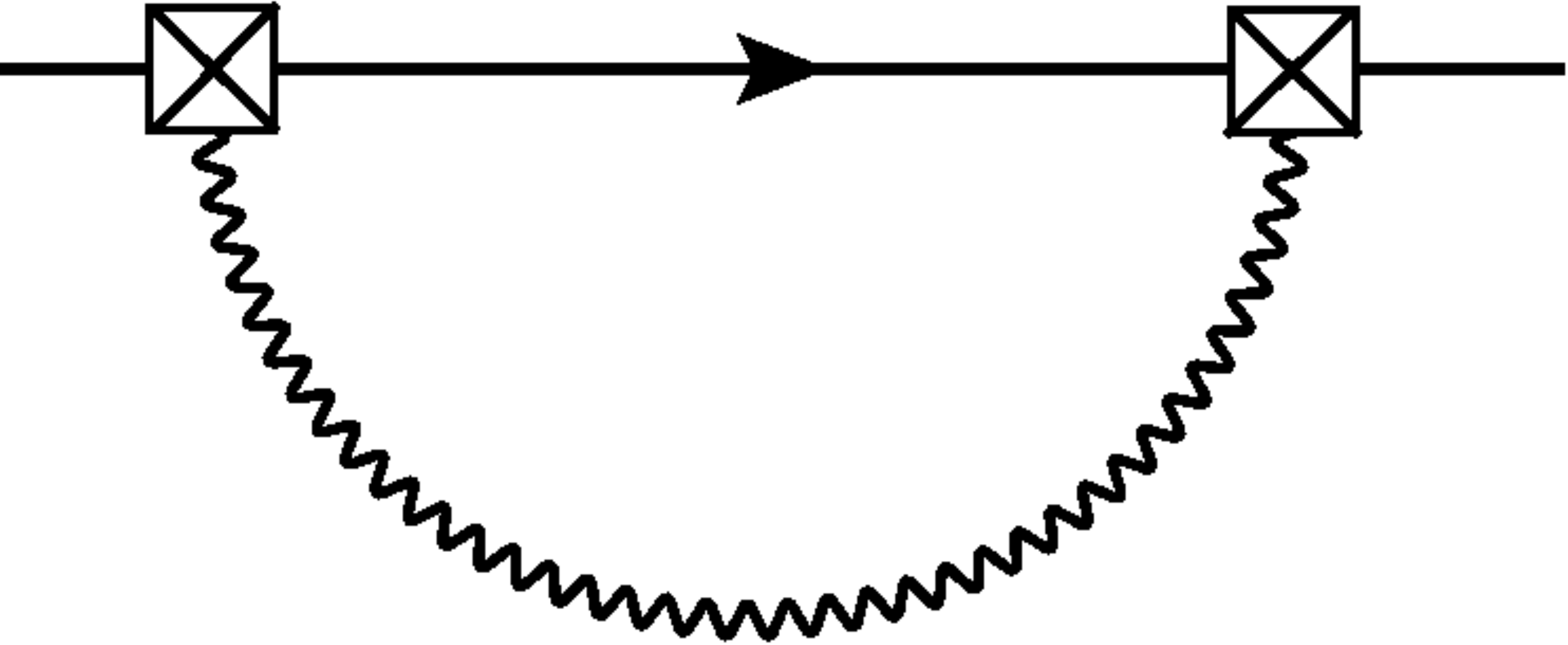}
\caption{{\small The N$^2$LO, ${\cal O}\left( \alpha_e/M_\psi^2 L^3 \right)$, 
 FV QED correction to the mass of a baryon from its magnetic moment. 
 The crossed square denotes an insertion of the magnetic moment operator given in Eq.~(\protect\ref{eq:baryonL}).}}
\label{fig:magmagloop-NNLO}
\end{center}
\end{figure}

\begin{figure}[t]
  \centering
     \includegraphics[scale=0.335]{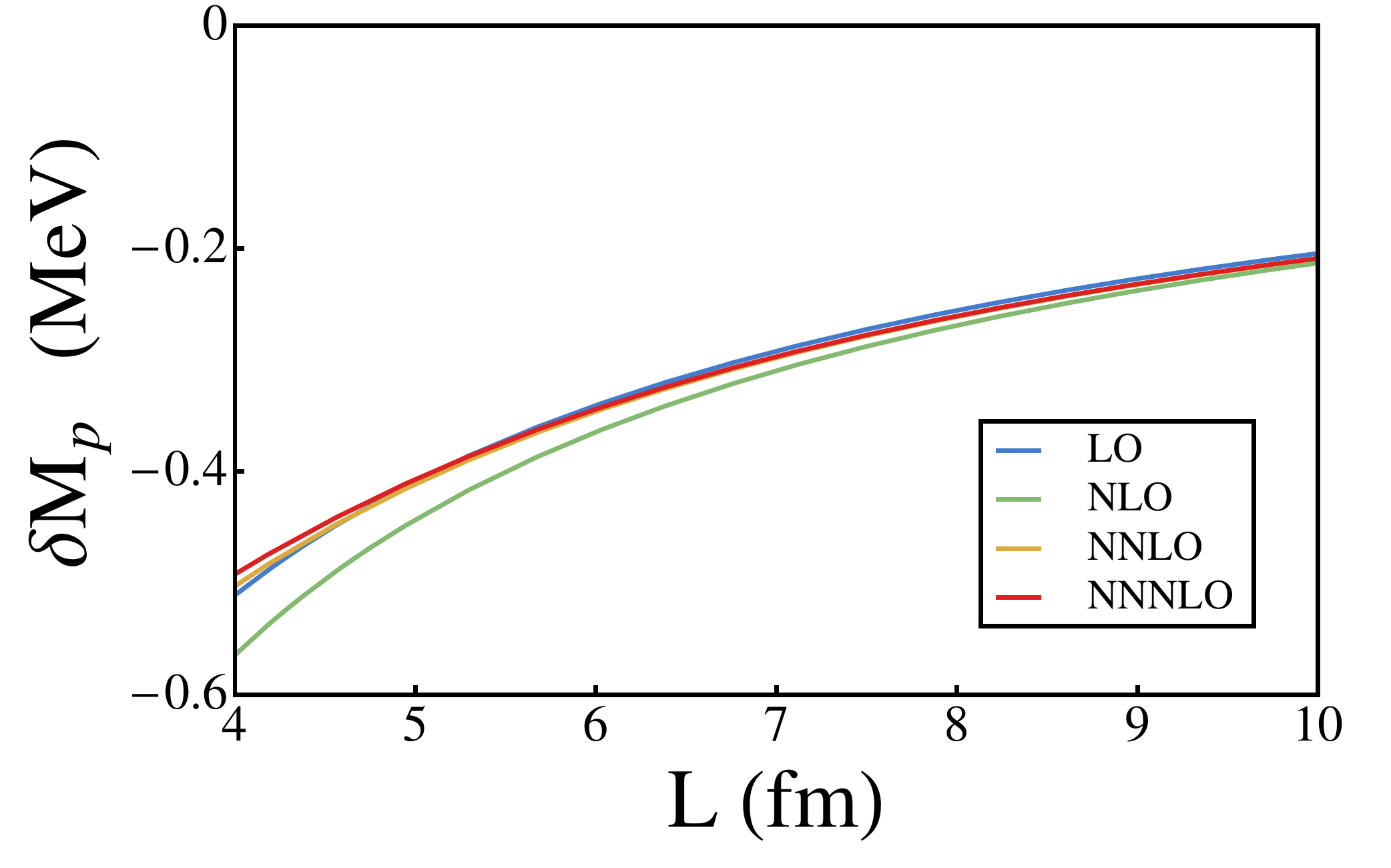}     \qquad
     \includegraphics[scale=0.335]{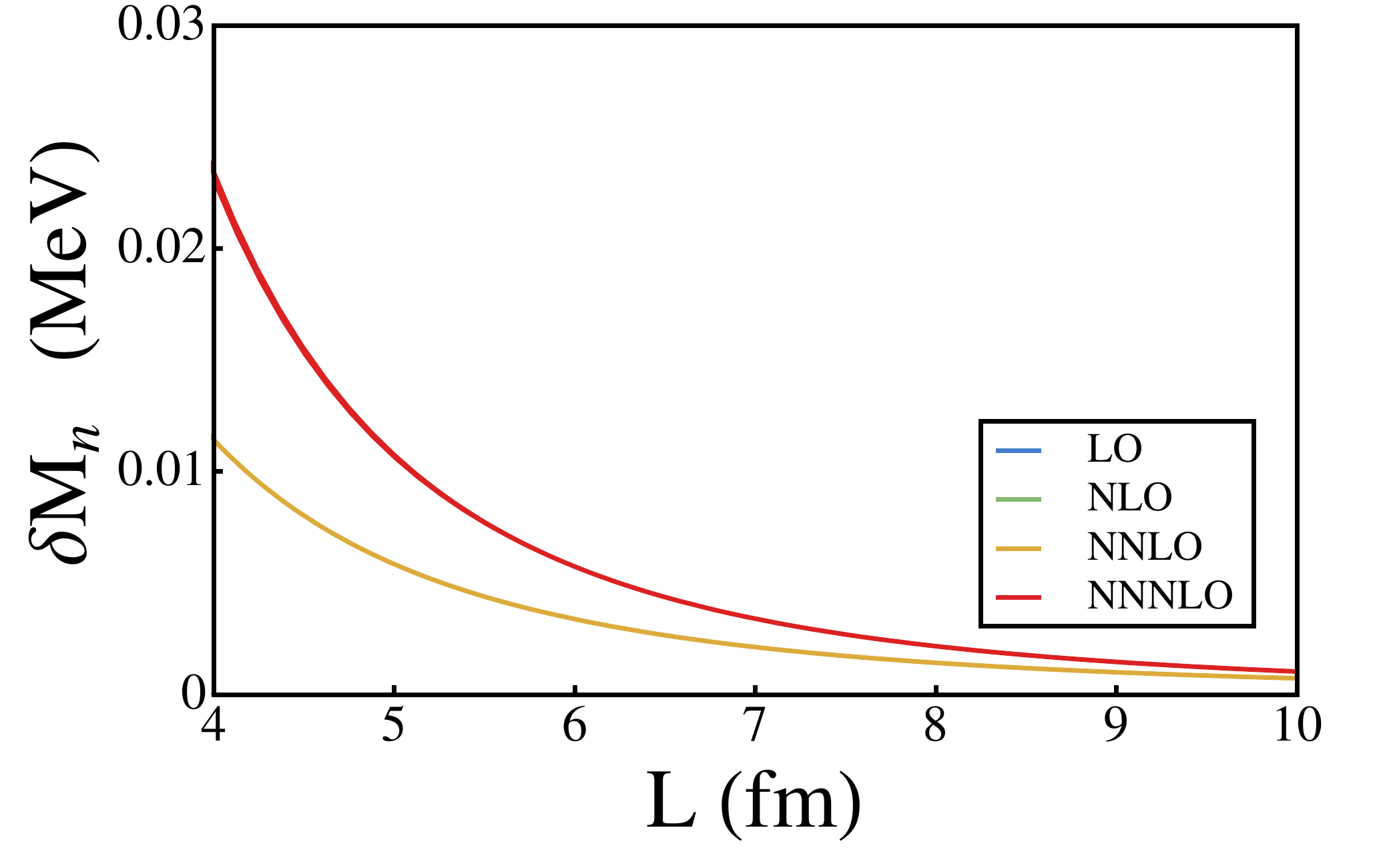}     
     \caption{{\small The FV QED correction to the mass of the proton (left panel) and neutron (right panel)
     at rest in a FV at the physical pion mass.
     The leading contribution to the proton mass shift is due to its electric charge, 
     and scales as $1/L$, while the leading contribution to the neutron mass shift is due to 
     its magnetic moment, and scales as $1/L^3$.
     The $1-\sigma$ uncertainty bands associated with each order in the expansion are determined from the uncertainties in the experimental and theoretical inputs.
    }}
  \label{fig:protonneutron}
\end{figure}

\noindent
In the case of baryons 
and $J={1\over 2}$ nuclei,
the method for determining the 
FV QED corrections is analogous to that 
for the mesons, described in the previous section, but modified to include the effects of spin and the 
reduction from a four-component  to a two-component spinor.
The low-energy EFT describing the interactions between the nucleons and the electromagnetic field is NRQED, but enhanced 
to include the compositeness of the nucleon.  
A nice review of NRQED, including the contributions from the non point-like structure of the nucleon, 
can be found in Ref.~\cite{Hill:2012rh}, and the relevant terms in the NRQED Lagrange density for 
a N$^3$LO calculation
are~\cite{Isgur:1989vq,Isgur:1989ed,Jenkins:1990jv,Jenkins:1991ne,Thacker:1990bm,Labelle:1992hd,Manohar:1997qy,Luke:1997ys,Chen:1999tn,Beane:2007es,Lee:2013lxa,Hill:2012rh}
\begin{eqnarray}
{\cal L}_\psi
& = & 
\psi^\dagger \left[
iD_0
\ +\ {|{\bf D}|^2\over 2 M_\psi}
\ + \ { |{\bf D}|^4 \over 8 M_\psi^3}\ 
\ +\ c_F {e\over 2M_\psi} {\bm\sigma}\cdot {\bf B}
\ +\  c_D {e\over 8 M_\psi^2} {\bm\nabla}\cdot {\bf E}
\right.\nonumber\\
&&\left. \qquad \qquad
\ +\ i c_S {e\over 8 M_\psi^2}\ {\bm\sigma}\cdot\left( {\bf D}\times {\bf E} -  {\bf E} \times {\bf D} \right)
\ +\ 2 \pi \tilde\alpha_E^{(\psi)}  |{\bf E}|^2
\ +\ 2\pi \tilde\beta_M^{(\psi)}  |{\bf B}|^2
\right.\nonumber\\
&&\left. \qquad \qquad
\ +\ e\ c_{W_1}\ {\{  {\bf D}^2 , {\bm\sigma}\cdot {\bf B} \} \over 8 M_\psi^3}
\ -\  e\ c_{W_2}\ { D^i {\bm\sigma}\cdot {\bf B} D^i\over 4 M_\psi^3}
\right.\nonumber\\
&&\left. \qquad \qquad
\ +\ e\ c_{p^\prime p}\ { {\bm\sigma}\cdot {\bf D} {\bf B}\cdot {\bf D} +  {\bf B}\cdot {\bf D} {\bm\sigma}\cdot {\bf D} \over  8 M_\psi^3}
\ +\ i e \ c_M\ { \{ D^i , ({\bm\nabla}\times {\bf B})^i \} \over  8 M_\psi^3}
\ +\  \cdots
 \right] \psi
,
\nonumber\\
\label{eq:baryonL}
\end{eqnarray}
where $c_F = Q + \kappa_\psi + {\cal O}(\alpha_e)$ is the coefficient of the magnetic-moment interaction, 
with $\kappa_\psi$ 
related to the anomalous magnetic moment of $\psi$,
$c_D =  Q + {4\over 3} M_\psi^2 \langle r^2\rangle_\psi + {\cal O}(\alpha_e)$ contains the leading 
charge-radius contribution, $c_S=2c_F-Q$ is the coefficient of the spin-orbit interaction and 
$c_M = (c_D-c_F)/2$.
The coefficients of the $|{\bf E}|^2$ and $|{\bf B}|^2$ terms contain the polarizabilities,
$1/M_\psi$
and $1/M_\psi^{3}$ corrections,
\begin{eqnarray}
\tilde\alpha_E^{(\psi)}  & = & 
\alpha_E^{(\psi)}  - {\alpha_e \over 4 M_\psi^3}\left(Q^2+\kappa_\psi^2\right) - {\alpha_e Q\over 3 M_\psi} \langle r^2\rangle_\psi
\ \ ,\ \ 
\tilde\beta_M^{(\psi)}  \ =  
\beta_M^{(\psi)}  +  {\alpha_e Q^2 \over 4 M_\psi^3}
.
\label{eq:pols}
\end{eqnarray}
The operators with coefficients $c_{W_1}$, $c_{W_2}$ and $c_{p^\prime p}$, 
given in Ref.~\cite{Hill:2012rh},
do not contribute to the FV corrections at this order.
The ellipses denote terms that are higher orders in $1/M_\psi$ and $1/\Lambda_\chi$.
Two insertions of the magnetic-moment operator provide its leading contribution, 
as shown in Fig.~\ref{fig:magmagloop-NNLO}, giving rise to $\mathcal{O}(\alpha_e/L^3)$ corrections to the mass of spin-$\frac{1}{2}$ particles.
 Without replicating the detail presented in the previous section,
the sum of the contributions to the FV self-energy modification of a composite fermion, up to N$^3$LO, is
\begin{eqnarray}
\delta M_\psi && = 
{\alpha_e Q^2\over 2 L} c_1 
\left( 1 + {2\over M_\psi L} \right)
\ +\ {2\pi \alpha_e Q\over 3 L^3} \langle r^2 \rangle_\psi 
\ +\ {\pi \alpha_e\over M_\psi^2 L^3}\ \left[\ {1\over 2} Q^2\ + \ (Q+\kappa_{\psi})^2 \right]
\nonumber\\
&& -  
{4\pi^2\over L^4} \left( \tilde\alpha_E^{(\psi)} + \tilde\beta_M^{(\psi)} \right) c_{-1}
+  {\pi^2 \alpha_e Q  \over M_\psi^3 L^4}\ \left( {4\over 3} M_\psi^2  \langle r^2 \rangle_\psi  - \kappa_\psi \right) c_{-1}-\frac{\alpha_e \pi^2}{M_{\psi}^3 L^4}\kappa_{\psi}(Q+\kappa_{\psi})c_{-1}.
\nonumber
\\
\end{eqnarray}
Therefore, for the proton and neutron, the FV QED mass shifts are
\begin{eqnarray}
\delta M_p & = & 
{\alpha_e \over 2 L} c_1 
\left( 1 + {2\over M_p L} \right)
+ {2 \pi \alpha_e  \over 3 L^3} \left( 1 + {4\pi\over M_p L} c_{-1} \right) \langle r^2 \rangle_p
+ {\pi \alpha_e\over M_p^2 L^3}\ \left( {1\over 2} + (1+\kappa_p)^2 \right)
\nonumber\\
& &
-  {4\pi^2\over L^4}\left( \alpha^{(p)}_E + \beta^{(p)}_M \right) c_{-1}
\ -\  {2\pi^2 \alpha_e  \kappa_p \over  M_p^3 L^4}\  c_{-1},
\nonumber\\
\delta M_n & = & 
\kappa_n^2\ {\pi \alpha_e \over  M_n^2 L^3}
\ -\  {4\pi^2\over L^4} \left( \alpha^{(n)}_E + \beta^{(n)}_M \right) c_{-1},
\end{eqnarray}
where the anomalous magnetic moments of the proton and neutron give
$\kappa_p = 1.792847356(23)$ and 
$\kappa_n = -1.9130427(5) M_n/M_p $, respectively~\cite{Beringer:1900zz}.
One of the N$^2$LO contributions to the proton  FV QED correction depends upon its charge radius, which is known experimentally to 
be,
$\langle r^2 \rangle_{p}  = 0.768\pm 0.012~{\rm fm}^2 $~\cite{Beringer:1900zz}.
Further, part of the N$^3$LO contribution depends upon the electric and magnetic polarizabilities, 
which are constrained by
the Baldin sum rule,~\cite{Holstein:2013kia}
\begin{align}
 &\alpha_E^{(p)} + \beta_M^{(p)} =  
 \left(13.69\pm 0.14\right)\times 10^{-4}~{\rm fm}^3
 \ ,\ 
 \alpha_E^{(n)} + \beta_M^{(n)}  =  
 \left(15.2\pm 0.5\right)\times 10^{-4}~{\rm fm}^3
 .
\end{align}
With these values for the properties of the proton and neutron, along with their experimentally measured masses, 
the expected FV modifications to their masses are shown in Fig.~\ref{fig:protonneutron}.

The   proton FV QED corrections are consistent with those of the charged scalar mesons.
However, the neutron corrections, while very small, of the order of a few keVs, exhibit more structure.  
The N$^2$LO contribution from the magnetic moment increases the mass in  FV, scaling as $1/M_n^2 L^3$, similar to the 
polarizabilities which make a positive contribution and scale as $1/L^4$ (N$^3$LO).  
Note that the polarizabilities of the nucleon are dominated by the
response of the pion cloud, while the magnetic moments are dominated by physics at the 
chiral symmetry breaking scale.  
Further the magnetic-moment contributions are suppressed by two powers of the nucleon mass.

There is an interesting difference between the meson and baryon FV modifications.  
As the nucleon mass is approximately seven times the pion mass, and twice the kaon mass, the recoil corrections 
are suppressed compared with those of the mesons. 
Further, the nucleons are significantly ``softer'' than the mesons, as evidenced by their polarizabilities.
However, the NLO recoil corrections to the proton mass 
are of approximately the same size as the N$^2$LO structure contributions, as seen in 
Fig.~\ref{fig:protonneutron}.

\section{Nuclei}
\noindent
A small number of LQCD collaborations have been calculating the binding of light nuclei and 
hypernuclei at unphysical light-quark masses in the isospin limit  and without 
QED~ \cite{Beane:2009py,Yamazaki:2009ua,Beane:2010hg,Inoue:2010es,Inoue:2011pg,Beane:2011iw,Yamazaki:2011nd,Yamazaki:2012hi,Yamazaki:2012fn,Beane:2012vq}.
However, it is known that 
as the atomic number of a nucleus increases, 
the Coulomb energy increases with the square of its charge, and   significantly reduces the 
binding of  large  nuclei.  
The simplest nucleus is the deuteron, but as it is weakly bound at the physical light-quark masses, and consequently unnaturally large, 
it is likely that it will be easier for LQCD collaborations 
to compute other light nuclei, such as $^4$He, rather than the deuteron.

\begin{figure}[!ht]
  \centering
     \includegraphics[scale=0.365]{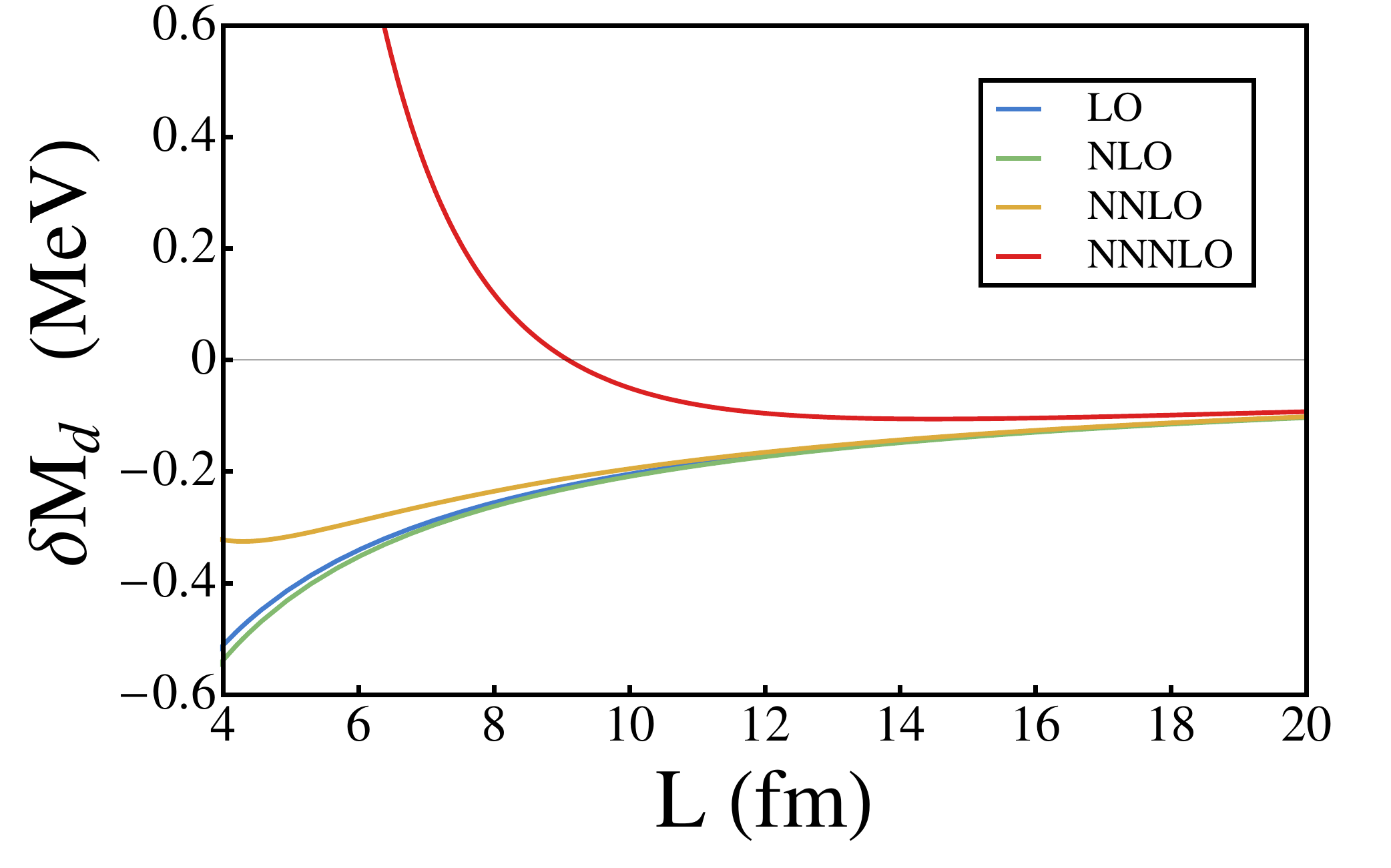}   \ \  
     \includegraphics[scale=0.365]{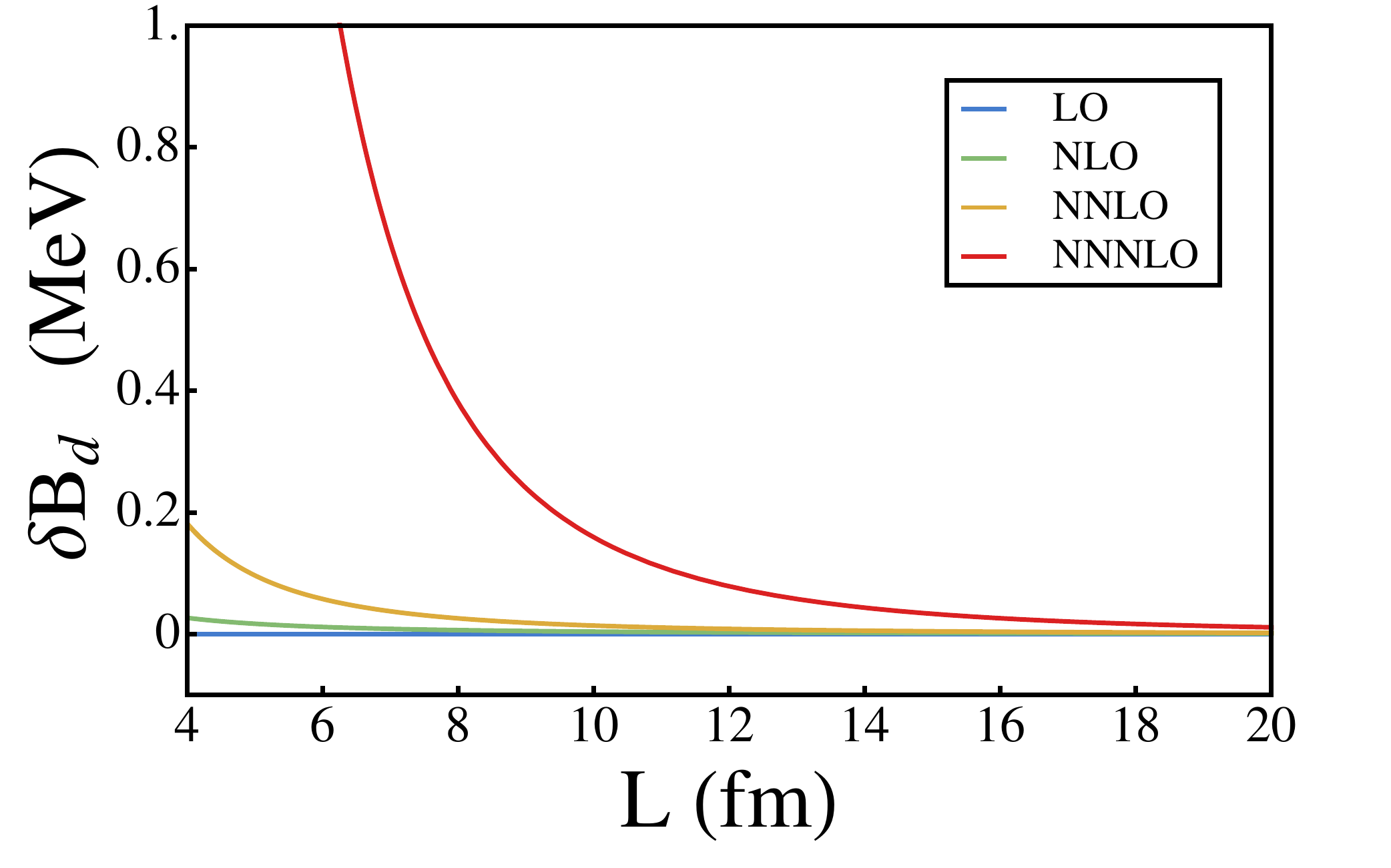}    
     \caption{The left panel shows the FV QED correction to the mass  of the deuteron at rest  in a FV at the physical pion mass.
     The leading contribution is from its electric charges, and scales as $1/{\rm L}$.
     The right panel shows the FV QED correction to the deuteron binding energy for which the  $1/{\rm L}$ contributions cancel.
     The $1\sigma$-uncertainty bands associated with each order in the expansion are determined from the uncertainties in the experimental and theoretical inputs.
         }
  \label{fig:deut}
\end{figure}

A NREFT for  vector QED shares the features of the NREFTs for scalars and fermions that are relevant for the current analysis. 
One difference is in the magnetic moment contribution,  and another is the contribution from the quadrupole interaction.
The FV corrections to the deuteron mass and binding energy, $\delta {\rm B}_d$, are shown in Fig.~\ref{fig:deut}, where the
experimentally determined charge radius, magnetic moment and polarizabilities have been used.
Due to the large size of the deuteron, and its large polarizability, the $1/{\rm L}$ expansion converges slowly in  modest volumes, and it
appears that ${\rm L}\gsim 12~{\rm fm}$ is required for a reliable determination of  the QED FV effects,
consistent with the size of volumes required to extract the binding and S-matrix parameters of the 
deuteron in the absence of QED~\cite{Briceno:2013bda}.
The QED FV corrections to the deuteron binding energy are seen to be significantly smaller than its total energy in large volumes, 
largely because the leading contribution to the deuteron and to the proton cancel.
As the deuteron has spin and parity of $J^\pi=1^+$, it also possesses a quadrupole moment which contributes to  the FV QED effects 
at ${\cal O}\left(1/{\rm L}^5\right)$ through two insertions.

\begin{figure}[!ht]
  \centering
     \includegraphics[scale=0.365]{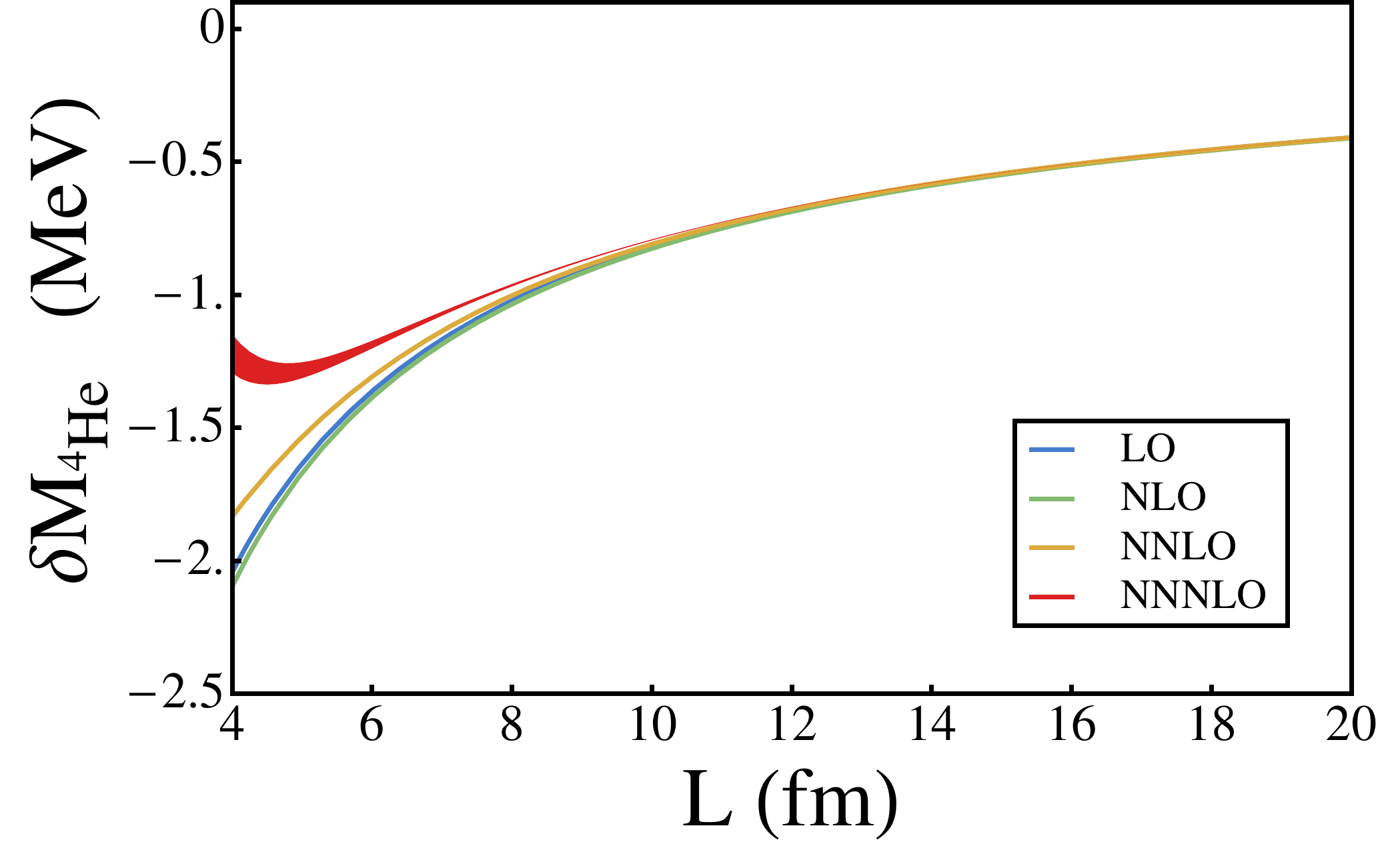}     \ \ 
     \includegraphics[scale=0.365]{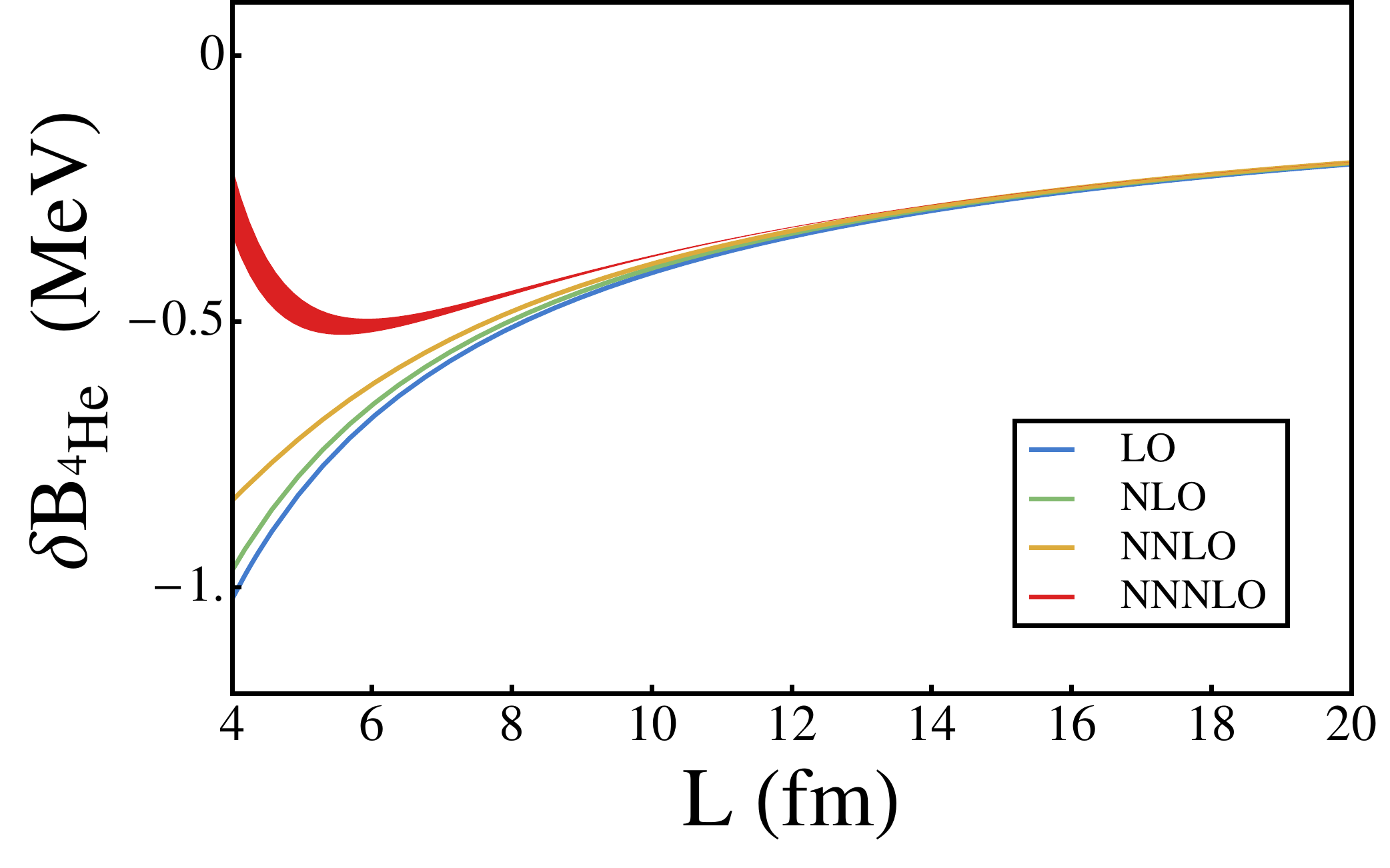}    
     \caption{The left panel shows the FV QED correction to the mass  of  $^4$He at rest in a FV
     at the physical pion  mass.
     The leading contribution is from its electric charge, and scales as $1/{\rm L}$.
     The right panel shows the FV QED correction to  the  $^4$He binding energy.
     The uncertainty bands associated with each order in the expansion are determined from the uncertainties in the experimental and theoretical inputs.
      }
  \label{fig:he4}
\end{figure}

The NREFTs used to study the FV contributions to the mass of the pions in the previous section also apply to the $^4$He nucleus,
and the FV corrections to  the mass of $^4$He and its binding energy, $\delta {\rm B}_{^4{\rm He}}$,  are shown in Fig.~\ref{fig:he4}.
Unlike the deuteron, the leading FV corrections to  $^4$He
do not cancel in the binding energy due to the interactions between the two protons, but are reduced by a factor of two.
%

\section{Anomalous Magnetic Moment of The Muon}
\noindent
Experimental and theoretical determinations of the
anomalous magnetic moment of the muon are providing  a stringent test of the Standard Model of particle physics. 
The current  discrepancy between the theoretical~\cite{Aoyama:2012wk,Gnendiger:2013pva} 
and experimental determinations~\cite{Bennett:2006fi},
at the level of $2.9$ to $3.6$ $\sigma$, but not $5\sigma$,
cannot yet be interpreted as a signal of new physics.
As  upcoming experiments, Fermilab  E989 and J-PARC E34, plan to reduce the 
experimental uncertainty down to 0.14 ppm,  
theoretical calculations of non-perturbative hadronic contributions must  be refined in the short term. 
LQCD is expected to contribute to improving the theoretical prediction of the standard model,
and several recent efforts have been directed 
at obtaining the hadronic vacuum-polarization 
and hadronic light-by-light contributions 
to the muon $g-2$~\cite{DellaMorte:2011aa,Feng:2011zk,Boyle:2011hu,Feng:2012gh,Blum:2013qu,Aubin:2013yba,Burger:2013jya,Aubin:2013daa,Blum:2013xva,Burger:2013jva,Aubin:2013oxa}. 
Theoretical challenges facing  these calculations have been identified 
and will be addressed during the next few years.

Here we show that the  most naive scheme to  obtain the magnetic moment of the muon 
by a direct calculation has volume effects that
scale as  $\mathcal{O}(\alpha_e/(m_{\mu}{\rm L}))$, requiring unrealistically large volumes to  achieve the  precision required to 
be sensitive to new physics. 
A detailed exploration of the issues related to extracting matrix elements of the electromagnetic current 
from LQCD calculations can be found in Ref.~\cite{Tiburzi:2007ep}.
Although it might appear that the leading contribution to the FV modification of the magnetic moment of the muon in NRQED
will arise from one-loop diagrams involving one insertion of the the magnetic moment operator, 
such contributions vanish. 
In fact, the leading $1/(m_{\mu}{\rm L})$ FV correction comes from the tree-level 
insertion of the magnetic-moment operator
multiplied by a factor of $E/m_{\mu}$, where $E$ is the energy of the muon, 
giving rise to, at ${\cal O}\left(\alpha_e\right)$, 
\begin{eqnarray}
\kappa_{\mu} \equiv \frac{g_{\mu}-2}{2}
\ =\ 
\frac{\alpha_e}{2\pi}\left[1+\frac{\pi c_1}{m_{\mu}{\rm L}}+\mathcal{O}\left(\frac{1}{m_{\mu}^2{\rm L}^2}\right)\right] 
\ \ .
\label{eq:kappa-mu}
\end{eqnarray}
The factor of $E/m_\mu$  arises in matching the NR theory to QED~\cite{Luke:1997ys},
in which each external leg in the NR theory must be accompanied  by a factor of $\sqrt{\frac{E}{m_{\mu}}}$.
Since $E=m_{\mu}+\frac{e^2}{8\pi}\frac{c_1}{{\rm L}}+\cdots$, it can be readily seen that the effective tree-level 
vertex multiplied by this normalization factor results in the $\kappa_{\mu}$  given in Eq. (\ref{eq:kappa-mu}). 
This contribution is present in the LO QED contribution to the anomalous magnetic moment 
(Schwinger term)
when calculated 
in a cubic FV  with PBCs and the photon zero mode removed.

To better understand the severity of the volume corrections to such a naive calculation, 
it is sufficient to note that in order  to 
reduce the FV correction to 1 ppm (comparable to the current  experimental error), 
 a volume of $\sim (60~{\rm nm})^3$ is required.
In the largest volumes that will be available to LQCD+QED calculations in the near future, with $L \lsim 10~{\rm fm} $,
the FV corrections to a direct calculation of the muon magnetic moment will flip the sign of the anomalous magnetic moment.
It is important to note that lattice practitioners are not attempting direct calculations of the  muon g-2, but rather are isolating the hadronic
contributions, which have enormously smaller FV effects compared to the one we have identified.  
In some sense, the result we have presented in this section is for entertainment purposes only.

\section{Lattice Artifacts}
\noindent
The results that we have presented in the previous sections have 
assumed a continuous spacetime, and have not yet considered the impact of a finite lattice spacing.
With the inclusion of QED, there are two distinct sources of lattice spacing artifacts that will modify the FV QED  corrections we have considered.
The coefficients of each of the higher dimension operators in the NREFTs will receive lattice spacing corrections, and 
for an ${\cal O}\left(a\right)$-improved action ($a$ is the lattice spacing) 
they are a polynomial in powers of $a$ of the form 
$d_i\sim d_{i0} +  d_{i2} a^2 + d_{i3} a^3 + ...$.
The coefficients $d_{ij}$ are determined by the strong interaction dynamics and the particular discretizations used in a given calculation.
In addition, the electromagnetic interaction will be modified in analogy with the strong sector, 
giving rise to further lattice spacing artifacts in the matching conditions between the full and the NR theories, and also in the value of one-loop 
diagrams~\footnote{The lattice artifacts will depend upon whether the compact or non-compact formulation of QED is employed - the former inducing
non-linearities in the electromagnetic field which vanish in the continuum limit.
The discussions we present in this section apply to both the compact and non-compact formulations.
}.  
For an improved action, the naive expectation is that such correction will first appear,
beyond the trivial correction from the modified hadron mass in the NLO term,
at
$\alpha_e a^2/{\rm L}^3$ in the $1/{\rm L}$ expansion.
They are a N$^2$LO contribution arising from
modifications to the one-loop  Coulomb self-energy diagram.
This is the same order as  contributions from the charge radius, recoil corrections and the magnetic moment, which are found to make a small contribution
to the mass shift in modest lattice volumes.  
As the lattice spacing is small compared to the size of the proton and the inverse mass of the proton, 
these lattice artifacts are expected to provide  a small modification to the N$^2$LO terms we have determined.
In addition, there are operators in the Symanzik action~\cite{Symanzik:1983dc,Symanzik:1983gh,Parisi:1985iv} 
that violate Lorentz symmetry as the calculations are performed on an underlying hypercubic grid.
Such operators require the contraction of at least four Lorentz vectors in order to form a hypercubically-invariant, but Lorentz-violating, operator,
for instance three derivatives and one electromagnetic field, or four derivatives. 
The suppression of Lorentz-violating contributions at small lattice spacings, along with smearing, 
has been discussed  in Ref.~\cite{Davoudi:2012ya}.

A second artifact  arises from the lattice volume.
The NREFTs are constructed as an expansion in derivatives acting on fields near their classical trajectory.  
As emphasized by Tiburzi~\cite{Tiburzi:2007ep} and others, this leads to modifications in calculated 
matrix elements because derivatives are approximated by finite differences in lattice calculations.  
For large momenta, this is a small effect because of the large density of states, but at low momenta,
particularly near zero, this can be a non-negligible effect that must be accounted for.  
This leads to a complication in determining, for instance,  magnetic moments from the forward limit of a form factor, relevant to the
discussion in the previous section.
However, this does not impact  the present calculations of FV QED corrections to the masses of the mesons, baryons and nuclei.

\section{Conclusions}
\noindent
For Lattice QCD calculations performed in 
volumes that are much larger than the inverse pion mass, 
the finite-volume electromagnetic corrections to hadron masses
 can be calculated systematically using a NREFT.
The leading two orders in the $1/{\rm L}$ expansion for mesons
have been previously calculated using chiral perturbation theory, and depend only upon their electric charge and  mass.  
We have shown that these two orders are universal FV QED corrections to the mass of charged particles.
Higher orders in the expansion are determined by 
recoil corrections and by the structure of the hadron, 
such as its electromagnetic multipole moments and polarizabilities, which we calculate using a NREFT.  
One advantage enjoyed by the NREFT  
is that the coefficients of the operators in the Lagrange density are directly related to the structure of the hadron, 
order by order in $\alpha_e$,
as opposed to being perturbative approximations as computed, for instance, in $\chi$PT.
For the mesons and baryons, the FV QED effects associated with their structure, beyond their charge, 
are  found to be small even in modest  lattice volumes.   
For nuclei, as long as the volume is large enough so that the non-QED effects are exponentially small, 
dictated by the nuclear radius,
their charge dominates the FV QED corrections, with only small modifications due to the structure of the nucleus.

\noindent
\subsection*{Acknowledgments}
We would like to thank Silas Beane, Michael Buchoff, Emmanuel Chang and Sanjay Reddy for interesting discussions.
We were supported in part by the DOE grant No. DE-FG02-97ER41014 and by 
DOE grant No. DE-FG02-00ER41132.

\bibliography{bibi}

\begin{thebibliography}{84}
\expandafter\ifx\csname natexlab\endcsname\relax\def\natexlab#1{#1}\fi
\expandafter\ifx\csname bibnamefont\endcsname\relax
  \def\bibnamefont#1{#1}\fi
\expandafter\ifx\csname bibfnamefont\endcsname\relax
  \def\bibfnamefont#1{#1}\fi
\expandafter\ifx\csname citenamefont\endcsname\relax
  \def\citenamefont#1{#1}\fi
\expandafter\ifx\csname url\endcsname\relax
  \def\url#1{\texttt{#1}}\fi
\expandafter\ifx\csname urlprefix\endcsname\relax\def\urlprefix{URL }\fi
\providecommand{\bibinfo}[2]{#2}
\providecommand{\eprint}[2][]{\url{#2}}

\bibitem[{\citenamefont{Aoki et~al.}(2010)}]{Aoki:2009ix}
\bibinfo{author}{\bibfnamefont{S.}~\bibnamefont{Aoki}} \bibnamefont{et~al.}
  (\bibinfo{collaboration}{PACS-CS Collaboration}),
  \bibinfo{journal}{Phys.Rev.} \textbf{\bibinfo{volume}{D81}},
  \bibinfo{pages}{074503} (\bibinfo{year}{2010}), \eprint{0911.2561}.

\bibitem[{\citenamefont{Durr et~al.}(2011)\citenamefont{Durr, Fodor, Hoelbling,
  Katz, Krieg et~al.}}]{Durr:2010vn}
\bibinfo{author}{\bibfnamefont{S.}~\bibnamefont{Durr}},
  \bibinfo{author}{\bibfnamefont{Z.}~\bibnamefont{Fodor}},
  \bibinfo{author}{\bibfnamefont{C.}~\bibnamefont{Hoelbling}},
  \bibinfo{author}{\bibfnamefont{S.}~\bibnamefont{Katz}},
  \bibinfo{author}{\bibfnamefont{S.}~\bibnamefont{Krieg}},
  \bibnamefont{et~al.}, \bibinfo{journal}{Phys.Lett.}
  \textbf{\bibinfo{volume}{B701}}, \bibinfo{pages}{265} (\bibinfo{year}{2011}),
  \eprint{1011.2403}.

\bibitem[{\citenamefont{Arthur et~al.}(2013)}]{Arthur:2012opa}
\bibinfo{author}{\bibfnamefont{R.}~\bibnamefont{Arthur}} \bibnamefont{et~al.}
  (\bibinfo{collaboration}{RBC Collaboration, UKQCD Collaboration}),
  \bibinfo{journal}{Phys.Rev.} \textbf{\bibinfo{volume}{D87}},
  \bibinfo{pages}{094514} (\bibinfo{year}{2013}), \eprint{1208.4412}.

\bibitem[{\citenamefont{Aoki et~al.}(2012)\citenamefont{Aoki, Ishikawa,
  Ishizuka, Kanaya, Kuramashi et~al.}}]{Aoki:2012st}
\bibinfo{author}{\bibfnamefont{S.}~\bibnamefont{Aoki}},
  \bibinfo{author}{\bibfnamefont{K.}~\bibnamefont{Ishikawa}},
  \bibinfo{author}{\bibfnamefont{N.}~\bibnamefont{Ishizuka}},
  \bibinfo{author}{\bibfnamefont{K.}~\bibnamefont{Kanaya}},
  \bibinfo{author}{\bibfnamefont{Y.}~\bibnamefont{Kuramashi}},
  \bibnamefont{et~al.}, \bibinfo{journal}{Phys.Rev.}
  \textbf{\bibinfo{volume}{D86}}, \bibinfo{pages}{034507}
  (\bibinfo{year}{2012}), \eprint{1205.2961}.

\bibitem[{\citenamefont{Durr et~al.}(2013)\citenamefont{Durr, Fodor, Hoelbling,
  Krieg, Kurth et~al.}}]{Durr:2013goa}
\bibinfo{author}{\bibfnamefont{S.}~\bibnamefont{Durr}},
  \bibinfo{author}{\bibfnamefont{Z.}~\bibnamefont{Fodor}},
  \bibinfo{author}{\bibfnamefont{C.}~\bibnamefont{Hoelbling}},
  \bibinfo{author}{\bibfnamefont{S.}~\bibnamefont{Krieg}},
  \bibinfo{author}{\bibfnamefont{T.}~\bibnamefont{Kurth}}, \bibnamefont{et~al.}
  (\bibinfo{year}{2013}), \eprint{1310.3626}.

\bibitem[{\citenamefont{Blum et~al.}(2007)\citenamefont{Blum, Doi, Hayakawa,
  Izubuchi, and Yamada}}]{Blum:2007cy}
\bibinfo{author}{\bibfnamefont{T.}~\bibnamefont{Blum}},
  \bibinfo{author}{\bibfnamefont{T.}~\bibnamefont{Doi}},
  \bibinfo{author}{\bibfnamefont{M.}~\bibnamefont{Hayakawa}},
  \bibinfo{author}{\bibfnamefont{T.}~\bibnamefont{Izubuchi}}, \bibnamefont{and}
  \bibinfo{author}{\bibfnamefont{N.}~\bibnamefont{Yamada}},
  \bibinfo{journal}{Phys.Rev.} \textbf{\bibinfo{volume}{D76}},
  \bibinfo{pages}{114508} (\bibinfo{year}{2007}), \eprint{0708.0484}.

\bibitem[{\citenamefont{Basak et~al.}(2008)}]{Basak:2008na}
\bibinfo{author}{\bibfnamefont{S.}~\bibnamefont{Basak}} \bibnamefont{et~al.}
  (\bibinfo{collaboration}{MILC Collaboration}), \bibinfo{journal}{PoS}
  \textbf{\bibinfo{volume}{LATTICE2008}}, \bibinfo{pages}{127}
  (\bibinfo{year}{2008}), \eprint{0812.4486}.

\bibitem[{\citenamefont{Blum et~al.}(2010)\citenamefont{Blum, Zhou, Doi,
  Hayakawa, Izubuchi et~al.}}]{Blum:2010ym}
\bibinfo{author}{\bibfnamefont{T.}~\bibnamefont{Blum}},
  \bibinfo{author}{\bibfnamefont{R.}~\bibnamefont{Zhou}},
  \bibinfo{author}{\bibfnamefont{T.}~\bibnamefont{Doi}},
  \bibinfo{author}{\bibfnamefont{M.}~\bibnamefont{Hayakawa}},
  \bibinfo{author}{\bibfnamefont{T.}~\bibnamefont{Izubuchi}},
  \bibnamefont{et~al.}, \bibinfo{journal}{Phys.Rev.}
  \textbf{\bibinfo{volume}{D82}}, \bibinfo{pages}{094508}
  (\bibinfo{year}{2010}), \eprint{1006.1311}.

\bibitem[{\citenamefont{Portelli et~al.}(2010)}]{Portelli:2010yn}
\bibinfo{author}{\bibfnamefont{A.}~\bibnamefont{Portelli}} \bibnamefont{et~al.}
  (\bibinfo{collaboration}{Budapest-Marseille-Wuppertal Collaboration}),
  \bibinfo{journal}{PoS} \textbf{\bibinfo{volume}{LATTICE2010}},
  \bibinfo{pages}{121} (\bibinfo{year}{2010}), \eprint{1011.4189}.

\bibitem[{\citenamefont{Portelli et~al.}(2011)\citenamefont{Portelli, Durr,
  Fodor, Frison, Hoelbling et~al.}}]{Portelli:2012pn}
\bibinfo{author}{\bibfnamefont{A.}~\bibnamefont{Portelli}},
  \bibinfo{author}{\bibfnamefont{S.}~\bibnamefont{Durr}},
  \bibinfo{author}{\bibfnamefont{Z.}~\bibnamefont{Fodor}},
  \bibinfo{author}{\bibfnamefont{J.}~\bibnamefont{Frison}},
  \bibinfo{author}{\bibfnamefont{C.}~\bibnamefont{Hoelbling}},
  \bibnamefont{et~al.}, \bibinfo{journal}{PoS}
  \textbf{\bibinfo{volume}{LATTICE2011}}, \bibinfo{pages}{136}
  (\bibinfo{year}{2011}), \eprint{1201.2787}.

\bibitem[{\citenamefont{de~Divitiis et~al.}(2013)\citenamefont{de~Divitiis,
  Frezzotti, Lubicz, Martinelli, Petronzio et~al.}}]{deDivitiis:2013xla}
\bibinfo{author}{\bibfnamefont{G.}~\bibnamefont{de~Divitiis}},
  \bibinfo{author}{\bibfnamefont{R.}~\bibnamefont{Frezzotti}},
  \bibinfo{author}{\bibfnamefont{V.}~\bibnamefont{Lubicz}},
  \bibinfo{author}{\bibfnamefont{G.}~\bibnamefont{Martinelli}},
  \bibinfo{author}{\bibfnamefont{R.}~\bibnamefont{Petronzio}},
  \bibnamefont{et~al.}, \bibinfo{journal}{Phys.Rev.}
  \textbf{\bibinfo{volume}{D87}}, \bibinfo{pages}{114505}
  (\bibinfo{year}{2013}), \eprint{1303.4896}.

\bibitem[{\citenamefont{Borsanyi et~al.}(2013)\citenamefont{Borsanyi, Durr,
  Fodor, Frison, Hoelbling et~al.}}]{Borsanyi:2013lga}
\bibinfo{author}{\bibfnamefont{S.}~\bibnamefont{Borsanyi}},
  \bibinfo{author}{\bibfnamefont{S.}~\bibnamefont{Durr}},
  \bibinfo{author}{\bibfnamefont{Z.}~\bibnamefont{Fodor}},
  \bibinfo{author}{\bibfnamefont{J.}~\bibnamefont{Frison}},
  \bibinfo{author}{\bibfnamefont{C.}~\bibnamefont{Hoelbling}},
  \bibnamefont{et~al.}, \bibinfo{journal}{Phys.Rev.Lett.}
  \textbf{\bibinfo{volume}{111}}, \bibinfo{pages}{252001}
  (\bibinfo{year}{2013}), \eprint{1306.2287}.

\bibitem[{\citenamefont{Drury et~al.}(2013)\citenamefont{Drury, Blum, Hayakawa,
  Izubuchi, Sachrajda et~al.}}]{Drury:2013sfa}
\bibinfo{author}{\bibfnamefont{S.}~\bibnamefont{Drury}},
  \bibinfo{author}{\bibfnamefont{T.}~\bibnamefont{Blum}},
  \bibinfo{author}{\bibfnamefont{M.}~\bibnamefont{Hayakawa}},
  \bibinfo{author}{\bibfnamefont{T.}~\bibnamefont{Izubuchi}},
  \bibinfo{author}{\bibfnamefont{C.}~\bibnamefont{Sachrajda}},
  \bibnamefont{et~al.} (\bibinfo{year}{2013}), \eprint{1312.0477}.

\bibitem[{\citenamefont{Borsanyi et~al.}(2014)\citenamefont{Borsanyi, Durr,
  Fodor, Hoelbling, Katz et~al.}}]{Borsanyi:2014jba}
\bibinfo{author}{\bibfnamefont{S.}~\bibnamefont{Borsanyi}},
  \bibinfo{author}{\bibfnamefont{S.}~\bibnamefont{Durr}},
  \bibinfo{author}{\bibfnamefont{Z.}~\bibnamefont{Fodor}},
  \bibinfo{author}{\bibfnamefont{C.}~\bibnamefont{Hoelbling}},
  \bibinfo{author}{\bibfnamefont{S.}~\bibnamefont{Katz}}, \bibnamefont{et~al.}
  (\bibinfo{year}{2014}), \eprint{1406.4088}.

\bibitem[{\citenamefont{Hilf and Polley}(1983)}]{Hilf1983412}
\bibinfo{author}{\bibfnamefont{E.}~\bibnamefont{Hilf}} \bibnamefont{and}
  \bibinfo{author}{\bibfnamefont{L.}~\bibnamefont{Polley}},
  \bibinfo{journal}{Physics Letters B} \textbf{\bibinfo{volume}{131}},
  \bibinfo{pages}{412 } (\bibinfo{year}{1983}), ISSN \bibinfo{issn}{0370-2693}.

\bibitem[{\citenamefont{Duncan et~al.}(1996)\citenamefont{Duncan, Eichten, and
  Thacker}}]{Duncan:1996xy}
\bibinfo{author}{\bibfnamefont{A.}~\bibnamefont{Duncan}},
  \bibinfo{author}{\bibfnamefont{E.}~\bibnamefont{Eichten}}, \bibnamefont{and}
  \bibinfo{author}{\bibfnamefont{H.}~\bibnamefont{Thacker}},
  \bibinfo{journal}{Phys.Rev.Lett.} \textbf{\bibinfo{volume}{76}},
  \bibinfo{pages}{3894} (\bibinfo{year}{1996}), \eprint{hep-lat/9602005}.

\bibitem[{\citenamefont{Hayakawa and Uno}(2008)}]{Hayakawa:2008an}
\bibinfo{author}{\bibfnamefont{M.}~\bibnamefont{Hayakawa}} \bibnamefont{and}
  \bibinfo{author}{\bibfnamefont{S.}~\bibnamefont{Uno}},
  \bibinfo{journal}{Prog.Theor.Phys.} \textbf{\bibinfo{volume}{120}},
  \bibinfo{pages}{413} (\bibinfo{year}{2008}), \eprint{0804.2044}.

\bibitem[{\citenamefont{Bardeen et~al.}(1989)\citenamefont{Bardeen, Bijnens,
  and G\'erard}}]{PhysRevLett.62.1343}
\bibinfo{author}{\bibfnamefont{W.~A.} \bibnamefont{Bardeen}},
  \bibinfo{author}{\bibfnamefont{J.}~\bibnamefont{Bijnens}}, \bibnamefont{and}
  \bibinfo{author}{\bibfnamefont{J.-M.} \bibnamefont{G\'erard}},
  \bibinfo{journal}{Phys. Rev. Lett.} \textbf{\bibinfo{volume}{62}},
  \bibinfo{pages}{1343} (\bibinfo{year}{1989}).

\bibitem[{\citenamefont{Isgur and Wise}(1989)}]{Isgur:1989vq}
\bibinfo{author}{\bibfnamefont{N.}~\bibnamefont{Isgur}} \bibnamefont{and}
  \bibinfo{author}{\bibfnamefont{M.~B.} \bibnamefont{Wise}},
  \bibinfo{journal}{Phys.Lett.} \textbf{\bibinfo{volume}{B232}},
  \bibinfo{pages}{113} (\bibinfo{year}{1989}).

\bibitem[{\citenamefont{Isgur and Wise}(1990)}]{Isgur:1989ed}
\bibinfo{author}{\bibfnamefont{N.}~\bibnamefont{Isgur}} \bibnamefont{and}
  \bibinfo{author}{\bibfnamefont{M.~B.} \bibnamefont{Wise}},
  \bibinfo{journal}{Phys.Lett.} \textbf{\bibinfo{volume}{B237}},
  \bibinfo{pages}{527} (\bibinfo{year}{1990}).

\bibitem[{\citenamefont{Jenkins and
  Manohar}(1991{\natexlab{a}})}]{Jenkins:1990jv}
\bibinfo{author}{\bibfnamefont{E.~E.} \bibnamefont{Jenkins}} \bibnamefont{and}
  \bibinfo{author}{\bibfnamefont{A.~V.} \bibnamefont{Manohar}},
  \bibinfo{journal}{Phys.Lett.} \textbf{\bibinfo{volume}{B255}},
  \bibinfo{pages}{558} (\bibinfo{year}{1991}{\natexlab{a}}).

\bibitem[{\citenamefont{Jenkins and
  Manohar}(1991{\natexlab{b}})}]{Jenkins:1991ne}
\bibinfo{author}{\bibfnamefont{E.~E.} \bibnamefont{Jenkins}} \bibnamefont{and}
  \bibinfo{author}{\bibfnamefont{A.~V.} \bibnamefont{Manohar}},
  \bibinfo{journal}{Effective field theories of the standard model.
  Proceedings, Workshop, Dobogoko, Hungary, August 22-26, 1991}
  (\bibinfo{year}{1991}{\natexlab{b}}).

\bibitem[{\citenamefont{Thacker and Lepage}(1991)}]{Thacker:1990bm}
\bibinfo{author}{\bibfnamefont{B.}~\bibnamefont{Thacker}} \bibnamefont{and}
  \bibinfo{author}{\bibfnamefont{G.~P.} \bibnamefont{Lepage}},
  \bibinfo{journal}{Phys.Rev.} \textbf{\bibinfo{volume}{D43}},
  \bibinfo{pages}{196} (\bibinfo{year}{1991}).

\bibitem[{\citenamefont{Labelle}(1992)}]{Labelle:1992hd}
\bibinfo{author}{\bibfnamefont{P.}~\bibnamefont{Labelle}}
  (\bibinfo{year}{1992}), \eprint{hep-ph/9209266}.

\bibitem[{\citenamefont{Manohar}(1997)}]{Manohar:1997qy}
\bibinfo{author}{\bibfnamefont{A.~V.} \bibnamefont{Manohar}},
  \bibinfo{journal}{Phys.Rev.} \textbf{\bibinfo{volume}{D56}},
  \bibinfo{pages}{230} (\bibinfo{year}{1997}), \eprint{hep-ph/9701294}.

\bibitem[{\citenamefont{Luke and Savage}(1998)}]{Luke:1997ys}
\bibinfo{author}{\bibfnamefont{M.~E.} \bibnamefont{Luke}} \bibnamefont{and}
  \bibinfo{author}{\bibfnamefont{M.~J.} \bibnamefont{Savage}},
  \bibinfo{journal}{Phys.Rev.} \textbf{\bibinfo{volume}{D57}},
  \bibinfo{pages}{413} (\bibinfo{year}{1998}), \eprint{hep-ph/9707313}.

\bibitem[{\citenamefont{Hill and Paz}(2011)}]{Hill:2011wy}
\bibinfo{author}{\bibfnamefont{R.~J.} \bibnamefont{Hill}} \bibnamefont{and}
  \bibinfo{author}{\bibfnamefont{G.}~\bibnamefont{Paz}},
  \bibinfo{journal}{Phys.Rev.Lett.} \textbf{\bibinfo{volume}{107}},
  \bibinfo{pages}{160402} (\bibinfo{year}{2011}), \eprint{1103.4617}.

\bibitem[{\citenamefont{Chen et~al.}(1999)\citenamefont{Chen, Rupak, and
  Savage}}]{Chen:1999tn}
\bibinfo{author}{\bibfnamefont{J.-W.} \bibnamefont{Chen}},
  \bibinfo{author}{\bibfnamefont{G.}~\bibnamefont{Rupak}}, \bibnamefont{and}
  \bibinfo{author}{\bibfnamefont{M.~J.} \bibnamefont{Savage}},
  \bibinfo{journal}{Nucl.Phys.} \textbf{\bibinfo{volume}{A653}},
  \bibinfo{pages}{386} (\bibinfo{year}{1999}), \eprint{nucl-th/9902056}.

\bibitem[{\citenamefont{Kaplan et~al.}(1998{\natexlab{a}})\citenamefont{Kaplan,
  Savage, and Wise}}]{Kaplan:1998tg}
\bibinfo{author}{\bibfnamefont{D.~B.} \bibnamefont{Kaplan}},
  \bibinfo{author}{\bibfnamefont{M.~J.} \bibnamefont{Savage}},
  \bibnamefont{and} \bibinfo{author}{\bibfnamefont{M.~B.} \bibnamefont{Wise}},
  \bibinfo{journal}{Phys.Lett.} \textbf{\bibinfo{volume}{B424}},
  \bibinfo{pages}{390} (\bibinfo{year}{1998}{\natexlab{a}}),
  \eprint{nucl-th/9801034}.

\bibitem[{\citenamefont{Kaplan et~al.}(1998{\natexlab{b}})\citenamefont{Kaplan,
  Savage, and Wise}}]{Kaplan:1998we}
\bibinfo{author}{\bibfnamefont{D.~B.} \bibnamefont{Kaplan}},
  \bibinfo{author}{\bibfnamefont{M.~J.} \bibnamefont{Savage}},
  \bibnamefont{and} \bibinfo{author}{\bibfnamefont{M.~B.} \bibnamefont{Wise}},
  \bibinfo{journal}{Nucl.Phys.} \textbf{\bibinfo{volume}{B534}},
  \bibinfo{pages}{329} (\bibinfo{year}{1998}{\natexlab{b}}),
  \eprint{nucl-th/9802075}.

\bibitem[{\citenamefont{Butler and Chen}(2000)}]{Butler:1999sv}
\bibinfo{author}{\bibfnamefont{M.}~\bibnamefont{Butler}} \bibnamefont{and}
  \bibinfo{author}{\bibfnamefont{J.-W.} \bibnamefont{Chen}},
  \bibinfo{journal}{Nucl.Phys.} \textbf{\bibinfo{volume}{A675}},
  \bibinfo{pages}{575} (\bibinfo{year}{2000}), \eprint{nucl-th/9905059}.

\bibitem[{\citenamefont{Butler et~al.}(2001)\citenamefont{Butler, Chen, and
  Kong}}]{Butler:2000zp}
\bibinfo{author}{\bibfnamefont{M.}~\bibnamefont{Butler}},
  \bibinfo{author}{\bibfnamefont{J.-W.} \bibnamefont{Chen}}, \bibnamefont{and}
  \bibinfo{author}{\bibfnamefont{X.}~\bibnamefont{Kong}},
  \bibinfo{journal}{Phys.Rev.} \textbf{\bibinfo{volume}{C63}},
  \bibinfo{pages}{035501} (\bibinfo{year}{2001}), \eprint{nucl-th/0008032}.

\bibitem[{\citenamefont{Butler and Chen}(2001)}]{Butler:2001jj}
\bibinfo{author}{\bibfnamefont{M.}~\bibnamefont{Butler}} \bibnamefont{and}
  \bibinfo{author}{\bibfnamefont{J.-W.} \bibnamefont{Chen}},
  \bibinfo{journal}{Phys.Lett.} \textbf{\bibinfo{volume}{B520}},
  \bibinfo{pages}{87} (\bibinfo{year}{2001}), \eprint{nucl-th/0101017}.

\bibitem[{\citenamefont{Martinelli et~al.}(1982)\citenamefont{Martinelli,
  Parisi, Petronzio, and Rapuano}}]{Martinelli:1982cb}
\bibinfo{author}{\bibfnamefont{G.}~\bibnamefont{Martinelli}},
  \bibinfo{author}{\bibfnamefont{G.}~\bibnamefont{Parisi}},
  \bibinfo{author}{\bibfnamefont{R.}~\bibnamefont{Petronzio}},
  \bibnamefont{and} \bibinfo{author}{\bibfnamefont{F.}~\bibnamefont{Rapuano}},
  \bibinfo{journal}{Phys.Lett.} \textbf{\bibinfo{volume}{B116}},
  \bibinfo{pages}{434} (\bibinfo{year}{1982}).

\bibitem[{\citenamefont{Fiebig et~al.}(1989)\citenamefont{Fiebig, Wilcox, and
  Woloshyn}}]{Fiebig:1988en}
\bibinfo{author}{\bibfnamefont{H.}~\bibnamefont{Fiebig}},
  \bibinfo{author}{\bibfnamefont{W.}~\bibnamefont{Wilcox}}, \bibnamefont{and}
  \bibinfo{author}{\bibfnamefont{R.}~\bibnamefont{Woloshyn}},
  \bibinfo{journal}{Nucl.Phys.} \textbf{\bibinfo{volume}{B324}},
  \bibinfo{pages}{47} (\bibinfo{year}{1989}).

\bibitem[{\citenamefont{Bernard et~al.}(1982)\citenamefont{Bernard, Draper,
  Olynyk, and Rushton}}]{Bernard:1982yu}
\bibinfo{author}{\bibfnamefont{C.~W.} \bibnamefont{Bernard}},
  \bibinfo{author}{\bibfnamefont{T.}~\bibnamefont{Draper}},
  \bibinfo{author}{\bibfnamefont{K.}~\bibnamefont{Olynyk}}, \bibnamefont{and}
  \bibinfo{author}{\bibfnamefont{M.}~\bibnamefont{Rushton}},
  \bibinfo{journal}{Phys.Rev.Lett.} \textbf{\bibinfo{volume}{49}},
  \bibinfo{pages}{1076} (\bibinfo{year}{1982}).

\bibitem[{\citenamefont{Lee et~al.}(2005)\citenamefont{Lee, Kelly, Zhou, and
  Wilcox}}]{Lee:2005ds}
\bibinfo{author}{\bibfnamefont{F.}~\bibnamefont{Lee}},
  \bibinfo{author}{\bibfnamefont{R.}~\bibnamefont{Kelly}},
  \bibinfo{author}{\bibfnamefont{L.}~\bibnamefont{Zhou}}, \bibnamefont{and}
  \bibinfo{author}{\bibfnamefont{W.}~\bibnamefont{Wilcox}},
  \bibinfo{journal}{Phys.Lett.} \textbf{\bibinfo{volume}{B627}},
  \bibinfo{pages}{71} (\bibinfo{year}{2005}), \eprint{hep-lat/0509067}.

\bibitem[{\citenamefont{Christensen et~al.}(2005)\citenamefont{Christensen,
  Wilcox, Lee, and Zhou}}]{Christensen:2004ca}
\bibinfo{author}{\bibfnamefont{J.~C.} \bibnamefont{Christensen}},
  \bibinfo{author}{\bibfnamefont{W.}~\bibnamefont{Wilcox}},
  \bibinfo{author}{\bibfnamefont{F.~X.} \bibnamefont{Lee}}, \bibnamefont{and}
  \bibinfo{author}{\bibfnamefont{L.-m.} \bibnamefont{Zhou}},
  \bibinfo{journal}{Phys.Rev.} \textbf{\bibinfo{volume}{D72}},
  \bibinfo{pages}{034503} (\bibinfo{year}{2005}), \eprint{hep-lat/0408024}.

\bibitem[{\citenamefont{Lee et~al.}(2006)\citenamefont{Lee, Zhou, Wilcox, and
  Christensen}}]{Lee:2005dq}
\bibinfo{author}{\bibfnamefont{F.~X.} \bibnamefont{Lee}},
  \bibinfo{author}{\bibfnamefont{L.}~\bibnamefont{Zhou}},
  \bibinfo{author}{\bibfnamefont{W.}~\bibnamefont{Wilcox}}, \bibnamefont{and}
  \bibinfo{author}{\bibfnamefont{J.~C.} \bibnamefont{Christensen}},
  \bibinfo{journal}{Phys.Rev.} \textbf{\bibinfo{volume}{D73}},
  \bibinfo{pages}{034503} (\bibinfo{year}{2006}), \eprint{hep-lat/0509065}.

\bibitem[{\citenamefont{Engelhardt}(2007)}]{Engelhardt:2007ub}
\bibinfo{author}{\bibfnamefont{M.}~\bibnamefont{Engelhardt}}
  (\bibinfo{collaboration}{LHPC Collaboration}), \bibinfo{journal}{Phys.Rev.}
  \textbf{\bibinfo{volume}{D76}}, \bibinfo{pages}{114502}
  (\bibinfo{year}{2007}), \eprint{0706.3919}.

\bibitem[{\citenamefont{Detmold et~al.}(2009)\citenamefont{Detmold, Tiburzi,
  and Walker-Loud}}]{Detmold:2009dx}
\bibinfo{author}{\bibfnamefont{W.}~\bibnamefont{Detmold}},
  \bibinfo{author}{\bibfnamefont{B.~C.} \bibnamefont{Tiburzi}},
  \bibnamefont{and}
  \bibinfo{author}{\bibfnamefont{A.}~\bibnamefont{Walker-Loud}},
  \bibinfo{journal}{Phys.Rev.} \textbf{\bibinfo{volume}{D79}},
  \bibinfo{pages}{094505} (\bibinfo{year}{2009}), \eprint{0904.1586}.

\bibitem[{\citenamefont{Alexandru and Lee}(2009)}]{Alexandru:2009id}
\bibinfo{author}{\bibfnamefont{A.}~\bibnamefont{Alexandru}} \bibnamefont{and}
  \bibinfo{author}{\bibfnamefont{F.~X.} \bibnamefont{Lee}},
  \bibinfo{journal}{PoS} \textbf{\bibinfo{volume}{LAT2009}},
  \bibinfo{pages}{144} (\bibinfo{year}{2009}), \eprint{0911.2520}.

\bibitem[{\citenamefont{Detmold et~al.}(2010)\citenamefont{Detmold, Tiburzi,
  and Walker-Loud}}]{Detmold:2010ts}
\bibinfo{author}{\bibfnamefont{W.}~\bibnamefont{Detmold}},
  \bibinfo{author}{\bibfnamefont{B.}~\bibnamefont{Tiburzi}}, \bibnamefont{and}
  \bibinfo{author}{\bibfnamefont{A.}~\bibnamefont{Walker-Loud}},
  \bibinfo{journal}{Phys.Rev.} \textbf{\bibinfo{volume}{D81}},
  \bibinfo{pages}{054502} (\bibinfo{year}{2010}), \eprint{1001.1131}.

\bibitem[{\citenamefont{Primer et~al.}(2013)\citenamefont{Primer, Kamleh,
  Leinweber, and Burkardt}}]{Primer:2013pva}
\bibinfo{author}{\bibfnamefont{T.}~\bibnamefont{Primer}},
  \bibinfo{author}{\bibfnamefont{W.}~\bibnamefont{Kamleh}},
  \bibinfo{author}{\bibfnamefont{D.}~\bibnamefont{Leinweber}},
  \bibnamefont{and} \bibinfo{author}{\bibfnamefont{M.}~\bibnamefont{Burkardt}}
  (\bibinfo{year}{2013}), \eprint{1307.1509}.

\bibitem[{\citenamefont{Lee and Tiburzi}(2013)}]{Lee:2013lxa}
\bibinfo{author}{\bibfnamefont{J.-W.} \bibnamefont{Lee}} \bibnamefont{and}
  \bibinfo{author}{\bibfnamefont{B.~C.} \bibnamefont{Tiburzi}}
  (\bibinfo{year}{2013}), \eprint{1312.3969}.

\bibitem[{\citenamefont{{Castro} et~al.}(2000)\citenamefont{{Castro}, {Rubio},
  and {Stott}}}]{2000physics..12024C}
\bibinfo{author}{\bibfnamefont{A.}~\bibnamefont{{Castro}}},
  \bibinfo{author}{\bibfnamefont{A.}~\bibnamefont{{Rubio}}}, \bibnamefont{and}
  \bibinfo{author}{\bibfnamefont{M.~J.} \bibnamefont{{Stott}}},
  \bibinfo{journal}{ArXiv Physics e-prints}  (\bibinfo{year}{2000}),
  \eprint{physics/0012024}.

\bibitem[{\citenamefont{Kronfeld and Wiese}(1993)}]{Kronfeld:1992ae}
\bibinfo{author}{\bibfnamefont{A.~S.} \bibnamefont{Kronfeld}} \bibnamefont{and}
  \bibinfo{author}{\bibfnamefont{U.}~\bibnamefont{Wiese}},
  \bibinfo{journal}{Nucl.Phys.} \textbf{\bibinfo{volume}{B401}},
  \bibinfo{pages}{190} (\bibinfo{year}{1993}), \eprint{hep-lat/9210008}.

\bibitem[{\citenamefont{Luscher}(1986)}]{Luscher:1986pf}
\bibinfo{author}{\bibfnamefont{M.}~\bibnamefont{Luscher}},
  \bibinfo{journal}{Commun.Math.Phys.} \textbf{\bibinfo{volume}{105}},
  \bibinfo{pages}{153} (\bibinfo{year}{1986}).

\bibitem[{\citenamefont{Hasenfratz and Leutwyler}(1990)}]{Hasenfratz:1989pk}
\bibinfo{author}{\bibfnamefont{P.}~\bibnamefont{Hasenfratz}} \bibnamefont{and}
  \bibinfo{author}{\bibfnamefont{H.}~\bibnamefont{Leutwyler}},
  \bibinfo{journal}{Nucl.Phys.} \textbf{\bibinfo{volume}{B343}},
  \bibinfo{pages}{241} (\bibinfo{year}{1990}).

\bibitem[{\citenamefont{Luscher}(1991)}]{Luscher:1990ux}
\bibinfo{author}{\bibfnamefont{M.}~\bibnamefont{Luscher}},
  \bibinfo{journal}{Nucl.Phys.} \textbf{\bibinfo{volume}{B354}},
  \bibinfo{pages}{531} (\bibinfo{year}{1991}).

\bibitem[{\citenamefont{Beane et~al.}(2008)\citenamefont{Beane, Detmold, Luu,
  Orginos, Savage et~al.}}]{Beane:2007es}
\bibinfo{author}{\bibfnamefont{S.~R.} \bibnamefont{Beane}},
  \bibinfo{author}{\bibfnamefont{W.}~\bibnamefont{Detmold}},
  \bibinfo{author}{\bibfnamefont{T.~C.} \bibnamefont{Luu}},
  \bibinfo{author}{\bibfnamefont{K.}~\bibnamefont{Orginos}},
  \bibinfo{author}{\bibfnamefont{M.~J.} \bibnamefont{Savage}},
  \bibnamefont{et~al.}, \bibinfo{journal}{Phys.Rev.Lett.}
  \textbf{\bibinfo{volume}{100}}, \bibinfo{pages}{082004}
  (\bibinfo{year}{2008}), \eprint{0710.1827}.

\bibitem[{\citenamefont{Beringer et~al.}(2012)}]{Beringer:1900zz}
\bibinfo{author}{\bibfnamefont{J.}~\bibnamefont{Beringer}} \bibnamefont{et~al.}
  (\bibinfo{collaboration}{Particle Data Group}), \bibinfo{journal}{Phys.Rev.}
  \textbf{\bibinfo{volume}{D86}}, \bibinfo{pages}{010001}
  (\bibinfo{year}{2012}).

\bibitem[{\citenamefont{Holstein and Scherer}(2013)}]{Holstein:2013kia}
\bibinfo{author}{\bibfnamefont{B.~R.} \bibnamefont{Holstein}} \bibnamefont{and}
  \bibinfo{author}{\bibfnamefont{S.}~\bibnamefont{Scherer}}
  (\bibinfo{year}{2013}), \eprint{1401.0140}.

\bibitem[{\citenamefont{Hill et~al.}(2013)\citenamefont{Hill, Lee, Paz, and
  Solon}}]{Hill:2012rh}
\bibinfo{author}{\bibfnamefont{R.~J.} \bibnamefont{Hill}},
  \bibinfo{author}{\bibfnamefont{G.}~\bibnamefont{Lee}},
  \bibinfo{author}{\bibfnamefont{G.}~\bibnamefont{Paz}}, \bibnamefont{and}
  \bibinfo{author}{\bibfnamefont{M.~P.} \bibnamefont{Solon}},
  \bibinfo{journal}{Phys.Rev.} \textbf{\bibinfo{volume}{D87}},
  \bibinfo{pages}{053017} (\bibinfo{year}{2013}), \eprint{1212.4508}.

\bibitem[{\citenamefont{Beane et~al.}(2010)}]{Beane:2009py}
\bibinfo{author}{\bibfnamefont{S.~R.} \bibnamefont{Beane}} \bibnamefont{et~al.}
  (\bibinfo{collaboration}{NPLQCD Collaboration}), \bibinfo{journal}{Phys.Rev.}
  \textbf{\bibinfo{volume}{D81}}, \bibinfo{pages}{054505}
  (\bibinfo{year}{2010}), \eprint{0912.4243}.

\bibitem[{\citenamefont{Yamazaki et~al.}(2010)\citenamefont{Yamazaki,
  Kuramashi, and Ukawa}}]{Yamazaki:2009ua}
\bibinfo{author}{\bibfnamefont{T.}~\bibnamefont{Yamazaki}},
  \bibinfo{author}{\bibfnamefont{Y.}~\bibnamefont{Kuramashi}},
  \bibnamefont{and} \bibinfo{author}{\bibfnamefont{A.}~\bibnamefont{Ukawa}}
  (\bibinfo{collaboration}{PACS-CS Collaboration}),
  \bibinfo{journal}{Phys.Rev.} \textbf{\bibinfo{volume}{D81}},
  \bibinfo{pages}{111504} (\bibinfo{year}{2010}), \eprint{0912.1383}.

\bibitem[{\citenamefont{Beane et~al.}(2011)}]{Beane:2010hg}
\bibinfo{author}{\bibfnamefont{S.}~\bibnamefont{Beane}} \bibnamefont{et~al.}
  (\bibinfo{collaboration}{NPLQCD Collaboration}),
  \bibinfo{journal}{Phys.Rev.Lett.} \textbf{\bibinfo{volume}{106}},
  \bibinfo{pages}{162001} (\bibinfo{year}{2011}), \eprint{1012.3812}.

\bibitem[{\citenamefont{Inoue et~al.}(2011)}]{Inoue:2010es}
\bibinfo{author}{\bibfnamefont{T.}~\bibnamefont{Inoue}} \bibnamefont{et~al.}
  (\bibinfo{collaboration}{HAL QCD Collaboration}),
  \bibinfo{journal}{Phys.Rev.Lett.} \textbf{\bibinfo{volume}{106}},
  \bibinfo{pages}{162002} (\bibinfo{year}{2011}), \eprint{1012.5928}.

\bibitem[{\citenamefont{Inoue}(2012)}]{Inoue:2011pg}
\bibinfo{author}{\bibfnamefont{T.}~\bibnamefont{Inoue}}
  (\bibinfo{collaboration}{HAL QCD Collaboration}), \bibinfo{journal}{AIP
  Conf.Proc.} \textbf{\bibinfo{volume}{1441}}, \bibinfo{pages}{335}
  (\bibinfo{year}{2012}), \eprint{1109.1620}.

\bibitem[{\citenamefont{Beane et~al.}(2012)}]{Beane:2011iw}
\bibinfo{author}{\bibfnamefont{S.}~\bibnamefont{Beane}} \bibnamefont{et~al.}
  (\bibinfo{collaboration}{NPLQCD Collaboration}), \bibinfo{journal}{Phys.Rev.}
  \textbf{\bibinfo{volume}{D85}}, \bibinfo{pages}{054511}
  (\bibinfo{year}{2012}), \eprint{1109.2889}.

\bibitem[{\citenamefont{Yamazaki et~al.}(2011)\citenamefont{Yamazaki,
  Kuramashi, and Ukawa}}]{Yamazaki:2011nd}
\bibinfo{author}{\bibfnamefont{T.}~\bibnamefont{Yamazaki}},
  \bibinfo{author}{\bibfnamefont{Y.}~\bibnamefont{Kuramashi}},
  \bibnamefont{and} \bibinfo{author}{\bibfnamefont{A.}~\bibnamefont{Ukawa}}
  (\bibinfo{collaboration}{Collaboration for the PACS-CS}),
  \bibinfo{journal}{Phys.Rev.} \textbf{\bibinfo{volume}{D84}},
  \bibinfo{pages}{054506} (\bibinfo{year}{2011}), \eprint{1105.1418}.

\bibitem[{\citenamefont{Yamazaki
  et~al.}(2012{\natexlab{a}})\citenamefont{Yamazaki, Ishikawa, Kuramashi, and
  Ukawa}}]{Yamazaki:2012hi}
\bibinfo{author}{\bibfnamefont{T.}~\bibnamefont{Yamazaki}},
  \bibinfo{author}{\bibfnamefont{K.-i.} \bibnamefont{Ishikawa}},
  \bibinfo{author}{\bibfnamefont{Y.}~\bibnamefont{Kuramashi}},
  \bibnamefont{and} \bibinfo{author}{\bibfnamefont{A.}~\bibnamefont{Ukawa}},
  \bibinfo{journal}{Phys.Rev.} \textbf{\bibinfo{volume}{D86}},
  \bibinfo{pages}{074514} (\bibinfo{year}{2012}{\natexlab{a}}),
  \eprint{1207.4277}.

\bibitem[{\citenamefont{Yamazaki
  et~al.}(2012{\natexlab{b}})\citenamefont{Yamazaki, Ishikawa, Kuramashi, and
  Ukawa}}]{Yamazaki:2012fn}
\bibinfo{author}{\bibfnamefont{T.}~\bibnamefont{Yamazaki}},
  \bibinfo{author}{\bibfnamefont{K.-i.} \bibnamefont{Ishikawa}},
  \bibinfo{author}{\bibfnamefont{Y.}~\bibnamefont{Kuramashi}},
  \bibnamefont{and} \bibinfo{author}{\bibfnamefont{A.}~\bibnamefont{Ukawa}},
  \bibinfo{journal}{PoS} \textbf{\bibinfo{volume}{LATTICE2012}},
  \bibinfo{pages}{143} (\bibinfo{year}{2012}{\natexlab{b}}),
  \eprint{1211.4334}.

\bibitem[{\citenamefont{Beane et~al.}(2013)\citenamefont{Beane, Chang, Cohen,
  Detmold, Lin et~al.}}]{Beane:2012vq}
\bibinfo{author}{\bibfnamefont{S.}~\bibnamefont{Beane}},
  \bibinfo{author}{\bibfnamefont{E.}~\bibnamefont{Chang}},
  \bibinfo{author}{\bibfnamefont{S.}~\bibnamefont{Cohen}},
  \bibinfo{author}{\bibfnamefont{W.}~\bibnamefont{Detmold}},
  \bibinfo{author}{\bibfnamefont{H.}~\bibnamefont{Lin}}, \bibnamefont{et~al.},
  \bibinfo{journal}{Phys.Rev.} \textbf{\bibinfo{volume}{D87}},
  \bibinfo{pages}{034506} (\bibinfo{year}{2013}), \eprint{1206.5219}.

\bibitem[{\citenamefont{Briceno et~al.}(2013)\citenamefont{Briceno, Davoudi,
  Luu, and Savage}}]{Briceno:2013bda}
\bibinfo{author}{\bibfnamefont{R.~A.} \bibnamefont{Briceno}},
  \bibinfo{author}{\bibfnamefont{Z.}~\bibnamefont{Davoudi}},
  \bibinfo{author}{\bibfnamefont{T.}~\bibnamefont{Luu}}, \bibnamefont{and}
  \bibinfo{author}{\bibfnamefont{M.~J.} \bibnamefont{Savage}},
  \bibinfo{journal}{Phys.Rev.} \textbf{\bibinfo{volume}{D88}},
  \bibinfo{pages}{114507} (\bibinfo{year}{2013}), \eprint{1309.3556}.

\bibitem[{\citenamefont{Aoyama et~al.}(2012)\citenamefont{Aoyama, Hayakawa,
  Kinoshita, and Nio}}]{Aoyama:2012wk}
\bibinfo{author}{\bibfnamefont{T.}~\bibnamefont{Aoyama}},
  \bibinfo{author}{\bibfnamefont{M.}~\bibnamefont{Hayakawa}},
  \bibinfo{author}{\bibfnamefont{T.}~\bibnamefont{Kinoshita}},
  \bibnamefont{and} \bibinfo{author}{\bibfnamefont{M.}~\bibnamefont{Nio}},
  \bibinfo{journal}{Phys.Rev.Lett.} \textbf{\bibinfo{volume}{109}},
  \bibinfo{pages}{111808} (\bibinfo{year}{2012}), \eprint{1205.5370}.

\bibitem[{\citenamefont{Gnendiger et~al.}(2013)\citenamefont{Gnendiger,
  Stockinger, and Stockinger-Kim}}]{Gnendiger:2013pva}
\bibinfo{author}{\bibfnamefont{C.}~\bibnamefont{Gnendiger}},
  \bibinfo{author}{\bibfnamefont{D.}~\bibnamefont{Stockinger}},
  \bibnamefont{and}
  \bibinfo{author}{\bibfnamefont{H.}~\bibnamefont{Stockinger-Kim}},
  \bibinfo{journal}{Phys.Rev.} \textbf{\bibinfo{volume}{D88}},
  \bibinfo{pages}{053005} (\bibinfo{year}{2013}), \eprint{1306.5546}.

\bibitem[{\citenamefont{Bennett et~al.}(2006)}]{Bennett:2006fi}
\bibinfo{author}{\bibfnamefont{G.}~\bibnamefont{Bennett}} \bibnamefont{et~al.}
  (\bibinfo{collaboration}{Muon G-2 Collaboration}),
  \bibinfo{journal}{Phys.Rev.} \textbf{\bibinfo{volume}{D73}},
  \bibinfo{pages}{072003} (\bibinfo{year}{2006}), \eprint{hep-ex/0602035}.

\bibitem[{\citenamefont{Della~Morte et~al.}(2012)\citenamefont{Della~Morte,
  Jager, Juttner, and Wittig}}]{DellaMorte:2011aa}
\bibinfo{author}{\bibfnamefont{M.}~\bibnamefont{Della~Morte}},
  \bibinfo{author}{\bibfnamefont{B.}~\bibnamefont{Jager}},
  \bibinfo{author}{\bibfnamefont{A.}~\bibnamefont{Juttner}}, \bibnamefont{and}
  \bibinfo{author}{\bibfnamefont{H.}~\bibnamefont{Wittig}},
  \bibinfo{journal}{JHEP} \textbf{\bibinfo{volume}{1203}}, \bibinfo{pages}{055}
  (\bibinfo{year}{2012}), \eprint{1112.2894}.

\bibitem[{\citenamefont{Feng et~al.}(2011)\citenamefont{Feng, Jansen,
  Petschlies, and Renner}}]{Feng:2011zk}
\bibinfo{author}{\bibfnamefont{X.}~\bibnamefont{Feng}},
  \bibinfo{author}{\bibfnamefont{K.}~\bibnamefont{Jansen}},
  \bibinfo{author}{\bibfnamefont{M.}~\bibnamefont{Petschlies}},
  \bibnamefont{and} \bibinfo{author}{\bibfnamefont{D.~B.}
  \bibnamefont{Renner}}, \bibinfo{journal}{Phys.Rev.Lett.}
  \textbf{\bibinfo{volume}{107}}, \bibinfo{pages}{081802}
  (\bibinfo{year}{2011}), \eprint{1103.4818}.

\bibitem[{\citenamefont{Boyle et~al.}(2012)\citenamefont{Boyle, Del~Debbio,
  Kerrane, and Zanotti}}]{Boyle:2011hu}
\bibinfo{author}{\bibfnamefont{P.}~\bibnamefont{Boyle}},
  \bibinfo{author}{\bibfnamefont{L.}~\bibnamefont{Del~Debbio}},
  \bibinfo{author}{\bibfnamefont{E.}~\bibnamefont{Kerrane}}, \bibnamefont{and}
  \bibinfo{author}{\bibfnamefont{J.}~\bibnamefont{Zanotti}},
  \bibinfo{journal}{Phys.Rev.} \textbf{\bibinfo{volume}{D85}},
  \bibinfo{pages}{074504} (\bibinfo{year}{2012}), \eprint{1107.1497}.

\bibitem[{\citenamefont{Feng et~al.}(2012)\citenamefont{Feng, Hotzel, Jansen,
  Petschlies, and Renner}}]{Feng:2012gh}
\bibinfo{author}{\bibfnamefont{X.}~\bibnamefont{Feng}},
  \bibinfo{author}{\bibfnamefont{G.}~\bibnamefont{Hotzel}},
  \bibinfo{author}{\bibfnamefont{K.}~\bibnamefont{Jansen}},
  \bibinfo{author}{\bibfnamefont{M.}~\bibnamefont{Petschlies}},
  \bibnamefont{and} \bibinfo{author}{\bibfnamefont{D.~B.}
  \bibnamefont{Renner}}, \bibinfo{journal}{PoS}
  \textbf{\bibinfo{volume}{LATTICE2012}}, \bibinfo{pages}{174}
  (\bibinfo{year}{2012}), \eprint{1211.0828}.

\bibitem[{\citenamefont{Blum et~al.}(2012)\citenamefont{Blum, Hayakawa, and
  Izubuchi}}]{Blum:2013qu}
\bibinfo{author}{\bibfnamefont{T.}~\bibnamefont{Blum}},
  \bibinfo{author}{\bibfnamefont{M.}~\bibnamefont{Hayakawa}}, \bibnamefont{and}
  \bibinfo{author}{\bibfnamefont{T.}~\bibnamefont{Izubuchi}},
  \bibinfo{journal}{PoS} \textbf{\bibinfo{volume}{LATTICE2012}},
  \bibinfo{pages}{022} (\bibinfo{year}{2012}), \eprint{1301.2607}.

\bibitem[{\citenamefont{Aubin et~al.}(2013{\natexlab{a}})\citenamefont{Aubin,
  Blum, Golterman, Maltman, and Peris}}]{Aubin:2013yba}
\bibinfo{author}{\bibfnamefont{C.}~\bibnamefont{Aubin}},
  \bibinfo{author}{\bibfnamefont{T.}~\bibnamefont{Blum}},
  \bibinfo{author}{\bibfnamefont{M.}~\bibnamefont{Golterman}},
  \bibinfo{author}{\bibfnamefont{K.}~\bibnamefont{Maltman}}, \bibnamefont{and}
  \bibinfo{author}{\bibfnamefont{S.}~\bibnamefont{Peris}}
  (\bibinfo{year}{2013}{\natexlab{a}}), \eprint{1311.5504}.

\bibitem[{\citenamefont{Burger et~al.}(2013{\natexlab{a}})\citenamefont{Burger,
  Feng, Hotzel, Jansen, Petschlies et~al.}}]{Burger:2013jya}
\bibinfo{author}{\bibfnamefont{F.}~\bibnamefont{Burger}},
  \bibinfo{author}{\bibfnamefont{X.}~\bibnamefont{Feng}},
  \bibinfo{author}{\bibfnamefont{G.}~\bibnamefont{Hotzel}},
  \bibinfo{author}{\bibfnamefont{K.}~\bibnamefont{Jansen}},
  \bibinfo{author}{\bibfnamefont{M.}~\bibnamefont{Petschlies}},
  \bibnamefont{et~al.} (\bibinfo{year}{2013}{\natexlab{a}}),
  \eprint{1308.4327}.

\bibitem[{\citenamefont{Aubin et~al.}(2013{\natexlab{b}})\citenamefont{Aubin,
  Blum, Golterman, and Peris}}]{Aubin:2013daa}
\bibinfo{author}{\bibfnamefont{C.}~\bibnamefont{Aubin}},
  \bibinfo{author}{\bibfnamefont{T.}~\bibnamefont{Blum}},
  \bibinfo{author}{\bibfnamefont{M.}~\bibnamefont{Golterman}},
  \bibnamefont{and} \bibinfo{author}{\bibfnamefont{S.}~\bibnamefont{Peris}},
  \bibinfo{journal}{Phys.Rev.} \textbf{\bibinfo{volume}{D88}},
  \bibinfo{pages}{074505} (\bibinfo{year}{2013}{\natexlab{b}}),
  \eprint{1307.4701}.

\bibitem[{\citenamefont{Blum et~al.}(2013)\citenamefont{Blum, Denig,
  Logashenko, de~Rafael, Lee~Roberts et~al.}}]{Blum:2013xva}
\bibinfo{author}{\bibfnamefont{T.}~\bibnamefont{Blum}},
  \bibinfo{author}{\bibfnamefont{A.}~\bibnamefont{Denig}},
  \bibinfo{author}{\bibfnamefont{I.}~\bibnamefont{Logashenko}},
  \bibinfo{author}{\bibfnamefont{E.}~\bibnamefont{de~Rafael}},
  \bibinfo{author}{\bibfnamefont{B.}~\bibnamefont{Lee~Roberts}},
  \bibnamefont{et~al.} (\bibinfo{year}{2013}), \eprint{1311.2198}.

\bibitem[{\citenamefont{Burger et~al.}(2013{\natexlab{b}})\citenamefont{Burger,
  Feng, Hotzel, Jansen, Petschlies et~al.}}]{Burger:2013jva}
\bibinfo{author}{\bibfnamefont{F.}~\bibnamefont{Burger}},
  \bibinfo{author}{\bibfnamefont{X.}~\bibnamefont{Feng}},
  \bibinfo{author}{\bibfnamefont{G.}~\bibnamefont{Hotzel}},
  \bibinfo{author}{\bibfnamefont{K.}~\bibnamefont{Jansen}},
  \bibinfo{author}{\bibfnamefont{M.}~\bibnamefont{Petschlies}},
  \bibnamefont{et~al.}, \bibinfo{journal}{PoS}
  \textbf{\bibinfo{volume}{LATTICE2013}}, \bibinfo{pages}{301}
  (\bibinfo{year}{2013}{\natexlab{b}}), \eprint{1311.3885}.

\bibitem[{\citenamefont{Aubin et~al.}(2013{\natexlab{c}})\citenamefont{Aubin,
  Blum, Golterman, and Peris}}]{Aubin:2013oxa}
\bibinfo{author}{\bibfnamefont{C.}~\bibnamefont{Aubin}},
  \bibinfo{author}{\bibfnamefont{T.}~\bibnamefont{Blum}},
  \bibinfo{author}{\bibfnamefont{M.}~\bibnamefont{Golterman}},
  \bibnamefont{and} \bibinfo{author}{\bibfnamefont{S.}~\bibnamefont{Peris}}
  (\bibinfo{year}{2013}{\natexlab{c}}), \eprint{1311.1078}.

\bibitem[{\citenamefont{Tiburzi}(2008)}]{Tiburzi:2007ep}
\bibinfo{author}{\bibfnamefont{B.~C.} \bibnamefont{Tiburzi}},
  \bibinfo{journal}{Phys.Rev.} \textbf{\bibinfo{volume}{D77}},
  \bibinfo{pages}{014510} (\bibinfo{year}{2008}), \eprint{0710.3577}.

\bibitem[{\citenamefont{Symanzik}(1983{\natexlab{a}})}]{Symanzik:1983dc}
\bibinfo{author}{\bibfnamefont{K.}~\bibnamefont{Symanzik}},
  \bibinfo{journal}{Nucl.Phys.} \textbf{\bibinfo{volume}{B226}},
  \bibinfo{pages}{187} (\bibinfo{year}{1983}{\natexlab{a}}).

\bibitem[{\citenamefont{Symanzik}(1983{\natexlab{b}})}]{Symanzik:1983gh}
\bibinfo{author}{\bibfnamefont{K.}~\bibnamefont{Symanzik}},
  \bibinfo{journal}{Nucl.Phys.} \textbf{\bibinfo{volume}{B226}},
  \bibinfo{pages}{205} (\bibinfo{year}{1983}{\natexlab{b}}).

\bibitem[{\citenamefont{Parisi}(1985)}]{Parisi:1985iv}
\bibinfo{author}{\bibfnamefont{G.}~\bibnamefont{Parisi}},
  \bibinfo{journal}{Nucl.Phys.} \textbf{\bibinfo{volume}{B254}},
  \bibinfo{pages}{58} (\bibinfo{year}{1985}).

\bibitem[{\citenamefont{Davoudi and Savage}(2012)}]{Davoudi:2012ya}
\bibinfo{author}{\bibfnamefont{Z.}~\bibnamefont{Davoudi}} \bibnamefont{and}
  \bibinfo{author}{\bibfnamefont{M.~J.} \bibnamefont{Savage}},
  \bibinfo{journal}{Phys.Rev.} \textbf{\bibinfo{volume}{D86}},
  \bibinfo{pages}{054505} (\bibinfo{year}{2012}), \eprint{1204.4146}.

\end{thebibliography}

\end{document}